\documentclass[twocolumn,dvipsnames,twocolappendix]{aastex631}
\usepackage{amsmath,amsfonts,amssymb,graphicx,chngcntr,multirow, float,booktabs}
\usepackage[]{hyperref}
\hypersetup{colorlinks=true}
\usepackage{threeparttablex}
\usepackage{wasysym}
\usepackage{arydshln}
\usepackage{mathrsfs}
\usepackage{amssymb}
\usepackage{verbatim}

\newcommand{\Msun}{M$_\odot$}

\newcommand{\cgsflux}{erg\,s$^{-1}$\,cm$^{-2}$}

\newcommand{\pfrac}[2]{\left( \frac{#1}{#2} \right)}

\begin{document}

\title{Even a precessing clock is right twice per orbit - The super-periods of eRO-QPE2 and challenges for quasi-periodic eruption orbital models}

\author[0000-0003-4054-7978]{R. Arcodia}\thanks{NASA Einstein Fellow}
\affiliation{Kavli Institute for Astrophysics and Space Research, Massachusetts Institute of Technology, Cambridge, MA 02139, USA}
\affiliation{Black Hole Initiative at Harvard University, 20 Garden Street, Cambridge, MA 02138, USA}

\author[0000-0003-0707-4531]{G. Miniutti}
\affiliation{Centro de Astrobiolog\'ia (CAB), CSIC-INTA, Camino Bajo del Castillo s/n, 28692 Villanueva de la Ca\~nada, Madrid, Spain}

\author[0000-0002-0568-6000]{J. Chakraborty}
\affiliation{Kavli Institute for Astrophysics and Space Research, Massachusetts Institute of Technology, Cambridge, MA 02139, USA}

\author[0000-0002-8400-0969]{A. Franchini}
\affiliation{Dipartimento di Fisica, Università degli Studi di Milano, Via Celoria 16, 20133 Milano, Italy}

\author{M. Giustini}
\affiliation{Centro de Astrobiolog\'ia (CAB), CSIC-INTA, Camino Bajo del Castillo s/n, 28692 Villanueva de la Ca\~nada, Madrid, Spain}

\author[0000-0002-8304-1988]{I. Linial}\thanks{NASA Einstein Fellow}
\affiliation{Department of Physics and Columbia Astrophysics Laboratory, Columbia University, New York, NY 10027, USA}
\affiliation{Center for Cosmology and Particle Physics, Physics Department, New York University, New York, NY 10003, USA}

\author{A. Mummery}
\affiliation{School of Natural Sciences, Institute for Advanced Study, 1 Einstein Drive, Princeton, NJ 08540, USA}

\author{L. Bertassi}
\affiliation{Universit\`a degli Studi di Milano-Bicocca, Piazza della Scienza 3, I-20126 Milano, Italy}
\affiliation{INFN, Sezione di Milano-Bicocca, Piazza della Scienza 3, I-20126 Milano, Italy}
\affiliation{INAF - Osservatorio Astronomico di Brera, via Brera 20, I-20121 Milano, Italy}

\author{M. Bonetti}
\affiliation{Universit\`a degli Studi di Milano-Bicocca, Piazza della Scienza 3, I-20126 Milano, Italy}
\affiliation{INFN, Sezione di Milano-Bicocca, Piazza della Scienza 3, I-20126 Milano, Italy}

\author{E. Kara}
\affiliation{Kavli Institute for Astrophysics and Space Research, Massachusetts Institute of Technology, Cambridge, MA 02139, USA}

\author{A. Merloni}
\affiliation{Max-Planck-Institut für extraterrestrische Physik (MPE), Gießenbachstraße 1, 85748 Garching bei München, Germany}

\author[0009-0008-6234-6062]{A. Motta}
\affiliation{Universit\`a degli Studi di Milano-Bicocca, Piazza della Scienza 3, I-20126 Milano, Italy}

\author{G. Ponti}
\affiliation{INAF-Osservatorio Astronomico di Brera, Via E. Bianchi 46, I-23807 Merate (LC), Italy}
\affiliation{Max-Planck-Institut für extraterrestrische Physik (MPE), Gießenbachstraße 1, 85748 Garching bei München, Germany}
\affiliation{Como Lake Center for Astrophysics (CLAP), DiSAT, Università degli Studi dell’Insubria, via Valleggio 11, I-22100 Como, Italy}

\author{E. Quintin}
\affiliation{European Space Astronomy Centre (ESAC), European Space Agency (ESA), Madrid, Spain}

\author{R. Soria}
\affiliation{INAF-Osservatorio Astrofisico di Torino, Strada Osservatorio 20, I-10025 Pino Torinese, Italy}

\author{P. Baldini}
\affiliation{Max-Planck-Institut für extraterrestrische Physik (MPE), Gießenbachstraße 1, 85748 Garching bei München, Germany}

\author{J. Buchner}
\affiliation{Max-Planck-Institut für extraterrestrische Physik (MPE), Gießenbachstraße 1, 85748 Garching bei München, Germany}

\author{M. Dotti}
\affiliation{Universit\`a degli Studi di Milano-Bicocca, Piazza della Scienza 3, I-20126 Milano, Italy}
\affiliation{INFN, Sezione di Milano-Bicocca, Piazza della Scienza 3, I-20126 Milano, Italy}
\affiliation{INAF - Osservatorio Astronomico di Brera, via Brera 20, I-20121 Milano, Italy}

\author[0000-0002-5786-186X]{P. C. Fragile}
\affiliation{Department of Physics and Astronomy, College of Charleston, 66 George Street, Charleston, SC 29424, USA}

\author[0000-0002-5311-9078]{A. Ingram}
\affiliation{School of Mathematics, Statistics, and Physics, Newcastle University, Newcastle upon Tyne NE1 7RU, UK}

\author{M. Middleton}
\affiliation{School of Physics and Astronomy, University of Southampton, University Road, Southampton SO17 1BJ, UK}

\author{C. Panagiotou}
\affiliation{Kavli Institute for Astrophysics and Space Research, Massachusetts Institute of Technology, Cambridge, MA 02139, USA}

\author{A. Sesana}
\affiliation{Universit\`a degli Studi di Milano-Bicocca, Piazza della Scienza 3, I-20126 Milano, Italy}
\affiliation{INFN, Sezione di Milano-Bicocca, Piazza della Scienza 3, I-20126 Milano, Italy}
\affiliation{INAF - Osservatorio Astronomico di Brera, via Brera 20, I-20121 Milano, Italy}

\author{P. Yao}
\affiliation{Department of Astrophysical Sciences, Princeton University, Peyton Hall, Princeton, NJ 08544, USA}

\author{A. Rau}
\affiliation{Max-Planck-Institut für extraterrestrische Physik (MPE), Gießenbachstraße 1, 85748 Garching bei München, Germany}

\author{F. M.  Vincentelli}
\affiliation{Fluid and Complex Systems Centre, Coventry University, CV1 5FB, UK}
\affiliation{School of Physics and Astronomy, University of Southampton, University Road, Southampton SO17 1BJ, UK}

\author[0000-0002-5063-0751]{M. Guolo}
\affiliation{Bloomberg Center for Physics and Astronomy, Johns Hopkins University, 3400 N. Charles St., Baltimore, MD 21218, USA}

\author{R. Saxton}
\affiliation{European Space Astronomy Centre (ESAC), European Space Agency (ESA), Madrid, Spain}



\begin{abstract}

We present O$-$C (``observed minus calculated'') timing analysis of the quasi-periodic eruption (QPE) source eRO-QPE2 with a multi-mission X-ray campaign, which includes 32 observed eruptions spanning a month (i.e. 325 QPE cycles). In relation to accretion (e.g. disk instability) models, the O-C is consistent with a damped random walk of the QPE recurrence, albeit with highly uncertain parameters. If instead an underlying orbital clock is present, eRO-QPE2 is consistent with a period of $P \sim 2.24$\,h and two hierarchical super-periodic modulations, with periods of $\sim 4.4$\,d ($\sim47$\,P) and $\approx 95$\,d ($\approx 1000$\,P). We found no negative period derivative, with $|\dot{P}| \lesssim 2 \times 10^{-6}$\,s/s at $3\sigma$. Interpreting this as a limit on GW decay disfavors white dwarfs on very high-eccentricity orbits, and intermediate-mass black holes at high mass and/or eccentricity. For disk-collision models, where the $\dot{P}$ from gas drag and the QPE integrated energy provide a higher and lower bound to the local disk density, a main-sequence star is disfavored as EMRI secondary unless stellar debris streams are present, while stripped stars remain allowed. The correlated odd and even O-C data disfavor scenarios in which the two disk crossings per orbit are both observed. Interpreting the data with one \emph{observed} event per orbit, the short modulation is consistent with apsidal precession for $a \sim 140\,R_g$, $e \approx 0.1$, and $M_{\rm BH} \approx 1.5 \times 10^{5}\,M_\odot$. The longer modulation (much less constrained) is inconsistent with EMRI nodal precession and disk precession is allowed for a limited parameter volume, while there is a solution with a stable hierarchical triple system with an outer massive black hole at $\sim 0.4\,\mathrm{mpc}$ and mass $\sim(0.1-1) \times M_{\rm BH}$. However, no reliable solution can be found with more robust EMRI trajectory models, perhaps due to a multi-dimensional model with narrow likelihood peaks applied to relatively sparse data. 
\end{abstract}

\keywords{}


\section{Introduction}
\label{sec:intro}

Quasi-periodic eruptions (QPEs) are soft X-ray flares observed from galactic nuclei evolving on the timescales of hours to days \citep{Miniutti+2019:qpe1,Giustini+2020:qpe2,Arcodia+2021:eroqpes,Arcodia+2024:eroqpes,Chakraborty+2021:qpe5cand,Chakraborty+2025:upj,Quintin+2023:tormund,Bykov+2024:tormund,Nicholl+2024:qiz,Hernandez-Garcia+2025:ansky,Arcodia+2025:ero5,Baldini+2026:J2344}. When in quiescence, the emission from an accretion disk is detected \citep{Nicholl+2024:qiz,Wevers+2025:ero2hst,Guolo+2025:sed,Guolo+2025:gsnlongterm} which, at least for some sources, is due to a previous tidal disruption event \citep[TDE; e.g.,][]{Quintin+2023:tormund,Bykov+2024:tormund,Nicholl+2024:qiz,Chakraborty+2025:upj,Guolo+2025:gsnlongterm}. The eruptions detected on top of this continuum show a characteristic spectral evolution with a harder rise compared to the flare decay \citep{Arcodia+2022:ero1_timing,Miniutti+2023:gsnrebr,Arcodia+2024:eroqpes,Chakraborty+2025:upj,Nicholl+2024:qiz,Hernandez-Garcia+2025:ansky}. While the average duty cycle is around $\approx10-20\%$ across all sources, their timing properties are somewhat diverse in terms of the scatter around the average QPE peak-to-peak recurrence time ($t_{\rm recur}$) and flare amplitude. 

The presence of an underlying quasi-periodic clock is however ubiquitous and reminiscent of an orbital phenomenon, thus inspiring many related ideas. Among the various orbital models\footnote{But see \citet{Pan+2023:instab,Kaur+2023:Bdisks}, for some examples of accretion disk instability models.}, QPEs have been modeled as binary self-lensing \citep{Ingram+2021:lensing}, accretion disk tearing \citep{Raj+2021:tearing}, Lense-Thirring precession in a super-Eddington flow \citep{Middleton+2025:LT}, partial tidal disruptions (pTDEs) from high-eccentricity white dwarfs \citep[WDs; e.g.,][]{King2020:WD,Wang+2022:wd,Chen+2023:wd} or from mildly eccentric stellar orbits \citep{Linial+2023:transfer,Lu+2023:transfer}, quasi-circular pTDEs in a circumbinary disk \citep{D'Orazio+2025:cbdtde}, or disk spiral density waves excited by an orbiter \citep{Dodd+2025:waves}. More prominently, QPEs have been modeled as a stellar-mass object in a low- or moderate-eccentricity orbit crossing the accretion disk around the primary massive black hole (MBH), generally twice per orbit \citep{Xian+2021:collisions,Linial+2023:qpemodel,Franchini+2023:qpemodel,Tagawa+2023:qpemodel,Zhou+2024:qpemodel}. Many of these interpretations imply that QPEs are the electromagnetic precursor or counterpart of extreme mass-ratio inspirals (EMRIs) which, depending on the precise nature of the orbiter, may be detectable by the recently-adopted Laser Interferometer Space Antenna \citep[LISA,][]{2024arXiv240207571C} and TianQin \citep{Luo2016:tianqin}, or by future $\mu$-Hz gravitational wave detectors (\citealp{Sesana+2021:muAres}; see also \citealp{Suzuguchi+2026:qpegw,Zhan+2026:qpegw}). 

The latest developments of these disk collision models suggest that flares may either result from shocked disk gas ejected during the collision \citep[e.g.,][]{Linial+2023:qpemodel,Franchini+2023:qpemodel} or by the interaction between stellar debris streams and the disk \citep[e.g.,][]{Linial+2025:streams,Yao+2025:simul}. The latter emission mechanism may solve the tension of the elevated integrated energy per burst seen in some QPE sources, which is higher than what a stellar-mass orbiter could eject piercing through the disk \citep{Chakraborty+2025:spec,Mummery+2025:collisions,Guo+2025:coll}. In general, most current QPE disk collision models predict that eruptions would be almost equally observable both during ingress into and egress out of the disk \citep[but see][]{Linial+2025:streams}, although radiation hydrodynamic simulation show very asymmetric ejecta and significantly different luminosities (\citealp{Huang+2025:simul,Jankovic2026:sims}; but see \citealp{Lam+2025:IMBHscoll}). Having two observable events per orbit implies that the orbital period of the system is traced by $\sim 2 t_{\rm recur}$ (i.e., by the separation between eruptions of the same even/odd parity), which provides an estimate of the orbit size if the black hole mass is known. Thus, this model predicts that the inferred orbit size is smaller than the accretion disk, which is an important consistency check passed at least for the QPE sources AT2019qiz, GSN\,069, AT2022upj, and eRO-QPE2 \citep{Nicholl+2024:qiz,Guolo+2025:sed,Wevers+2025:ero2hst,Chakraborty+2025:upj}. However, in GSN\,069 the accretion disk was estimated to be large enough already in 2014 data when QPEs were not observed \citep{Guolo+2025:gsnlongterm}, thus leaving the observed appearance and disappearance of QPEs as a function of disk flux currently unexplained by QPE disk collision models \citep[but see, e.g.,][]{Franchini+2023:qpemodel,Miniutti+2023:alive}.

In the proposed EMRI orbital models, the observed eruptions would trace the orbital evolution of the system, which may open new avenues for BH parameter inference such as BH mass ($M_{\rm BH}$) and spin ($\chi$; as discussed in more detail in \citealp{Chakraborty+2025:qpefit,Zhou+2025:mass}). Some first attempts have been made using EMRI trajectory models, finding plausible solutions in both single epochs and multi-epoch observations of QPE sources \citep[e.g.,][]{Zhou+2024:longterm}. More empirical data analysis tests have also been made to study the quasi-periodicity of QPE sources in the context of EMRI models \citep[e.g.,][]{Chakraborty+2024:ero1,Arcodia+2024:ero2_ticking,Pasham+2024:ero2,Miniutti+2025:OC}. In particular, \citet{Miniutti+2025:OC} studied time delays of the eruptions in GSN\,069 and found that delays for odd and even eruptions are correlated, while a key prediction of EMRI models is that apsidal precession would induce an anti-correlation (by imprinting a negative delay to the ingress collision and a positive one to the egress collision with respect to the observer, and vice versa). Nonetheless, the sparse X-ray data did not allow  \citet{Miniutti+2025:OC} to draw unambiguous conclusions on the timing behavior of the source. In summary, while the EMRI disk collision models have revealed a promising way forward for the QPE field, no single flavor is able to explain all observables of all QPE sources \citep[e.g.,][]{Mummery+2025:collisions,Guo+2025:coll,Miniutti+2025:OC}.

The goal of this work is to improve on the current state-of-the-art of QPE timing campaigns and on their interpretation with current orbital models. For this, we designed and carried out a multi-mission X-ray campaign of eRO-QPE2, which has been the most regular QPE source discovered to date in terms of its timing and spectral properties \citep{Arcodia+2024:ero2_ticking,Pasham+2024:ero2}. We collected data with \emph{XMM-Newton}, \emph{Swift}/XRT, NICER and Einstein Probe, with the main observing baseline being enclosed within four \emph{XMM-Newton} observations spanning about a month in July 2024 and ancillary data taken more sparsely until December 2024 (see Section~\ref{sec:data}). We performed O$-$C (``observed minus calculated'') timing analysis (Section~\ref{sec:OC}) and comparisons against existing orbital models outlining their successes and failures in Section~\ref{sec:discussion} and~\ref{sec:disc_EMRImodels}.

\section{X-ray analysis and results}
\label{sec:data}

eRO-QPE2 has been observed by XMM-Newton \citep[][hereafter XMM]{Jansen+2001:xmm} several times starting from the discovery of QPEs in the system in 2020 \citep{Arcodia+2021:eroqpes}. This work focuses on a campaign performed by XMM in June-July 2024 (PI: Arcodia) designed to test quantitatively the periodic behavior of eRO-QPE2. This campaign includes four XMM observations spanning about a month of baseline (IDs 0941270101 to -401, taken on 28 June, 4 July, 16 July, and 28 July 2024, respectively), referred to hereafter as ‘XMM1’, ‘XMM2’, ‘XMM3’, and ‘XMM4’, respectively. In addition, due to XMM4 being incomplete compared to the requested exposure, part of it was rescheduled on 5 December 2024 (ID -501, hereafter `XMM5'), thus more observations were later added in between XMM4 and XMM5. The total exposure (with the pn camera) for XMM1-4 is $\sim134.4\,$ks, with $7.2\,$ks more in XMM5. We show XMM1-4 observations with their 19 observed eruptions in Fig.~\ref{fig:xmm1-4_lcs}, and report more details on the processing in Appendix~\ref{sec:app_processing}. For comparison with timing properties, we extracted and fitted XMM quiescence spectra at all epochs following precisely the analysis reported in \citet{Arcodia+2024:ero2_ticking}, to which we refer for more details.

\begin{figure*}[t]
     \includegraphics[width=\textwidth]{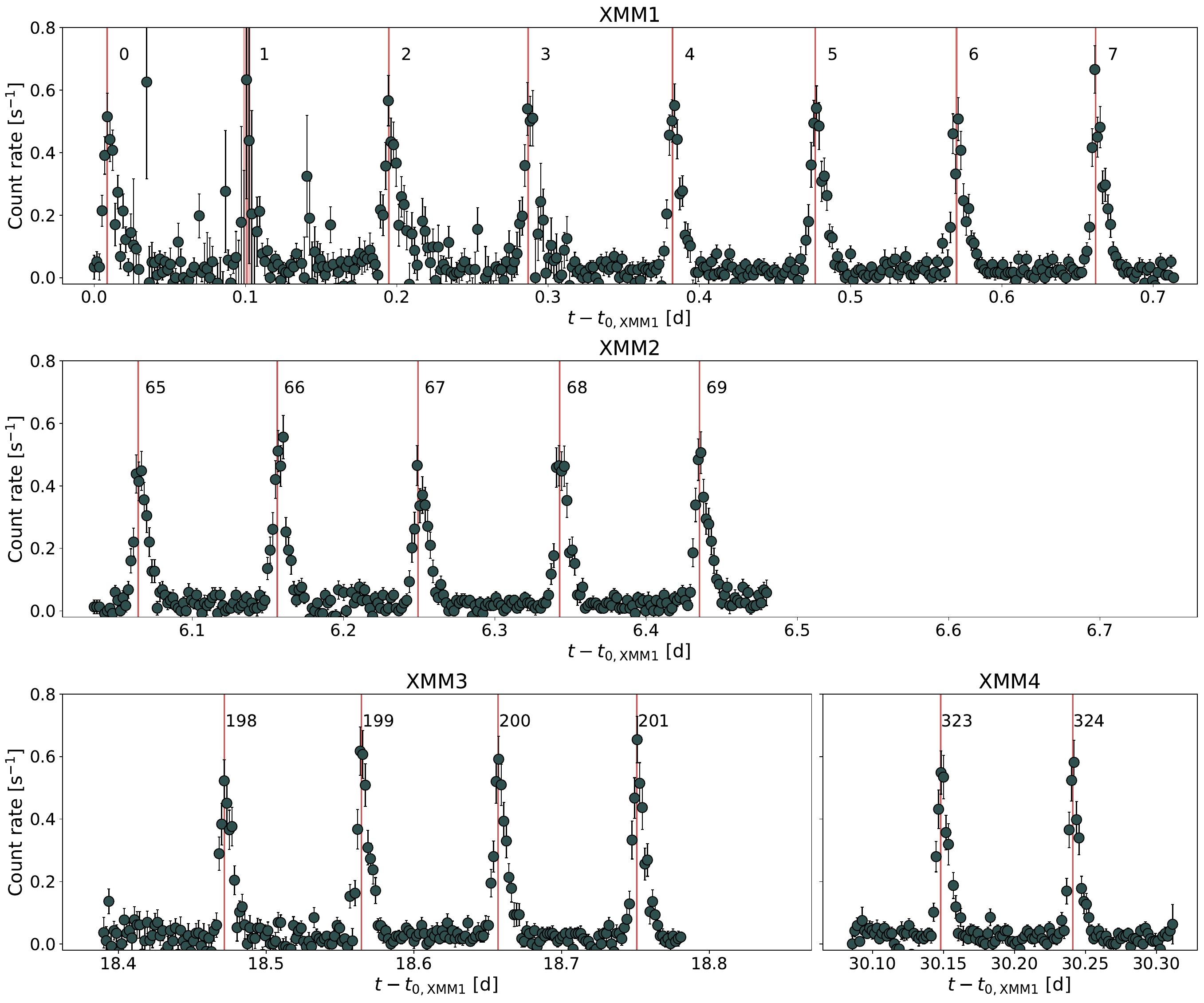}
     \caption{X-ray light curves of the four \emph{XMM-Newton} epochs of the main campaign, with the elapsed time relative to the start of the first observation ($t_{\rm 0,XMM1}$), which corresponds to $\rm MJD=60489.6419$. Red lines and contours show the eruption peak times and $1\sigma$ uncertainties fitted with an eruption parametric model. We show the identification number $N_{\rm QPE}$ (see Table~\ref{tab:OCdata}).}
     \label{fig:xmm1-4_lcs}
\end{figure*}

\begin{figure*}
     \includegraphics[width=\textwidth]{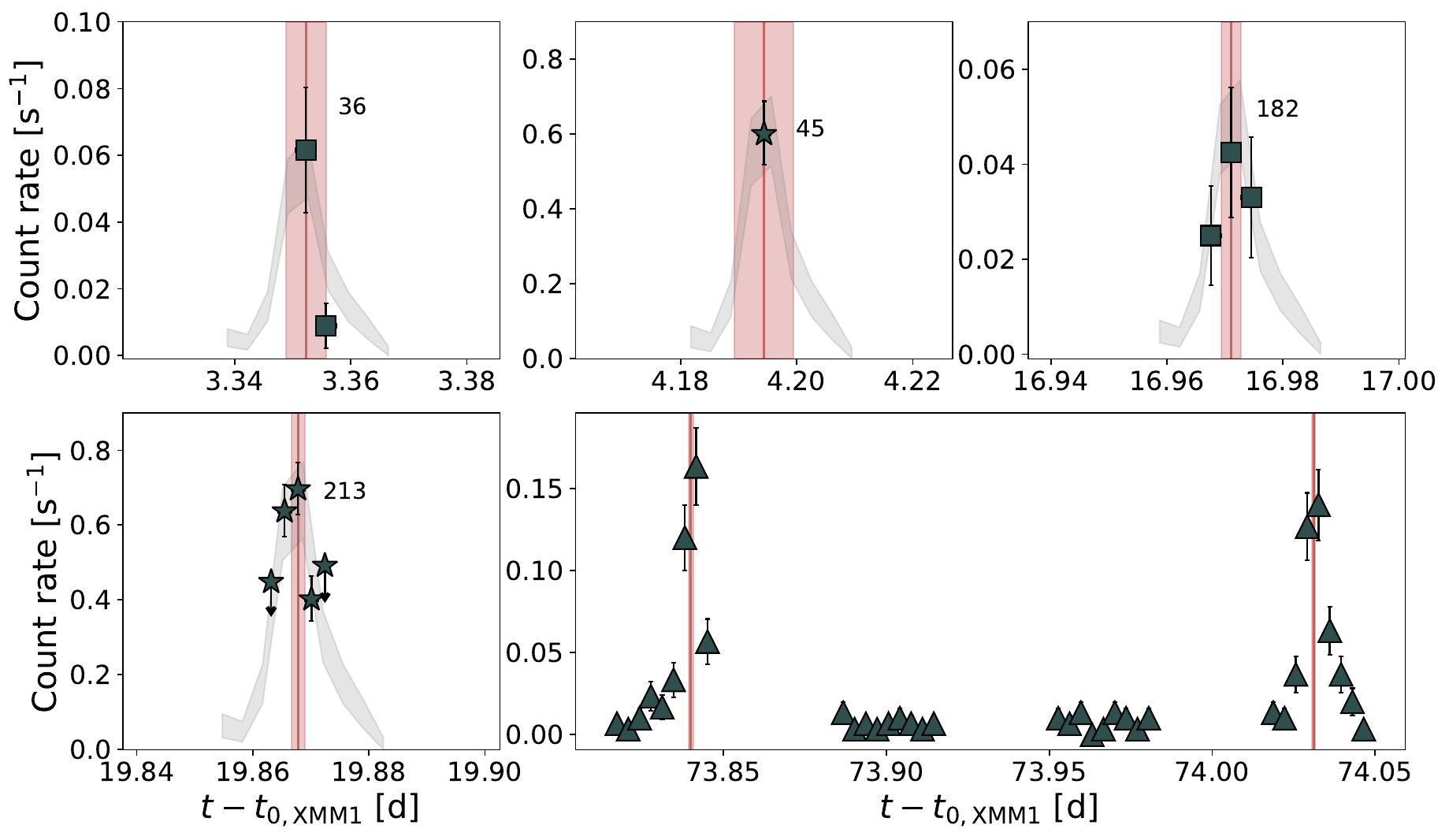}
     \caption{X-ray light curves of illustrative NICER and XRT eruptions (top panels, bottom left panel), and the EP observation (bottom right panel). Different instruments are shown with different symbols: squares for XRT (top left, top right), stars for NICER (top middle, bottom left), triangles for EP (bottom right). As in Fig.~\ref{fig:xmm1-4_lcs}, the QPE arrival time (red line, with shaded regions representing $1\sigma$ uncertainties, see Appendix~\ref{sec:app_processing}) and identification numbers are shown (see Table~\ref{tab:OCdata}). 
     The shaded gray profile is not a fit, but shows a silhouette of the second eruption observed by EP (renormalized arbitrarily) to guide the eye.}
     \label{fig:xmm1-4_nicerxrt_lcs}
\end{figure*}

In addition to XMM data, we scheduled Target of Opportunity observations with Swift-XRT (hereafter XRT; PI: Arcodia), NICER (PI: Arcodia) and Einstein Probe FXT (hereafter EP; PI: Rau), throughout the XMM1-4 campaign and beyond, for a total of $\sim 42.8$\,ks, $\sim 81.3$\,ks and $\sim 11.9$\,ks, for XRT, NICER, and EP, respectively. We report details on data processing in Appendix~\ref{sec:app_processing}. We show examples of XRT, NICER and EP data in Fig.~\ref{fig:xmm1-4_nicerxrt_lcs}. In general, adding these ancillary datasets results in the detection of 13 eruptions peaks during the XMM1-4 campaign (5 with XRT, 8 with NICER), plus an additional 6 between XMM4 and XMM5 (1 with XRT, 3 with NICER, 2 with EP, to add to the 2 observed in XMM5). Thus, the total number of eruptions observed during the XMM1-4 campaign is 32 (while extending to XMM5 adds 8 more). Finally, we note that for all light curves we performed a barycenter correction with DE405 ephemeris.

\subsection{X-ray eruptions arrival times} 

We estimated the arrival time of QPEs from their peak, but we have also extracted the start time of QPEs and verified a posteriori that this choice does not impact any of our results (see Appendix~\ref{sec:app_OC}). For XMM light curves, we estimated the peak arrival time fitting each eruption with a parametric double-exponential model, which has proven effective in modeling the fast rise and slow decay of QPE sources \citep[e.g.,][]{Arcodia+2022:ero1_timing,Arcodia+2024:eroqpes,Chakraborty+2024:ero1,Hernandez-Garcia+2025:ansky}. For eRO-QPE2, while the bursts may appear symmetric due to their shorter duration and lower signal-to-noise compared to other QPE sources, their phase-folded profile shows the same asymmetry (see Fig. A.1 in \citealp{Arcodia+2024:ero2_ticking}). We visually inspected all fits to ensure adequate residuals were obtained for all eruptions, and performed more simulations and tests to confirm that the parametric model choice for the QPE eruption profiles has no impact on our timing solutions (see Appendix~\ref{sec:app_OC}). For EP data, since the eruptions are fully resolved within the $\sim3\,$ks exposures (Fig.~\ref{fig:xmm1-4_nicerxrt_lcs}), we adopt the same fitting method applied to XMM data. For XRT and NICER eruptions, we note that their typical continuous snapshots are $\approx500-1000\,$s. Eruptions in eRO-QPE2 only last $\sim1700\,$s from rise to decay based on XMM data \citep{Arcodia+2024:ero2_ticking}, and since the eruptions absorbed (i.e. observed) peak flux of $F_{0.5-2.0\, \rm keV} \sim 4 \times 10^{13}\,$\cgsflux  ~\citep{Arcodia+2024:ero2_ticking} is only marginally above detectability for both XRT and NICER, the detectable time for XRT and NICER reduces even further. In addition, the eruption recurrence time of $\sim2.3\,$h \citep{Arcodia+2024:ero2_ticking} is also compatible with the orbital gaps of $\sim 1.5-2.0\,$h for XRT and NICER, thus detecting and identifying an eruption peak is much more challenging. However, the higher flux limit comes to our advantage, as any XRT or NICER snapshot detection would be conveniently associated to being close to the peak (e.g., $\gtrsim 0.02$\,c/s for XRT), while the wings of the eruptions can only be marginally detected (e.g., $\lesssim 0.006\,$c/s for XRT), and the quiescence is not detectable at all (e.g., $\lesssim 0.002\,$c/s for XRT). We report all the details of our XRT and NICER analyses to estimate the QPE peaks in Appendix~\ref{sec:app_OC}. We show the estimated arrival times for all instruments in Fig.~\ref{fig:xmm1-4_lcs} and~\ref{fig:xmm1-4_nicerxrt_lcs} with red vertical lines and related contours for their $1\sigma$ uncertainties. 


\section{O-C analysis}
\label{sec:OC}



We performed the O-C analysis, which compares the residuals between the observed arrival times (`O') with those computed (`C') starting from a reference event ($T_{\rm 0,est}$) and assuming a constant period ($P_{\rm est}$, estimated from available data) as a function of the elapsed number of events $N_{\rm QPE}$. This method \citep[e.g.,][]{Sterken2005:OC} has been widely used for stellar systems and binaries \citep[e.g.,][]{Zasche+2009:oc,Hajdu+2015:oc} and, more recently, also for QPEs \citep{Chakraborty+2024:ero1,Chakraborty+26:ansky,Miniutti+2025:OC} and other repeating nuclear transients \citep{Payne+2021:14ko}. In its simplest form, namely for a strictly periodic system, the O-C residuals are linear with a slope given by the `inaccuracy' of the estimated period $P_{\rm est}$ compared to the true period $P$, and a constant of the order of a small offset between the estimated start of the reference event and the true one: $O-C= (T_0-T_{0,\rm est})+ (P-P_{\rm est}) N_{\rm QPE}$. Most quasi-periodic systems reveal additional residuals to this linear trend which, once the noise model is known (see Sect.~\ref{sec:OC_noise}), may be modeled with extra terms. For instance, a quadratic term due to a change in the true period, adding or subtracting $\pm(1/2\,\dot{\rm P}\,P)N_{\rm QPE}^2$, or sinusoidal modulations, adding $A_{\mathrm{mod}} \sin(2\pi N_{\rm QPE}/P_{\mathrm{mod}} + \phi_{\mathrm{mod}})$, where $P_{\mathrm{mod}}$ is defined in dimensionless units of $P$. 
For reference, for our XMM1-4 dataset the association with $N_{\rm QPE}$ was performed with $T_{0,\rm est}$ taken from the fitted peak time of the first eruption of XMM1 (i.e., $732.69\,$s after the start of XMM1, $\rm MJD=60489.6419$, which is the origin of the time axis, e.g. Fig.~\ref{fig:xmm1-4_lcs}), and $P_{\rm est}$ taken from the average peak-to-peak QPE recurrence time of the XMM1, XMM2, XMM3 and XMM4 observations (i.e., $8055.71$\,s). We report more details on the event identification with these ephemeris in Appendix~\ref{sec:app_OC}, which we stress is not ambiguous due to the XMM1-4 campaign being designed with this analysis in mind. We show the resulting O-C data in Fig.~\ref{fig:OConly} (and report the values in Table~\ref{tab:OCdata} in Appendix~\ref{sec:app_OC}). 

\begin{figure}
    \includegraphics[width=0.99\columnwidth]{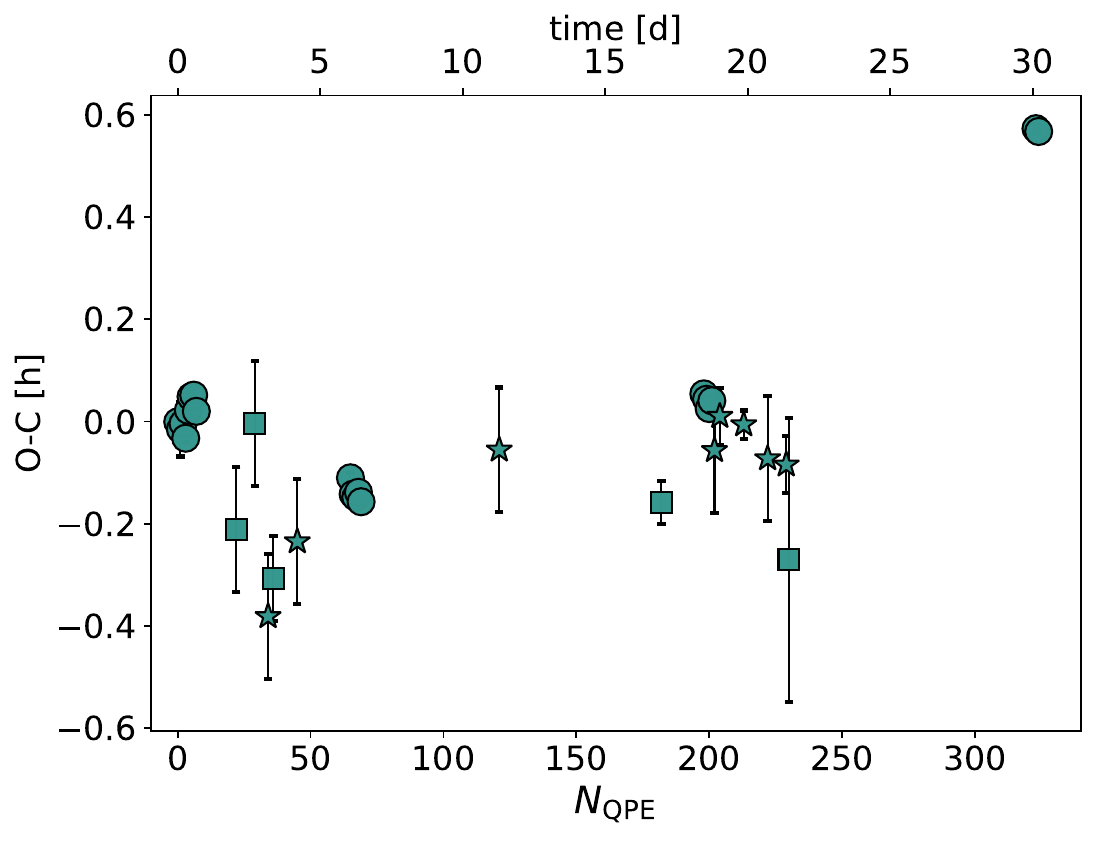}
    \includegraphics[width=0.6\columnwidth]{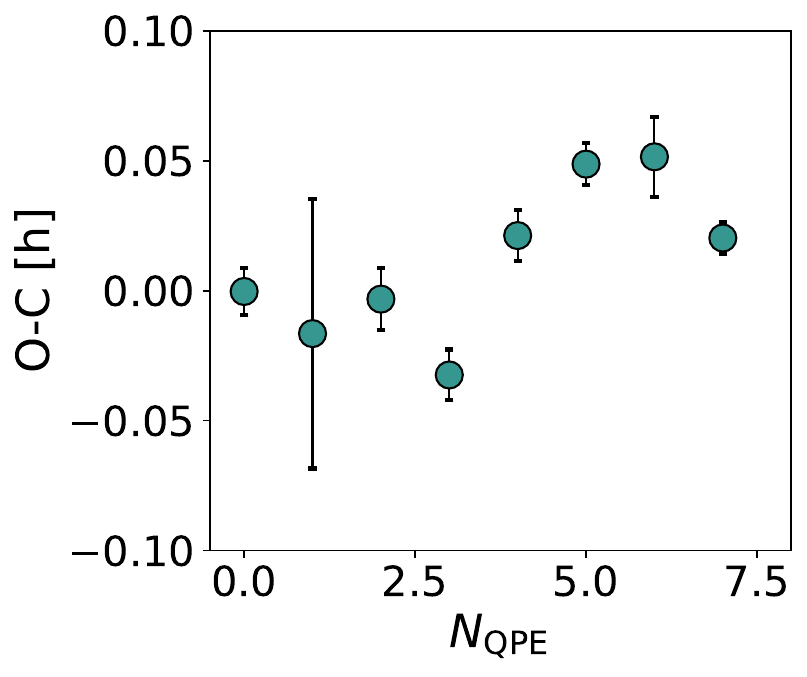}
    \caption{O-C data for all eruptions in the XMM1-4 campaign. Different symbols indicate different instruments (circles for XMM, stars for NICER, squares for XRT; see Fig.~\ref{fig:xmm1-4_lcs} and~\ref{fig:xmm1-4_nicerxrt_lcs}). The bottom subpanel shows a zoom-in of the first consecutive events in XMM1. }
    \label{fig:OConly}
\end{figure}

\begin{figure}
    \includegraphics[width=0.95\columnwidth]{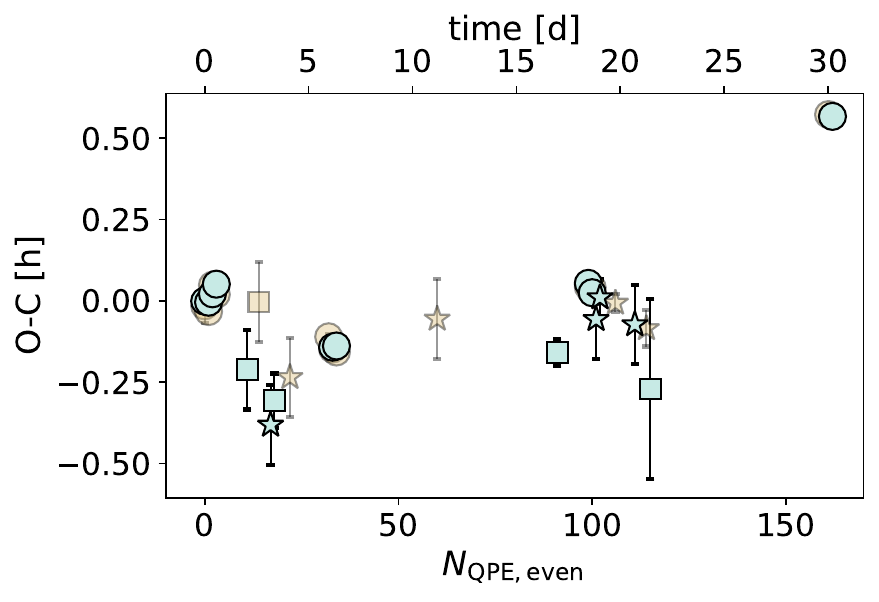}
    \includegraphics[width=0.95\columnwidth]{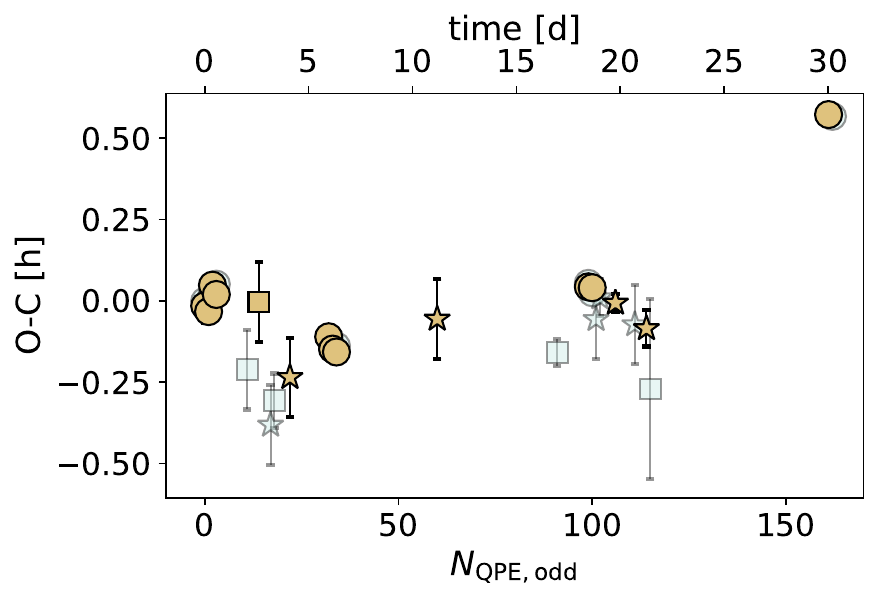}
    \caption{Same as Fig.~\ref{fig:OConly}, but for even (top) and odd (bottom) eruptions separately, with the opposite parity shown in transparency. In XMM data (i.e. circles), that are the only ones with consecutive events, data points with the opposite parity within a given epoch are covered as they have compatible delays.}
    \label{fig:OConly_oddeven}
\end{figure}


In this work, our main results are reported performing O-C analysis including all eruptions together, thus assuming that the periodic phenomenon driving the O-C delays is traced by each eruption and the separation $t_{\rm recur}$. The main reason for this choice is that it is the most agnostic, but for completeness, and for reasons that will become clear throughout this work, we also performed O-C analysis by separating odd and even eruptions (i.e., a scenario in which the period is traced by $\sim 2t_{\rm recur}$; Appendix~\ref{sec:app_OC_oddeven}). We show the same O-C data as in Fig.~\ref{fig:OConly}, but separating odd and even bursts, in Fig.~\ref{fig:OConly_oddeven}. 
The goal of this plot is to highlight how the delays of odd and even eruptions appear correlated, in that consecutive bursts within a single epoch have compatible delays despite the range covered by all epochs ($\approx 1\,$h). This suggests that separating odd and even eruptions does not provide more information than analyzing all eruptions together. 

\subsection{The source of noise in the O-C}
\label{sec:OC_noise}

From a first glance at the O-C data in Fig.~\ref{fig:OConly}, the best-sampled epochs ($N_{\rm QPE}\sim 0-250$) show that the data span delays between approximately $-0.4\,$h and $\sim 0 \,$h, and that XMM4 shows a positive delay significantly offset from the previous epochs, suggesting that on the timescale comparable to (or longer than) our XMM1-4 baseline there is a large-amplitude component driving the delays. Whether these O-C residuals are due to the deterministic quadratic and sinusoidal components shown above depends on the nature of the noise in the O-C plot. While the O-C technique is, per se, model independent and only assumes that a given event is quasi-periodic, determining the type of noise is instead somewhat model dependent (see e.g. Appendix A of \citealp{Chakraborty+26:ansky}). 

For instance, with a QPE origin from accretion disk instabilities one may expect that the quasi-period wobbles in a red-noise fashion. For an underlying orbital clock, one may instead expect the noise to be white, and if that of a given event is fully independent from the others, so would be the noise in the O-C delay data. However, it has been shown that if instead the jitter in the arrival time of a given event depends on the jitter of the previous ones, this white noise in time space cumulates with the number of events similarly to red noise in O-C delay space \citep{Koen+2006:OC}. Thus, establishing a noise model for the O-C of eRO-QPE2 implies making assumptions on the QPE origin.

We report in Appendix~\ref{sec:app_OC} our attempts at modeling the O-C data of eRO-QPE2 with a damped random walk (DRW) model, mimicking a red-noise process modulating the QPE recurrence time. An important result is that red-noise can reproduce the O-C data as well as or better than deterministic components, which is interesting on its own, although perhaps unsurprising given that it is a two-parameter model spanning orders of magnitude in its posteriors (see Fig.~\ref{fig:drw} in Appendix~\ref{sec:app_OC}), and considering the relatively sparse dataset (even if it is perhaps the best one yet for QPE sources). Furthermore, the fitted damping timescale $\tau$ spans orders of magnitude and mostly values greater than our baseline, thus predicting that on short timescales QPE recurrence times would be highly correlated. However, the autocorrelation coefficient of consecutive recurrence times in the XMM1-4 epoch is compatible with $\sim0$. Perhaps most fundamentally, this uncertain DRW model has little constraining or predictive power for existing QPE models. Thus, we defer further discussion to future work.

In this work, our main focus will be on testing existing QPE orbital models. In this case, we make the assumption that the underlying O-C noise is white, which is equivalent to assuming that the astrophysical uncertainties of each QPE arrival time are independent from those of other events. For disk collision models, each crossing is due to the deterministic orbital clock and the astrophysical uncertainty comes from the offset between disk crossing and emission, which is independent for each collision. In particular, the mean of this offset can be zero given that O-C delays are scaled from a reference epoch. A proxy for the standard deviation of this zero-mean offset during the XMM1-4 epoch could be obtained by subtracting in quadrature the measurement error of eruption peak times ($\sim 79\,$s) from the dispersion in QPE recurrences ($\sim 82\,$s), obtaining $\sim23\,$s. Neglecting this small offset dispersion does not affect our conclusions significantly, given that the measurement error of XMM data is a few times higher, and that of XRT/NICER points is $\gtrsim10-20$ times higher.

\subsection{O-C fitting with deterministic components}
\label{sec:OC_setup}

We implemented different O-C models by using some or all of the components described at the start of this Section. We derived posterior probability distributions and the Bayesian evidence of each fit with the nested sampling algorithm MLFriends \citep{Buchner2019:mlf} using the \texttt{UltraNest} package \citep{Buchner2021:ultranest} v 4.4.0, adopting uniform priors for all parameters. 
The linear component may be subtracted for more intuitive visualization of the additional components, although we note that it is always fitted in simultaneity with all other parameters.


\begin{figure}
    \includegraphics[width=0.99\columnwidth]{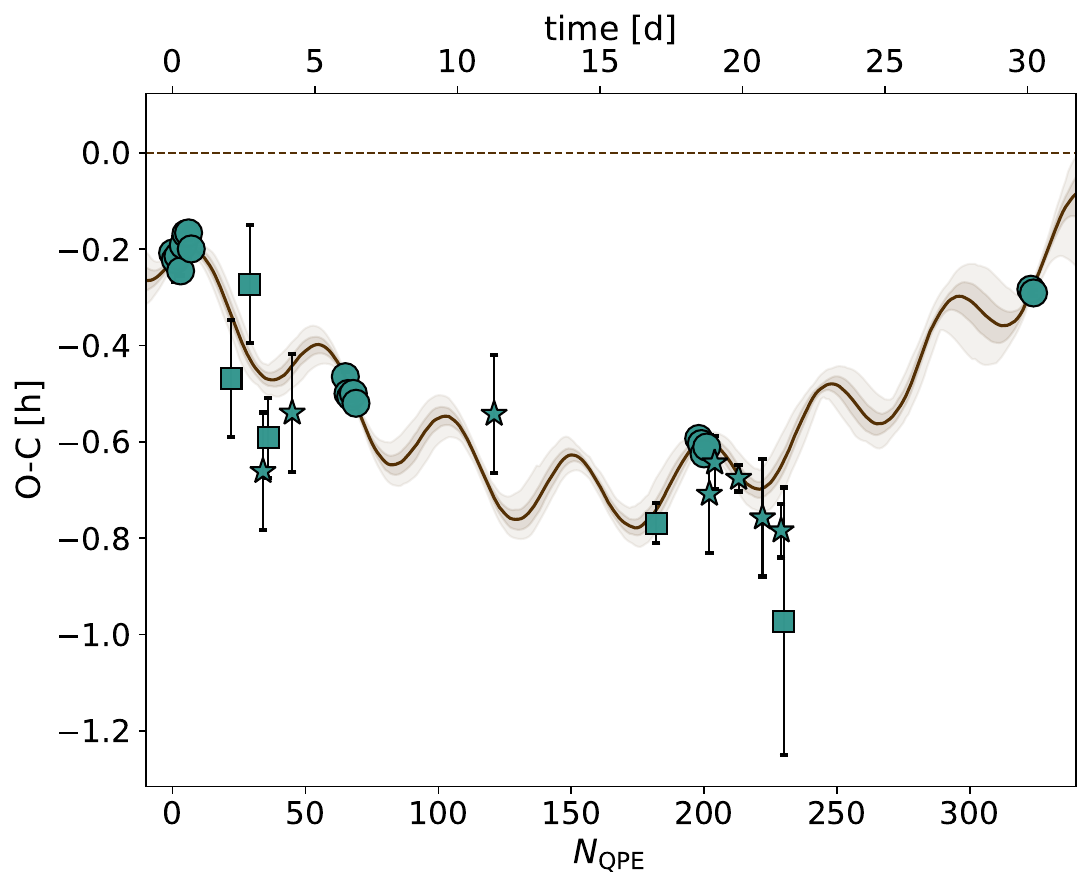}
    \includegraphics[width=0.99\columnwidth]{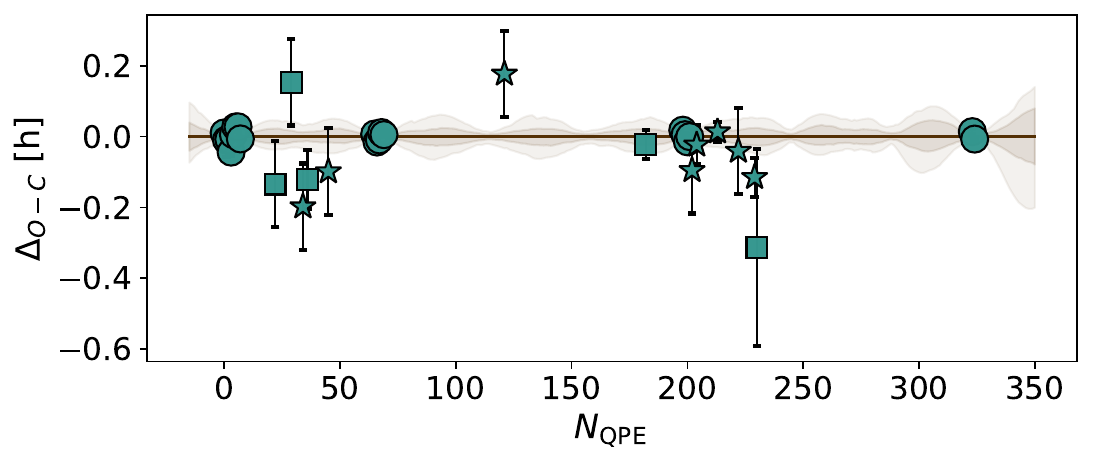}
    \caption{O-C data (as in Fig.~\ref{fig:OConly}) and best-fit model (solid line for median, shaded contours for $1\sigma$ and $3\sigma$), in which the fitted linear trend is subtracted for visualization.}
    \label{fig:OCbestfit}
\end{figure}

From a quick glance at Fig.~\ref{fig:OConly}, it is evident that a linear term alone, namely a model like $O-C= (T_0-T_{0,\rm est})+ (P-P_{\rm est}) N_{\rm QPE}$, is inadequate to describe all the deviations in the delays. In Appendix~\ref{sec:app_OC} we report details on our model comparison, in which we added model components iteratively stopping at the simplest model that significantly improves the Bayesian evidence by many orders of magnitude. Here, we only show results from the selected best-fit model (\texttt{lin+mod$_{\rm1}$+mod$_{\rm2}$}). The equation used was: 
\begin{align}
\nonumber
O - C =\;& (T_0 - T_{0,\rm est}) 
+ (P - P_{\rm est})\, N_{\rm QPE} \\
\nonumber
&+ A_{\mathrm{mod,1}} 
\sin\!\left( \frac{2\pi N_{\rm QPE}}{P_{\mathrm{mod,1}}} 
+ \phi_{\mathrm{mod,1}} \right) \\
\nonumber
&+ A_{\mathrm{mod,2}} 
\sin\!\left( \frac{2\pi N_{\rm QPE}}{P_{\mathrm{mod,2}}} 
+ \phi_{\mathrm{mod,2}} \right)\,,
\end{align}
which has 8 free parameters, namely $T_0$, $P$, $A_{\mathrm{mod,1}}$, $P_{\mathrm{mod,1}}$, $\phi_{\mathrm{mod,1}}$, $A_{\mathrm{mod,2}}$, $P_{\mathrm{mod,2}}$, and $\phi_{\mathrm{mod,2}}$. The prior bounds of the periods of the two modulations, given their obvious degeneracy, are separated at $N\sim100$. We show this fit in Fig.~\ref{fig:OCbestfit} and the related corner plot in Appendix~\ref{sec:app_OC} (Fig.~\ref{fig:best-fit corner}). The fitted true period of the XMM1-4 epoch is $P=(8064\pm7)\,$s, or $\sim2.24\,$h, which is subtracted from both data and model in Fig.~\ref{fig:OCbestfit}. 

Since the linear component has been subtracted, if eRO-QPE2 were a perfectly periodic system, all data points would lie close to the horizontal line at $O-C=0$. 
The two super-period modulations appear hierarchical, in that one with a smaller amplitude and period is superimposed to another with larger amplitude and period. Both fitted periods are far from the prior bounds chosen to separate them. The faster super period is $P_{\mathrm{mod,1}}=(47.0\pm0.6)\,$P, which corresponds to $P_{\mathrm{mod,1}}\sim4.4\,$d, with an amplitude $A_{\mathrm{mod,1}}=289^{+30}_{-31}$\,s. The longer modulation is less constrained, and the formal fit from the XMM1-4 campaign best-fit model yields a $1\sigma$ ($3\sigma$) lower limit on $P_{\mathrm{mod,2}}>917.6\,$P ($>600.1\,$P) and $A_{\mathrm{mod,2}}=4425_{-1456}^{+1299}\,$s. In Appendix~\ref{sec:OC_modulations_real}, we reported our attempts at using ancillary data taken after XMM4 to assess the possible bound constraint on $P_{\mathrm{mod,2}}$. Using two further NICER eruptions $\sim(12-13)\,$d after XMM4 suggests $P_{\mathrm{mod,2}}$ is bound around $\sim 1016\,$P ($\sim95$\,d). 

Based on the best-fit \texttt{lin+mod$_{\rm1}$+mod$_{\rm2}$} model (and using archival data in support, see Appendix~\ref{sec:app_OC}), the XMM1-4 epoch is consistent with the absence of a strong dissipative term, namely a large period decrease or increase $\dot{\rm P}$. We note that, as extensively discussed in \citet{Arcodia+2024:ero2_ticking} and \citet{Miniutti+2025:OC}, comparing recurrence times across epochs that are far apart in time is not a faithful probe of a true $\dot{\rm P}$, unless all the precise evolution terms are known. With our O-C baseline, we are able to put a sensitive constraint on the possible period decay. We subtracted a quadratic component, $-(1/2\,\dot{\rm P}\,P)N_{\rm QPE}^2$, to the best-fit model to infer an upper limit on the period decrease $\dot{\rm P}$ while marginalizing over the other model components describing the evolution of the system. We note that the sign in the $\dot{P}$ model component is switched to negative so that positive values of $\dot{\rm P}$ can be sampled log-uniformly. First, we left all model components free to vary within the original uniform prior ranges. The fit is of comparable quality, with $\Delta \log Z \sim 1$, thus the additional dissipative component is not statistically required. In fact, the $\dot{\rm P}$ is bound to the lowest prior limit, which was chosen to be $10^{-8}$\,s/s given that it corresponds to a few seconds for the $-(1/2\,\dot{\rm P}\,P)N_{\rm QPE}^2$ term with 325 eruptions (below which it is not meaningful to sample). This fit yields a $3\sigma$ upper limit at $\dot{\rm P}<3.9 \times 10^{-6}$\,s/s. We also performed a fit allowing the parameters to vary only within the 10th-90th interquantile range of the best-fit model, so that $\dot{\rm P}$ is marginalized over the best-fit uncertainties and additional degeneracies are kept under control. The $3\sigma$ upper limit of this fit is compatible with the previous, at $\dot{\rm P}<2.3 \times 10^{-6}$\,s/s.

Finally, we note that performing the same exact fitting procedure on odd and even eruptions separately (jointly fitting all parameters except constants and phases) obtains consistent results modulo a factor two in the super-periods related to the $P$ (now traced by $\sim2 t_{\rm recur}$). However, as discussed more at length in Sect.~\ref{sec:disc_EMRImodels}, we found it does not appear to be the correct setup to study QPE sources, thus we only report this analysis in Appendix~\ref{sec:app_OC_oddeven} for future reference.

\section{Interpretation of the O-C analysis within the EMRI framework}
\label{sec:discussion}

One natural way to interpret super-period modulations within orbital phenomena is with precession terms. Of course, a realistic orbital evolution would not be described by a perfect sine even for a simple source like eRO-QPE2, thus these tests should be considered a first exploratory data-model comparison.
For proper comparisons with the EMRI framework, we note that adopting a single QPE recurrence as the periodic cycle implies interpreting QPEs as either a single disk collision per orbit (e.g. with eccentric orbits and/or disks), or two, but only one being electromagnetically observable \citep[e.g., see the discussions in][]{Linial+2025:streams,Huang+2025:simul}. Sect.~\ref{sec:disc_all_short} and Sect.~\ref{sec:disc_all_long} discuss possible origins of the shorter and longer modulations, respectively. The $\dot{P}$ constraint, and its implications for the type of the possible orbiter, is discussed in Sect.~\ref{sec:disc_all_type}.


\subsection{The shorter super-period modulation: apsidal precession?}
\label{sec:disc_all_short}

One of the possible super-orbital periods in an EMRI system is due to GR apsidal precession of the orbiter within the orbital plane:
\begin{equation}
\label{eq:apsidal}
\frac{P_{\rm aps}}{P_{\rm orb}} = \frac{(1-e^2)\,a}{3\,R_g} \sim  \frac{a}{3\,R_g}
\end{equation}
where we ignore higher order corrections due to BH spin, and where the last term holds for low eccentricities (with a $\lesssim 10\%$ correction up to $e\lesssim0.3$). This regime seems appropriate for a system like eRO-QPE2 showing rather regular timings and flare amplitudes \citep{Arcodia+2024:ero2_ticking,Pasham+2024:ero2} implying relatively consistent collision radii with respect to the MBH. 

If the orbital period is known independently, the apsidal precession period $P_{\rm aps}$ only depends on the primary $M_{\rm BH}$, thus it can be used to constrain it: 
\begin{equation}
\label{eq:apsidal_mass}
M_{\rm BH} = \frac{c^3}{6\,\sqrt{3}\pi G}\,\,P_{\rm orb}^{5/2}\,\,P_{\rm aps}^{-3/2}
\end{equation}

We first try to identify the shortest of the fitted super-orbital modulations with apsidal precession. As the O-C simultaneously constrained $P=P_{\rm orb}$ and $P_{\rm mod,1}=P_{\rm aps}$, it provides direct constraints for $M_{\rm BH}$ (see more discussion in recent theoretical works, e.g., \citealp{Chakraborty+2025:qpefit} and~\citealp{Zhou+2025:mass}). Our best-fit model parameters yield $M_{\rm BH} = (1.53\pm 0.03) \times 10^5$\,\Msun, by scaling the observed periods to rest frame using $z=0.0175$ \citep{Arcodia+2021:eroqpes}. Naturally, the quoted uncertainties are only the statistical ones from the joint posterior of $P$ and $P_{\rm mod,1}$, while the systematics are unknown at this stage. Given the simplistic nature of the model components in the O-C analysis and the finite data sampling, they may be large. Thus, we consider this constraint as a rough estimate $M_{\rm BH} \sim 1.5 \times 10^5$\,\Msun. We note that eRO-QPE2 has two estimates in the literature from standard techniques: $M_{\rm BH}-\sigma$ host-galaxy scaling relations yielded $M_{\rm BH}\sim10^5$\,\Msun (with the usual $\sim0.4-0.5$\,dex uncertainties; \citealp{Wevers+2024:host}), while X-ray-to-UV accretion disk modeling yielded $M_{\rm BH}=7.9^{+7.9}_{-3.9}\times 10^5$\,\Msun \citep{Wevers+2025:ero2hst}. Given the simplistic nature of the O-C modeling, the overall agreement with the values from the literature is encouraging (keeping the unknown systematics in mind).

As a further consistency check, we used the modulation amplitude to verify that it is self consistent with $P_{\rm aps}$. From Eq.~\ref{eq:apsidal}, the expected semi-major axis $a$ of the orbit is $\sim 141 R_g$. The maximum O-C amplitude associated with light travel time induced by apsidal precession (which is obtained for an EMRI orbital plane close to edge-on) is $\sim a/c$, thus $\sim106\,$s. Another effect of apsidal precession is that the true anomaly, thus the velocity, changes at the collision point. For low eccentricity, the difference between the velocity at its maximum and minimum (i.e. at pericenter and apocenter) relative to an analogous circular orbit (i.e. the mean value about which apsidal precession would modulate the delays) is of order $\sim 2e$. The difference between time delays at these maximum and minimum is thus of order $\sim 2e/(2\pi/P)\sim e P/\pi$, which for e within $\sim0.01-0.1$ would imply delays of $\sim25-250\,$s. Thus, since the observed $A_{\rm mod,1}\sim289\,$s the system would require an eccentricity of $\sim 0.1$ to produce the observed time delays via apsidal precession. Given that the scenario discussed here refers to a single observable event per orbit (even if two disk crossing occur), an eccentricity of $\sim0.1$ is reasonable for a regular system like eRO-QPE2 (e.g., in terms of maintaining a regular flare amplitude).

\subsection{The origin of the longer modulation}
\label{sec:disc_all_long}

As discussed in the previous Section and Appendix~\ref{sec:app_OC}, the most conservative estimate for the longer modulation is a lower limit, while our analysis with ancillary data strengthens the constraints to being $\approx 1000\,$P (or $\approx 93\,$d). Regardless, these values can already provide meaningful constraints. Both apsidal and nodal EMRI precessions are bound to delay amplitudes of order $\sim a/c$, far smaller than the $\sim 3-4\,$ks observed for the longer modulation. 
Following the reasoning presented in \citet{Miniutti+2025:OC}, two possible sources of super-period modulations are disk precession and an outer MBH binary (MBHB), discussed in the subsections below. 

For clarity, we note that the statistical requirement of this longer modulation is partially driven by the assumption of underlying white noise, which seemed appropriate for an origin from disk collisions. If future modeling will indicate otherwise, and state that the astrophysical uncertainty on each QPE arrival time depends on the previous events, then a cumulative white-noise mimicking red-noise would be more appropriate \citep[e.g.,][]{Koen+2006:OC,Chakraborty+26:ansky}. However, we note that the period jitter scatter would need to be $>3-4$ times larger than the estimated $\sim23\,$s in order to reproduce the O-C delays of XMM4 at large $N\sim325$.

\subsubsection{Disk precession}

\begin{figure}
     \includegraphics[width=0.99\columnwidth]{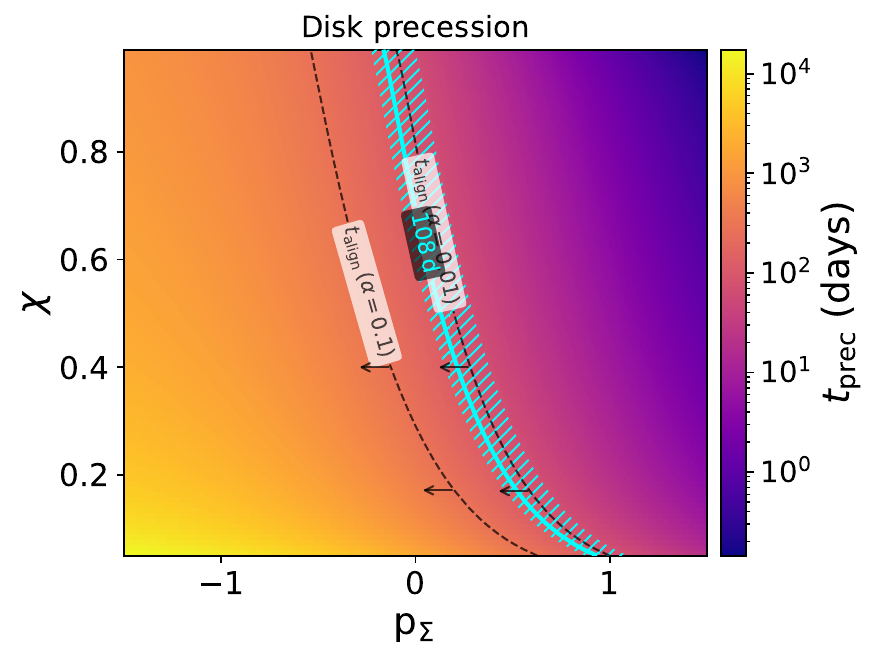}
     \includegraphics[width=0.99\columnwidth]{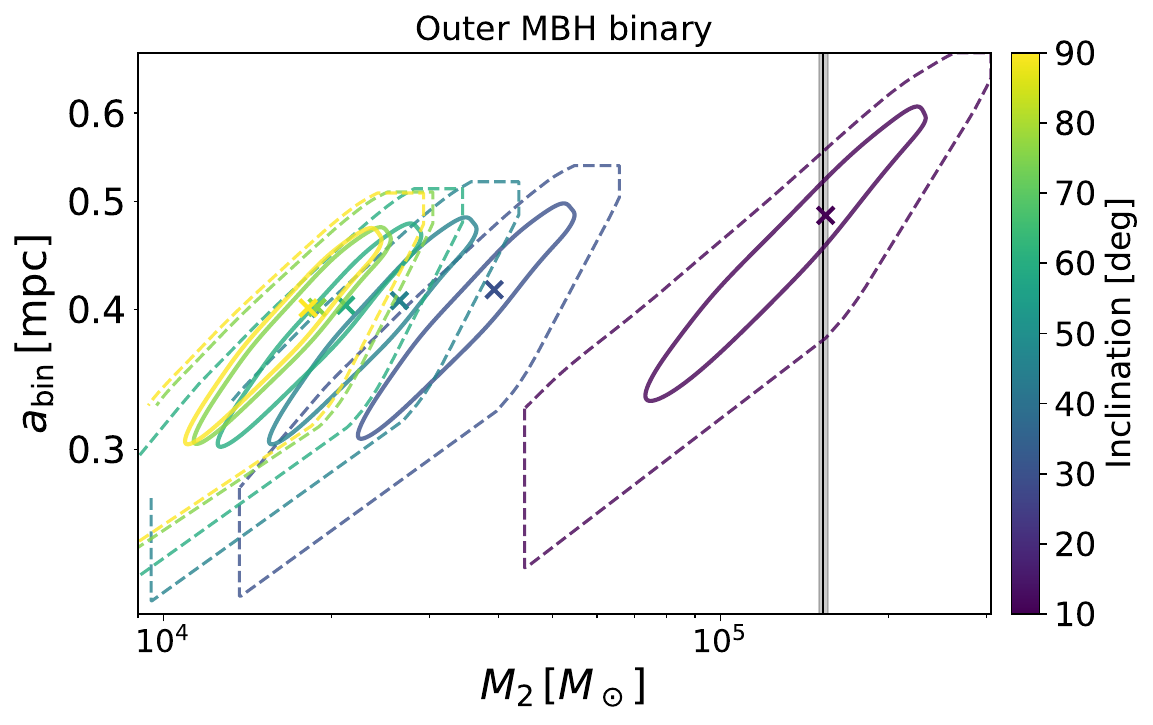}
     \caption{\emph{Top panel:} Parameters space for the precession period of a rigidly precessing compact disk \citep{Franchini+2016:prec} as a function of BH spin ($\chi$) and the slope of the disk surface density profile ($p_{\Sigma}$). While the fitted $P_{\rm mod,2}$ (cyan contour) can be reproduced with realistic combinations of $\chi$ and $p_{\Sigma}$, the same solutions provide predictions for the alignment timescale, which appears shorter than the current QPE/disk lifetime of $\sim5.5\,$y for a disk viscosity $\alpha=0.1$, and barely compatible with $\alpha=0.01$. \emph{Bottom panel:} Parameter space for an outer MBH binary (separation $a_{\rm bin}$, outer black hole mass $M_2$ and outer binary inclination) given the fitted $A_{\rm mod,2}$ and $P_{\rm mod,2}$ from the O-C analysis (Fig.~\ref{fig:OCbestfit}). The EMRI primary black hole mass was fixed to $M_1\sim1.5\times 10^5$\Msun (estimated from the possible apsidal precession constraint), which is shown with a black vertical line. Allowed $M_2\lesssim M_1$ solutions indicate $a_{\rm bin}\sim0.4\,$mpc, while no solutions exist for $M_2>M_1$.}
     \label{fig:mbhb_all}
\end{figure}

For the disk precession case, we computed the precession period $t_{\rm prec}$ adopting the rigid-precession framework from \citet{Franchini+2016:prec}. We computed $t_{\rm prec}$ for a range of spin $\chi$ of the primary MBH between $\chi=(0.05, 1)$ and slope of the disk surface density profile, $\Sigma \propto \pfrac{R}{R_g}^{-p_{\Sigma}}$, within $p_{\Sigma} = (-1.5, 1.5)$. We adopted as fiducial values a disk aspect ratio of 0.1 and $500 R_g$ for the disk outer disk radius\footnote{An estimate of the disk outer radius has been obtained in \citet{Wevers+2025:ero2hst} via disk SED fitting, albeit self-consistently with a larger $M_{\rm BH}$ estimate. Once converted to a physical size, it would correspond to $\sim 1760\,R_g$ for a $M_{\rm BH}\sim 1.5 \times 10^5\,$\Msun\,and for such extended disks the rigid precession framework would cease to be appropriate \citep{Franchini+2016:prec}, unless disk tearing/breaking occurs out to a sufficiently large radius.}. We show this 2D ($\chi$,\,$p_{\Sigma}$) map in the top panel of Fig.~\ref{fig:mbhb_all}, which shows that $t_{\rm prec}$ decreases for increasing spin and steeper disk density profiles. The fitted putative precession period from the O-C data is $P_{\rm mod,2}$, and we show its median value and $3\sigma$ range in cyan. 
At face value, the top panel of Fig.~\ref{fig:mbhb_all} would show that there is ample margin for a disk precession solution to provide the observed $P_{\rm mod,2}$. However, we have also considered the disk alignment problem. As summarized in \citet{Franchini+2016:prec}, alignment would occur on the timescale over which the disk cools and thins \citep[$t_{\rm thin}$; e.g.,][]{Stone+2012:tthin} and also at a rate inversely proportional to the disk viscosity $\alpha$ \citep[$t_{\rm align}$;][]{Franchini+2016:prec}. Both these timescales can be computed for values of the 2D map in the top panel of Fig.~\ref{fig:mbhb_all}, assuming a disk aspect ratio and viscosity parameter, and compared to the observed disk lifetime. 

The disk lifetime for eRO-QPE2 is at least $\sim4.3\,$y, from Aug. 2020 \citep{Arcodia+2021:eroqpes} to Dec. 2024 (this work, XMM5), and to the authors' knowledge the most recent eruptions observed from eRO-QPE2 are from an EP observation taken on August 12 2025 (PI: Rau) and by XMM in December 2025 (PI: Guolo), extending the lifetime to $\gtrsim5.5\,$y. As \citet{Franchini+2016:prec} concluded $t_{\rm align}$ is usually faster than $t_{\rm thin}$, so we focused on that quantity while varying $\alpha$ between 0.01 and 0.1. The dashed black lines in the top panel of Fig.~\ref{fig:mbhb_all} correspond to an alignment timescales equal to the lifetime constraint of $\sim 5.5\,$y, thus only disk solutions to the left of this line would induce precession ongoing until today (i.e., $>t_{\rm life}/t_{\rm prec}\sim18$ precession cycles). Comparing the black lines upper limits with the observed constraint in cyan shows tension with any $\alpha\gtrsim0.01$. As the $t_{\rm life}$ constraint increases over time with more detections, it would move the black line lower limits to the left, shrinking the parameter space further. We verified a posteriori that changing the outer disk radius value to $200 R_g$ or $800 R_g$ would not change the following conclusions significantly, but only the $\alpha$ value in tension: for a smaller $R_{\rm out}$ we obtained that $\alpha\gtrsim0.1$ are not allowed, while for the larger $R_{\rm out}$ no realistic $\alpha$ is allowed altogether. A more conclusive statement is hindered by the many assumptions and uncertainties in the parameter values (i.e. the disk radial extent, viscosity parameter and aspect ratio) involved in generating precession and alignment timescales \citep[e.g., see the possible effect of the gas stream in a TDE disk producing a quasi-stable warped configuration;][]{Xiang-Gruess+2016:streamwarp}.

Finally, we note that the parameter space required to reproduce the amplitude of delays (i.e. $A_{\rm mod,2}$) is also relatively small, as we estimated using the EMRI code \texttt{QPE-FIT} \citep{Chakraborty+2025:qpefit} that the relative inclination between EMRI orbit and accretion disk has to be in a relatively narrow range between $\approx15-25$ degrees to imprint time delays as large as the observed $\sim(3-4)\,$ks. Hence, while there is a disk precession solution with a rigidly precessing disk a la \citet{Franchini+2016:prec}, the parameter volume that it requires to satisfy observations is relatively small, and it will shrink further as the lifetime lower limit increases with future observations. Naturally, there is a clear prediction that the longer precession would disappear due to disk alignment within the next few years. 

\subsubsection{Hierarchical triple system}

As in the disk precession case, we can use the observed $A_{\rm mod,2}$ and $P_{\rm mod,2}$  to explore the allowed parameter space of a hierarchical triple with an inner EMRI ($M_1$ and $m_{2}$) and outer MBHB \citep[$M_1$ and $M_2$; e.g., see][]{Miniutti+2025:OC}. From the super-period modulation amplitude, we can constrain the outer binary separation,
\begin{equation}
a_{\rm bin} = \frac{c \, A_{\rm mod,2} \, (1+q)}{\sin i_{\rm out}},
\end{equation} given a mass ratio $q=M_1/M_2$ and inclination $i_{\rm out}$ with respect to the observer. From the apsidal precession constraint above, the primary of the EMRI system is $M_1\sim1.5\times 10^5$\Msun whereas $M_2$ is a free parameter. Under the assumption that the longer super-period modulation is due to a MBHB, from Kepler's third law we can numerically constrain the allowed $i_{\rm out}$ and $M_2$. We find that the observed $A_{\rm mod,2}$ can only be reproduced by an almost face-on binary with $M_2\sim M_1$, or by a progressively smaller $M_2$ and more inclined outer orbit, down to an edge-on $M_2\sim0.2\,M_1$. No solutions can be found for $M_2>M_1$ that satisfies the observed $A_{\rm mod,2}$ and $P_{\rm mod,2}$. We show these constraints in the bottom panel of Fig.~\ref{fig:mbhb_all} as $1\sigma$ and $3\sigma$ contours, where uncertainties are propagated from sampling the posteriors of $A_{\rm mod,2}$, $P_{\rm mod,2}$, and $M_1 = M_{\rm BH}$ (where we note the truncated $3\sigma$ contours due to the bound $P_{\rm mod,2}$ lower-limit posterior). Interestingly, the only solutions that reproduce $A_{\rm mod,2}$ and $P_{\rm mod,2}$ point toward an outer binary separation $a_{\rm bin}\sim 0.4$\,mpc. Compared to the EMRI binary separation of $\sim141 R_g\sim10^{-3}$\,mpc, the ratio between outer and inner binary separation is $\approx 400$, thus yielding a stable system \citep{2001MNRAS.321..398M}. The related merger timescales of the outer binary solutions are all $\gtrsim2\,$Gyr. We further investigated the stability of this hierarchical triple system from the Kozai-Lidov mechanism \citep{2016ARA&A..54..441N}. For the low eccentricities used in all the above estimates (e.g, $e\sim0.1$ for the EMRI and $e_{\rm out}\sim0.05$ for the outer binary) this effect is much slower than the other limiting factors present in the system (e.g., stellar ablation from collisions, disk dissipation). From the largest to the smallest $M_2$ in the bottom panel of Fig.~\ref{fig:mbhb_all}, the Kozai-Lidov timescale at the quadrupole-level approximation \citep{2016ARA&A..54..441N} spans from $\sim2300\,$y to $\sim19500\,$y. Finally, it is interesting that the constraint of $M_2<M_1$ to reproduce $A_{\rm mod,2}$ and $P_{\rm mod,2}$ implies that $M_{\rm tot}< 2 \, M_1 \lesssim 3\times 10^5$\Msun, which is compatible with the $M_{\rm BH}-\sigma$ mass \citep[$M_{\rm tot}\sim 10^5$\Msun;][]{Wevers+2024:host} within the $\sim 0.5\,$dex systematics.

Finally, if the system were part of a wider MBH binary it would induce Doppler boosting on the disk emission $\Delta F_X/F_X$. 
Following again the reasoning already applied to GSN\,069 data \citep{Miniutti+2025:OC}, we can write the predicted Doppler boost as:
\begin{equation}
\frac{\Delta F_X}{F_X}\simeq 2 \pi (3-\alpha_X) \frac{A_{\rm mod,2}}{P_{\rm mod,2}}\,\sim 0.03,
\end{equation}
evaluated at the maximum of the phase to yield the semi-amplitude, and with a spectral slope $\alpha_X\simeq -9$. 
Hence, even if the longer modulation were due to an outer binary, it would not be detectable via phase-dependent X-ray flux variability. This is self consistent with the fact that the observed quiescence flux of eRO-QPE2 has been remarkably constant within uncertainties for the last few years \citep{Arcodia+2024:ero2_ticking,Pasham+2024:ero2}. We also confirmed this comparing the quiescence flux across the new XMM1-5 epochs, which we found to be compatible within uncertainties.

\subsection{Constraints on the type of orbiter}
\label{sec:disc_all_type}

Our best-fit model does not require a quadratic dissipative component such as a period increase or decrease. 
Thus, we estimated an upper limit on the absolute value of the period decrease, which is bound at $\dot{P}< (2-3) \times 10^{-6}$ (Sect.~\ref{sec:OC}). 

\paragraph{GW decay}

First, we discuss the constraint from our $3\sigma$ upper limits on the GW orbital decay. We compute a range of $\dot{P}_{\rm GW}$ as a function of a range of EMRI secondary mass $m_{2}$ and EMRI eccentricity $e$, given the fitted $M_1\sim1.5\times 10^5$\Msun\,and $P=2.23\,$h from the O-C, using the \citet{Peters:gw} formula:
\begin{align}
\dot{P}_{\rm GW} &= -\frac{192 \pi}{5} \, 
          \frac{G^{5/3}}{c^5} 
          \left( \frac{2 \pi}{P} \right)^{5/3} 
          \frac{M_1 m_2}{(M_1 + m_2)^{1/3}} 
          \nonumber\\
        &\quad \times 
          \frac{1 + \frac{73}{24} e^2 + \frac{37}{96} e^4}{(1 - e^2)^{7/2}} \,.
\end{align}

\begin{figure}
     \includegraphics[width=0.99\columnwidth]{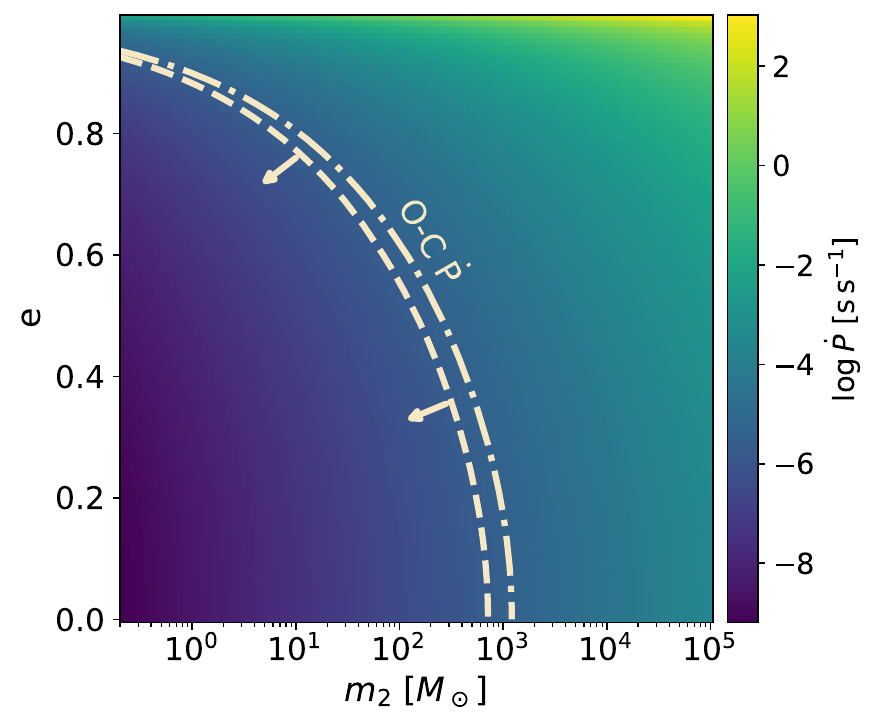}
     \caption{Map of the absolute value of the predicted period decrease ($\dot{P}$) due to GW emission for a range of EMRI secondary mass ($m_2$) and eccentricity, given the fitted $M_1\sim1.5\times 10^5$\Msun\,and $P=2.23\,$h. The fitted $3\sigma$ $\dot{P}$ upper limits from the O-C campaign are shown with yellow lines, highlighting the allowed parameter space. The dashed line is for model parameters fitted within 10-90th percentiles of the best-fit model, while the dot-dashed for free parameters within original prior bounds (see text for details). Their difference shows the (relatively low) impact of parameter degeneracy. Our constraints disfavor high eccentricities at all masses down to sub-solar orbiters, and moderate and low eccentricities for IMBH-like masses (e.g., $m_2\gtrsim250\,$\Msun\, is ruled out at $e\sim0.5$, and $m_2\gtrsim1130\,$\Msun\, at $e\sim0.1$).}
     \label{fig:pdot_all}
\end{figure}

Fig.~\ref{fig:pdot_all} shows the predicted $\dot{P}_{\rm GW}$ as a function of $m_2$ and $e$, with yellow lines showing our fitted $3\sigma$ upper limit constraints. Interestingly, our O-C campaign rules out high eccentricities at all secondary masses including sub-solar secondaries. For instance, $e$ is only allowed to be $<0.92$ at $3\sigma$ for $0.5\,$\Msun, and $<0.94$ at $3\sigma$ for $0.2\,$\Msun. Effectively, this would disfavor some of the proposed QPE models involving partial disruptions from a white dwarf on an extremely eccentric orbit \citep{King2020:WD,Wang+2022:wd,Chen+2023:wd}, leaving a very limited parameter space. Furthermore, orbiting intermediate mass BHs (IMBHs; \citealp{Lam+2025:IMBHscoll}) would also be disfavored following the yellow curves in Fig.~\ref{fig:pdot_all}. To give some values, $m_2\gtrsim250\,$\Msun\, is ruled out at $e\sim0.5$, and $m_2\gtrsim1130\,$\Msun\, at $e\sim0.1$. Perhaps this latter constraint is not surprising based on previous arguments disfavoring IMBH secondaries based on rates and accretion disk depletion arguments \citep[e.g.,][]{Linial+2023:qpemodel,Linial+2025:streams, Franchini+2023:qpemodel,Mummery+2025:collisions}. Nonetheless, our O-C analysis provides an independent empirical constraint for specific $(m_2, e)$ pairs. 

\paragraph{Gas drag} We carried out the same exercise for the orbital decay due to hydrodynamical gas drag from the disk collisions \citep[e.g.,][]{Linial+2023:qpemodel,Linial+2024:drag,Zhou+2024:longterm,Arcodia+2024:ero2_ticking}. For a stellar orbiter this effect is expected to be larger than that from GW emission. Ignoring possible long-term changes due to the change of relative inclination between the EMRI orbit and the disk \citep{Arcodia+2024:ero2_ticking}, the orbital decay of a star of mass $m_\star$ and radius $R_\star$ which impacts a disk of surface density $\Sigma$ is approximately:
\begin{align}
\label{eq:drag}
\dot{P}_{\rm drag}
&\approx - \frac{3\pi R_\star^2 \Sigma}{m_\star}
\\
&\approx -2.3\times 10^{-6}
\left( \frac{m_\star}{\mathrm{M}_\odot} \right)^{-1}
\left( \frac{R_\star}{\mathrm{R}_\odot} \right)^{2}
\left( \frac{\Sigma}{10^5 \, \mathrm{g\,cm^{-2}}} \right)\,. \nonumber
\end{align}

Thus, our $\dot{P}$ upper limit may provide simultaneous constraints for the orbiting star and the accretion disk density. We can restrict our analysis to a motivated range for $m_\star$ and $R_\star$ using both the QPE lifetime ($t_{\rm life}$), assuming that for a star the dominating factor for the period decrease are the collisions themselves \citep{Yao+2025:simul}, and the tidal radius, $r_T\sim R_\star\,(M_{\rm BH}/m_{\star})^{1/3}$, which is a lower bound to the collision radius \citep[e.g.,][]{Wevers+2025:ero2hst}. Assuming a main sequence star ($R_\star \propto m_\star^{0.8}$), the condition $a>r_T$ imposes $m_\star\lesssim0.72$\Msun, using the fitted $P=2.24\,$h as orbital period. At the same time, there is another constraint from the QPE lifetime $t_{\rm life}\gtrsim5.5\,$y. This imposes a condition $t_{\rm life} \sim t_{\rm recur} \, m_\star / \Delta m_\star\gtrsim 5.5\,$y, where the QPE period is the recurrence time $t_{\rm recur}$ for the O-C with all eruptions, and $\Delta m_\star$ is the mass lost through collisions \citep{Yao+2025:simul,Linial+2025:streams}. This imposes an upper limit on the fractional mass loss $ \Delta m_\star/m_\star \lesssim 4.6 \times 10^{-5}$. 

The fractional mass loss can be expressed in proportion to the ratio between the ram pressure experienced by the star during disk crossing and the star's mean internal pressure \citep{Linial+2023:qpemodel,Yao+2025:simul}. They are, respectively, $p_{\rm ram} \simeq \frac{1}{2} \rho_{\rm mid} v_{\rm sh}^2 \simeq \rho_{\rm mid}\, v^2_K$, where $\rho_{\rm mid}= \Sigma/(2H_{\rm disk})$ is the disk mass density and $v_K \sim \sqrt{G M_{\rm BH} / a}$ the Keplerian velocity at the collision radius $\sim a$, and $p_\star \sim (G m_\star^2)/(4 \pi R_\star^4)$. We adopt a range of possible star masses, $0.1\lesssim (m_\star/$\Msun$)\lesssim 0.7$, and use the fractional mass loss constraint from the QPE lifetime to restrict possible $\Sigma$ values for any given $m_\star$. At the collision radius $a \gg R_g$ and for accretion radiative efficiency of $\eta$ and dimensionless Eddington-normalized accretion rate $\dot{m}$, we can approximate $H_{\rm disk} \simeq \frac{3}{8 \eta} R_g \dot{m}$ \citep{Frank+2002:accr}. For the collision regime appropriate for a short period system like eRO-QPE2 (i.e. multiple shock, \citealp{Yao+2025:simul}) and noting that $H_{\rm disk}\approx 0.1 R_\odot$ (thus $H_{\rm disk}\lesssim R_\star$ for $m_\star$ range being studied), from Eq.~4 in \citet{Yao+2025:simul} we obtain:
\begin{align}
\label{eq:deltamass}
\Delta m_\star/m_\star
&\sim 0.03 \, \pfrac{H_{\rm disk}}{R_\star} \pfrac{p_{\rm ram}}{p_\star}
\end{align}
\begin{equation}
\sim 3 \times 10^{-5} \pfrac{\Sigma}{10^5\,\text{g\,cm}^{-2}} \pfrac{m_\star}{0.5\, M_\odot}^{2/5}\,.\nonumber
\end{equation}
Here we assumed a main sequence star, and given that $p_{\rm ram}\propto 1/H_{\rm disk}$, the only free parameters other than $R_\star$ and $m_\star$ are in the ratio $M_{BH}/a$ through the Keplerian velocity at the collision point, which is essentially the inverse of the semimajor axis in $R_g$ units (which we take from our fitted O-C values, i.e. $a/R_g\sim141$).

\begin{figure}
     \includegraphics[width=0.99\columnwidth]{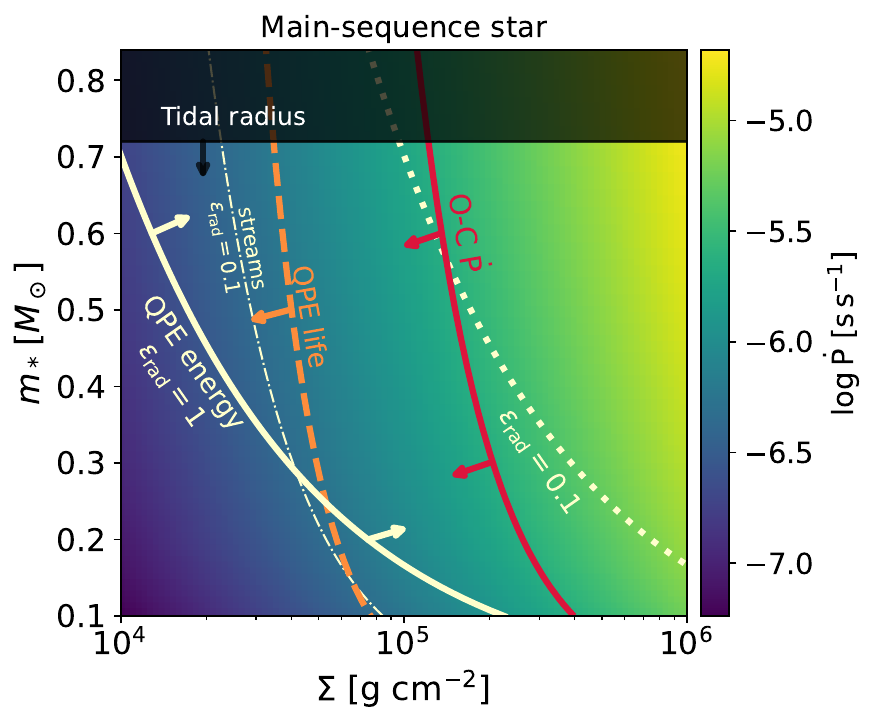}
     \caption{Map of the absolute value of the predicted period decrease ($\dot{P}$) induced by gas drag through disk collisions onto an orbiting main-sequence star, for a range of accretion disk surface density $\Sigma$ (at the collision radius) and mass of the secondary star $m_\star$. Additional constraints are: the tidal radius ($a>r_T$, black shaded region); the QPE lifetime ($\Delta m_\star/m_\star \sim t_{\rm recur}/t_{\rm life}$, with $t_{\rm life} \gtrsim 5.5\,$y; orange line); the QPE integrated energy (yellow) matched to the collision kinetic energy for the star as a ``bullet'', for two radiative efficiency values ($\varepsilon_{\rm rad}=1$ and 0.1, as labeled); the equivalent constraint for colliding stellar streams (dot dashed yellow line, for $\varepsilon_{\rm rad}=0.1$); the O-C $\dot{P}$ upper limit (red line). There are effectively no ($\Sigma$, $m_\star$) solutions for a ``bullet'' (for reasonable $\varepsilon_{\rm rad}\lesssim0.1$), while there are some for stellar streams.
     }
     \label{fig:pdot_drag}
\end{figure}

Both $\dot{P}$ and $\Delta m_\star/m_\star$ upper limits provide a range of allowed $m_\star$ and $\Sigma$ (at the collision radius) for the collision model with a stellar orbiter. We show this parameter space in Fig.~\ref{fig:pdot_drag} where the O-C $\dot{P}$ upper limit is shown in red, while the $\Delta m_\star/m_\star \lesssim 4.6 \times 10^{-5}$ constraint is shown in orange. The latter appears more stringent, although it requires more uncertain assumptions compared to the empirical O-C constraint so it should be interpreted with caution. Nonetheless, we note that the ablation constraint on $\Sigma$ will become more and more stringent as more eruptions are observed and the ratio $t_{\rm recur}/t_{\rm life}$ decreases. 

At the same time, there is a minimum disk density $\Sigma_{\rm min}$ required to produce the observed integrated energy per flare via disk collisions \citep{Mummery+2025:collisions}. For the case of a stellar ``bullet'' the relevant cross section to infer the ejected disk mass is given by $2 \pi R_\star^2$, where the factor two is due to the azimuthal flow of the disk. We note that the range of stellar masses allowed in Fig.~\ref{fig:pdot_drag} corresponds to $R_\star<0.77\,R_{\odot}$. The average observed integrated energy is $E_{\rm QPE}\sim 1.1\times 10^{45}$\,erg \citep{Arcodia+2024:ero2_ticking}, and for an emission process efficiency $\varepsilon_{\rm rad}$ bound to be smaller than one, we expect this to be a lower limit for the kinetic energy in the collision $E_{\rm col}=2\pi R_\star^2 \Sigma v_K^2>E_{\rm QPE}$. This relation provides the lower limit shown with a yellow line in Fig.~\ref{fig:pdot_drag}, while we show the same relation, but with $\varepsilon_{\rm rad}=0.1$, with a dotted line. If the true efficiency is close to $10\%$, there is virtually no parameter space for colliding main-sequence ``bullets''. Indeed, the efficiency is expected to be small ($\varepsilon_{\rm rad}\ll1$) if QPEs are powered by a stellar ``bullet'' crossing the disk \citep[e.g.,][]{Linial+2025:streams,Mummery+2025:collisions}. Hence, current estimates of $\Sigma_{\rm max}$ from the O-C $\dot{P}$ and $\Sigma_{\rm min}$ from the integrated energy would leave no parameter space for a main-sequence star ``bullet''.

Stellar streams have been invoked to reproduce the high integrated energy of the long-period QPE sources \citep{Yao+2025:simul,Mummery+2025:collisions,Linial+2025:streams}. Before our $\Sigma_{\rm max}$ constraint from the O-C $\dot{P}$, there was no reason to exclude a canonical ``bullet'' collision in eRO-QPE2 based on energetics. Our Fig.~\ref{fig:pdot_drag} now suggests otherwise and we investigate whether stellar streams would relax the tension. The model in \citet{Linial+2025:streams} indicates that for a system like eRO-QPE2 the stream would penetrate the disk with the effect of increasing the amount of disk mass contributing to the emission, thus significantly lowering $\Sigma_{\rm min}$. As a stream is by definition much more extended than the star, the gas drag $\dot{P}$ map would be equivalent to that shown in Fig.~\ref{fig:pdot_drag}, but the integrated energy constraint would be far less stringent. We show the related curve with a dot-dashed yellow line for $\varepsilon_{\rm rad}=0.1$, noting that the radiative efficiency for streams may be larger than in the ``bullet'' case \citep{Linial+2025:streams,Mummery+2025:collisions}. The lower $\Sigma_{\rm min}$ is due to a larger disk area interacting with the orbiter, increased to $\sim 4 r_H^2 \tilde{l}$, where $r_H = a \pfrac{m_\star}{3M_{\rm BH}}^{1/3}$ is the Hill's radius and for a stream elongation $\tilde{l}\sim 28$ appropriate for a system like eRO-QPE2 \citep{Linial+2025:streams}. Hence, the allowed parameter space is now somewhat larger, although we note that it is still limited to $\Sigma\approx (0.5-1) \times 10^5\,$g\,cm$^{-2}$, but neither much larger nor much smaller, and for $m_\star/M_{\odot} \approx (0.4-0.7)$.

We estimated whether realistic accretion disks in a system like eRO-QPE2 can satisfy this requirement assuming a surface density profile $\Sigma(R,t) = \Sigma_0(t) \,\pfrac{R}{R_g}^{-p_{\Sigma}}$. We assumed the disk was created by a TDE of a star with mass $m_\star^{\rm TDE}$ and circularization radius twice the tidal radius $2 r_T^{\rm TDE}$. We follow this approach given the mounting evidence connecting QPE accretion flows with previous TDEs \citep{Nicholl+2024:qiz,Chakraborty+2025:upj,Quintin+2023:tormund,Bykov+2024:tormund,Hernandez-Garcia+2025:ansky,Guolo+2025:gsnlongterm}. While to date no evidence of a precursor TDE has been found for eRO-QPE2 in particular, we note that adopting AGN-like steady state disks would produce higher $\Sigma$ values compared to an expanding TDE disk with a fixed mass budget. Thus, any parameter space excluded by the TDE disk profile would be ruled out for steady-state disks too. We constrain the surface density of a TDE disk at the collision radius $R=a=141\,R_g$ by setting mass and angular momentum conservation, $M_{\rm disk}\leq f^{TDE} m_\star^{\rm TDE}$ and $J_{\rm disk} = f^{TDE} J_\star^{\rm TDE}$, and keeping the smallest of the two $\Sigma_0$ estimates, namely between:
\begin{align}
\Sigma_0(t) = \;&
\frac{M_{\rm disk,0}\,(2-p_{\Sigma})}{2\pi R^2_g}
\nonumber\\
&\times
\left[
\left(\frac{R_{\rm out}(t)}{R_g}\right)^{2-p_{\Sigma}}
-
\left(\frac{R_{\rm in}}{R_g}\right)^{2-p_{\Sigma}}
\right]^{-1}
\end{align}
(which follows from mass conservation), and
\begin{align}
\Sigma_0(t)=\;&
\frac{M_{\rm disk,0}\,(5-2p_{\Sigma})}{4\pi R^2_g}
\left(\frac{2 r_T^{\rm TDE}}{R_g}\right)^{1/2} \times
\nonumber\\
&\times
\left[
\left(\frac{R_{\rm out}(t)}{R_g}\right)^{{5\over2}-p_{\Sigma}}
-
\left(\frac{R_{\rm in}}{R_g}\right)^{{5\over2}-p_{\Sigma}}
\right]^{-1},
\end{align}
which follows from angular momentum conservation.
We evaluated these profiles for a range of $f^{\rm TDE}=(0.01, 0.5)$, $m_\star^{\rm TDE}/M_{\odot}=(0.5,4)$ (i.e., combined spanning $M_{\rm disk,0}/M_{\odot}=0.005-2$), spin $\chi=(-0.998,0.998)$, which defines the inner radius $R_{\rm in}$, slopes $p_{\Sigma}=(0,1.5)$, and $R_{\rm out}/R_g=(200,1000)$. Disk solutions in the desired range $\Sigma\sim(0.5-1)\times 10^5\,$g\,cm$^{-2}$ exist for most $p_{\Sigma}$ and $\chi$, but with low $M_{\rm disk,0}$ and/or high $R_{\rm out}$. In particular, around $R_{\rm out} \approx 300 \,R_g$, the allowed $M_{\rm disk,0}$ range is $f_{\rm mass} m_\star^{\rm TDE}/M_{\odot}\lesssim 0.006\,$, while around $R_{\rm out} \approx 900\,R_g$ the $1\sigma$ range is $0.01 \lesssim f_{\rm mass} m_\star^{\rm TDE}/M_{\odot}\lesssim 0.07\,$. 

As an independent constraint on disk values, for $R_{\rm out} \gg a$ the local accretion rate equals $\dot{m} \approx L_{\rm Q}^{\rm bol}/\eta c^2$, as quasi steady-state is established in the inner disk, where $L_{\rm Q}^{\rm bol} \approx 10^{43} \, \rm erg \, s^{-1}$ is the bolometric quiescent luminosity of eRO-QPE2 \citep{Arcodia+2024:ero2_ticking,Wevers+2025:ero2hst}. We can therefore estimate the local disk density as $\Sigma(a)=L_{\rm Q}/(3\pi \eta c^2 \nu)$, for an $\alpha$-disk with viscosity $\nu=\alpha \sqrt{GM_{\rm BH} a} (H_{\rm disk}/a)^2$. Assuming the fiducial values used throughout this Section for $L_{\rm Q}$, $M_{\rm BH}$, $H_{\rm disk}/a$, we obtain:
\begin{equation}
\Sigma (a) \sim 1.2\times 10^6 \, {\rm g \, cm^{-2}}\,\pfrac{\alpha}{0.1}^{-1}\,\pfrac{H_{\rm disk}/a}{3.4 \times 10^{-3}}^{-2}\,,
\end{equation}
thus a disk density $\sim10$ times higher than the constraint from gas drag, namely $\Sigma\sim(0.5-1)\times 10^5\,$g\,cm$^{-2}$ (Fig.~\ref{fig:pdot_drag}). In turn, for self consistency this may provide some constraints on $\alpha$ and $H_{\rm disk}$.

\begin{figure}
     \includegraphics[width=0.99\columnwidth]{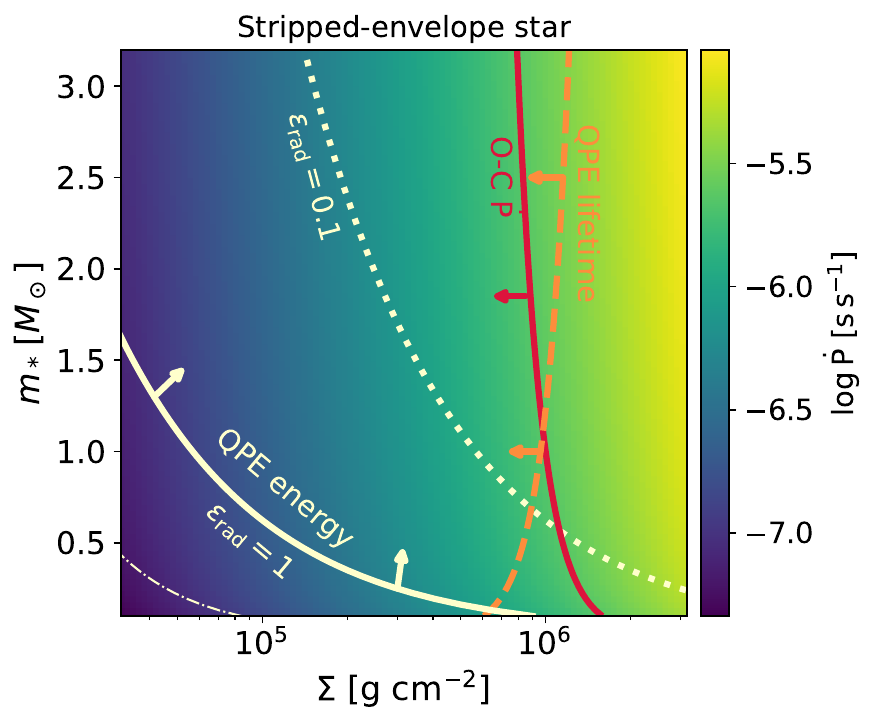}
     \caption{Same as Fig.~\ref{fig:pdot_drag}, but for stripped-envelope stars, namely for comparatively smaller $R_\star$.}
     \label{fig:pdot_drag_stripped}
\end{figure}

\paragraph{Stripped-envelope stars} We remind the reader that the range of disk density solutions shown in Fig.~\ref{fig:pdot_drag} only exists for stellar streams in the case of main-sequence stars. This is in part due to the gas drag $\dot{P}$ upper limit being too low compared to the $\Sigma_{\rm min}$ required by ``bullet'' collisions. Thus, we investigated whether there exists a larger parameter space for the ``bullet'' in the case of a different stellar mass-radius relation, in particular with an orbiter with comparably smaller $R_\star$. Interestingly, this would also relax the tension between collision distances inferred too close to the tidal radius or mass-transfer radius (or even within) for some QPE sources \citep[e.g.,][]{Franchini+2023:qpemodel,Dodd+2025:waves,Guo+2025:coll,Linial+2025:streams}. A putative stellar-EMRI with $R_\star\lesssim R_{\odot}$ for $m_\star\gtrsim M_{\odot}$ is akin to stripped-envelope stars during their phase of core helium fusion \citep[e.g.,][]{Goetberg+2018:stripped}. We fitted a functional form for the mass-radius relation to the models reported in \citet{Goetberg+2018:stripped} and obtained $\log \pfrac{R_\star}{R_\odot} = 0.6\,\log \left(\frac{m_\star}{M_\odot}\right) - 0.5$. 
With this, we produced an analog of Fig.~\ref{fig:pdot_drag} for stripped-envelope stars, for a range $m_\star/M_{\odot}=(0.3-3)$ and $R_\star/R_{\odot}=(0.15-0.70)$ (see Fig.~\ref{fig:pdot_drag_stripped}). Orbiters with lower $R_\star/m_\star$ have a range of solutions even for the ``bullet'' scenario for $\varepsilon_{\rm rad}\lesssim0.1$, with allowed disk densities $\Sigma\sim(0.5-1)\times 10^6\,$g\,cm$^{-2}$. This is due to a $\Sigma_{\rm max}\propto R_\star^{-3}$ dependency in the ablation (``QPE lifetime'') term from Eq.~\ref{eq:deltamass}, and $\Sigma_{\rm min}\propto R_\star^{-2}$ in the ``QPE energy'' lower limit. In particular, around $R_{\rm out} \approx 300 \,R_g$, the allowed $1\sigma$ $M_{\rm disk,0}$ is $0.02 \lesssim f_{\rm mass} m_\star^{\rm TDE}/M_{\odot}\lesssim 0.03\,$, while around $R_{\rm out} \approx 900\,R_g$ this range is $0.1 \lesssim f_{\rm mass} m_\star^{\rm TDE}/M_{\odot}\lesssim 0.3\,$. Naturally, similar to the main-sequence case, stellar streams from a stripped-envelope star would yield an even lower $\Sigma_{\rm min}$ constraint (e.g., see the dot-dashed line in Fig.~\ref{fig:pdot_drag_stripped} for $\varepsilon_{\rm rad}=0.1$), allowing even smaller $M_{\rm disk,0}$. However, as stripped-envelope stars are much rarer, we note that if a similar requirement will also be found for other QPE sources there would be tension with the QPE rates \citep{Arcodia+2024:rates}, unless it is the more abundant main sequence stars that evolve to $R_\star/m_\star$ values akin to stripped-envelope stars through the disk collisions themselves.

\paragraph{Stellar BH orbiters} The $\dot{P}$ upper limit from the O-C analysis is formally consistent with a stellar BH orbiter \citep[][]{Franchini+2023:qpemodel}. The $\Sigma_{\rm min}$ requirement would imply that the influence radius of the BH orbiter is $R_{\rm Bondi,\bullet}\approx R_\star$, so that the integrated energy is reproduced, and in fact \citet{Franchini+2023:qpemodel} inferred that for $m_\bullet=40-100\,$\Msun\, and low enough relative velocities (i.e. for grazing prograde collisions) may provide the required cross section. This means that for the sake of adapting the gas drag equation (Eq.~\ref{eq:drag}) to a BH orbiter with $R_{\rm Bondi,\bullet}\approx R_\star$, the only relevant dependence is the orbiter's mass $\dot{P}_{\rm drag} \propto \Sigma_{\rm max}/m_\bullet$, thus allowing the maximum disk density at the collision point to be $\sim40-100\times$ larger than the main-sequence star case in Fig.~\ref{fig:pdot_drag}. A BH orbiter at these masses is allowed at moderate and low eccentricity (e.g., $e\lessapprox 0.6$; see Fig.~\ref{fig:pdot_all}) for a system like eRO-QPE2. Hence, a stellar BH solution exists for a system like eRO-QPE2. In general, while EMRI rates are highly uncertain, recent work by \citet{Allievi+2026:emrirates} has shown that if grazing angles between the secondary stellar-mass BH and the disk are required \citep[e.g., see][]{Franchini+2023:qpemodel}, the stellar-mass BH rates are three orders of magnitude lower than the observed QPE rates \citep{Arcodia+2024:rates}. In addition, a cross section defined by their Bondi radii is unable to reproduce the integrated energy of the longer-period QPE sources \citep[e.g.,][]{Linial+2023:qpemodel,Linial+2025:streams,Mummery+2025:collisions}. Thus, if QPEs can be triggered by a variety of orbiters, stellar-mass BHs could still be invoked for short-period (and low integrated energy) sources like eRO-QPE2, or if the inclination constraint suppressing the rates is relaxed \citep{Allievi+2026:emrirates}. For instance, if orbiter-disk interactions from past TDEs or AGN phases and two-body scattering play a significant role in preferentially dragging the BH-EMRI towards low inclinations with the primary BH's equatorial plane \citep[e.g.,][]{Jiang+2025:qpe/agn,Zeng+2026:align}. However, relaxing the rate suppression does not affect the integrated-energy problem for the long-period QPEs \citep[e.g.,][]{Mummery+2025:collisions}. Finally, the low relative velocity between the accretion flow and the BH orbiter required to satisfy the collision cross section constraint poses an additional challenge, in that shock velocities $\ll 0.1 \, c$ seem to produce flare temperatures that are below the observed $\sim 100 \, \rm eV$, as shown in \cite{Vurm+2025:emission}.

\section{Fitting with EMRI trajectory models}
\label{sec:disc_EMRImodels}

Sect.~\ref{sec:discussion} provides constraints for EMRI models from the empirical O-C analysis. To test EMRI trajectory models directly we used two different codes, namely the timing analysis code \texttt{QPE-FIT}\footnote{Publicly available at \href{https://github.com/joheenc/QPE-FIT}{https://github.com/joheenc/QPE-FIT}.} v. 0.1.11 \citep{Chakraborty+2025:qpefit}, while the other is based on \citet{Franchini+2023:qpemodel}, and hereafter named \texttt{FB23} (but we note it will be published separately by Motta et al., in prep.). Both codes perform Bayesian parameter inference for EMRI and SMBH properties using QPE timings under the assumption that they are due to orbiter-disk collisions around a SMBH. The main difference is that while in \texttt{QPE-FIT} disk precession is disjoint from MBH parameters, in \texttt{FB23} it is linked to the BH spin. More details on these models and our fits are reported in Appendix~\ref{sec:app_EMRI}.

In summary, we found no reliable solution to the XMM1-4 and XMM1-5 eRO-QPE2 data. While EMRI code runs did converge to some solutions, the overall inability of modeling both short-term and long-term delays in QPE arrival times, and the high sensitivity of fitted values to both the number of data points used (e.g. censoring some epochs) and the prior volume, led us to consider all solutions as unreliable. We report more details on priors and fitted parameters in Appendix~\ref{sec:app_EMRI}, and report here only the useful lessons learned for QPE modeling. In particular, we note that obtaining good residuals and reasonable parameters with converged posteriors is not a reliable confirmation that a given adopted EMRI model setup is correct. Thus, we advise the community to use EMRI codes together with more empirical approaches like the O-C analysis performed here, when possible. 

\subsection{Two observed events per orbit are disfavored}

Section~\ref{sec:OC} reports analysis of O-C data where all eruptions are analyzed, thus $P$ is traced by $t_{\rm recur}$. In parallel, we performed analogous analysis and tests for the case of odd and even eruptions separately, where $P$ is traced by $2 t_{\rm recur}$ (see Appendix~\ref{sec:app_OC_oddeven}). This analysis is more appropriate to compare with models predicting two observable events per orbit, such as most current disk collision models for QPEs \citep[e.g.,][]{Linial+2023:qpemodel,Franchini+2023:qpemodel,Zhou+2024:qpemodel}. As discussed extensively in \citet{Miniutti+2025:OC} for GSN\,069, if QPEs are observed for both ingresses and egresses in the disk, a key prediction of apsidal precession is that odd and even data would be observed in anti-phase on this timescale. However, from our O-C analysis we found that for both super-period modulations (fitted at $\sim23\,P$ and $\approx 500\,P$ in this case) odd and even eruptions are in phase (e.g., Fig.~\ref{fig:OCbestfit_oddeven} and~\ref{fig:best-fit corner_oddeven} in Appendix~\ref{sec:app_OC_oddeven}). This is a fundamental inconsistency with current collision models with two observable events per orbit, strongly disfavoring such scenarios. The only way to overcome this tension would be to have an additional faster modulation with comparable or larger amplitude, that correlates odd and even data at the apsidal precession timescale. However, while this was still a possibility for GSN\,069 data \citep{Miniutti+2025:OC}, for our eRO-QPE2 dataset it is not possible to conceal a modulation on the timescales between P and $P_{\rm aps}\sim23\,P$ (Fig.~\ref{fig:OCbestfit_oddeven}). As a matter of fact, it would need to be of the order of $\sim2-5\,P$ in order to be sufficiently faster than apsidal and of the same amplitude ($\gtrsim300\,$s), which not only is unlikely to be a physical precession term, but is also not observed in the individual XMM epochs. For instance XMM1 is the observation with the most consecutive eruptions (eight, see bottom panel of Fig.~\ref{fig:OConly}), and it may show at best a hint of residuals at $\sim150\,s$ level.

However, given a solution with EMRI trajectory codes with two observed crossings per orbit had been found before for eRO-QPE2 \citep[e.g.,][see more details in Sect.~\ref{sec:literature}]{Zhou+2024:longterm}, we tried to fit our data with the EMRI trajectory models described above, assuming the standard two observable events per orbit. While both \texttt{QPE-FIT} and \texttt{FB23} models find a solution with good residuals and reasonable parameters, after deeper analysis they were found to be merely a numerical artifact. The most useful diagnostic was to generate the mock O-C from the best-fit posteriors of both \texttt{QPE-FIT} and \texttt{FB23} models and to overlay them to the O-C data. Quite worryingly, the model solutions fitted the higher signal-to-noise XMM data at the crossing of the anti-correlated odd and even predictions from the models (see Appendix~\ref{sec:app_EMRI} and Fig.~\ref{fig:emrimodel_oddeven}). This corresponds to a precession phase in which the semimajor axis is aligned to the disk plane (still with an inclined orbital plane), and collisions happen at pericenter and apocenter. Clearly, it is unrealistic that all the XMM observations, and only the XMM observations, were taken during this relatively rare orbital phase. 
In the attempt to quantify the absurdity of this result, we shifted the best-fit model keeping the XMM observations fixed to quantify the probability of obtaining this alignment by chance. Effectively, we computed the difference between odd and even mock O-C curves from \texttt{QPE-FIT} (top panel of Fig.~\ref{fig:emrimodel_oddeven}) and slid this array by $62$ integers towards lower $N_{\rm QPE}$ (i.e., up to $\sim 6\,$d). We obtained that in none of these 62 configurations, namely in no other configuration other than the one we have, we would obtain XMM1-3 datasets at epochs with such a small difference between odd and even curves (i.e. at the crossings). The result is the same changing any observation between XMM1, XMM2, and XMM3 by $\sim\pm 1\,$d. 

Given that these unphysical solutions comes in addition to the absence of the predicted anti-correlation between odd and even eruptions, we conclude that QPE EMRI models in which two disk crossings occur, and both are electromagnetically observed, are disfavored. 

\subsection{One observed event per orbit: no reliable solution is found}

Based on our O-C analysis, in order for QPE EMRI models to be correct they would require that only the ingress or egress induces an observed event. Naturally, it is also possible that one event per orbit occurs in the first place, for instance for elevated orbital eccentricity or disk eccentricity. However, for a regular source like eRO-QPE2 in terms of recurrence time and eruptions amplitude, we do not expect a large diversity in collision distances or high eccentricity. Thus, we adopt the former scenario for further tests, also considering that most current disk collision simulations show rather asymmetric forward and backwards ejecta \citep{Huang+2025:simul,Jankovic2026:sims}. We adopted \texttt{QPE-FIT} to carry these out, the fastest of the two algorithms used in this work. We report more details on the fitted parameters and comparisons with the observed O-C delays in Appendix~\ref{sec:app_EMRI} and again report here the main lessons learned. We attempted several fits varying the number of data points (e.g., using XMM1-4 data only, or extending out to XMM5) and the prior volume (see Table~\ref{tab:qpefit_priors} in Appendix~\ref{sec:app_EMRI}), and, worryingly, obtained inconsistent solutions. This remains true after exploring different volume sampling algorithms within \texttt{Ultranest}. 
Hence, for caution we do not consider any of these solutions as conclusive and robust. We suspect that this is due to the application of a model with a large, multi-dimensional, and highly-correlated parameter volume and relatively narrow likelihood peaks. Even a relatively transformational campaign such the XMM1-4 dataset presented in this work is too sparse in comparison, let alone the much sparser archival data of other QPE sources.

\section{Comparisons with the previous literature on eRO-QPE2}
\label{sec:literature}

\begin{figure*}
     \includegraphics[width=1.24\columnwidth]{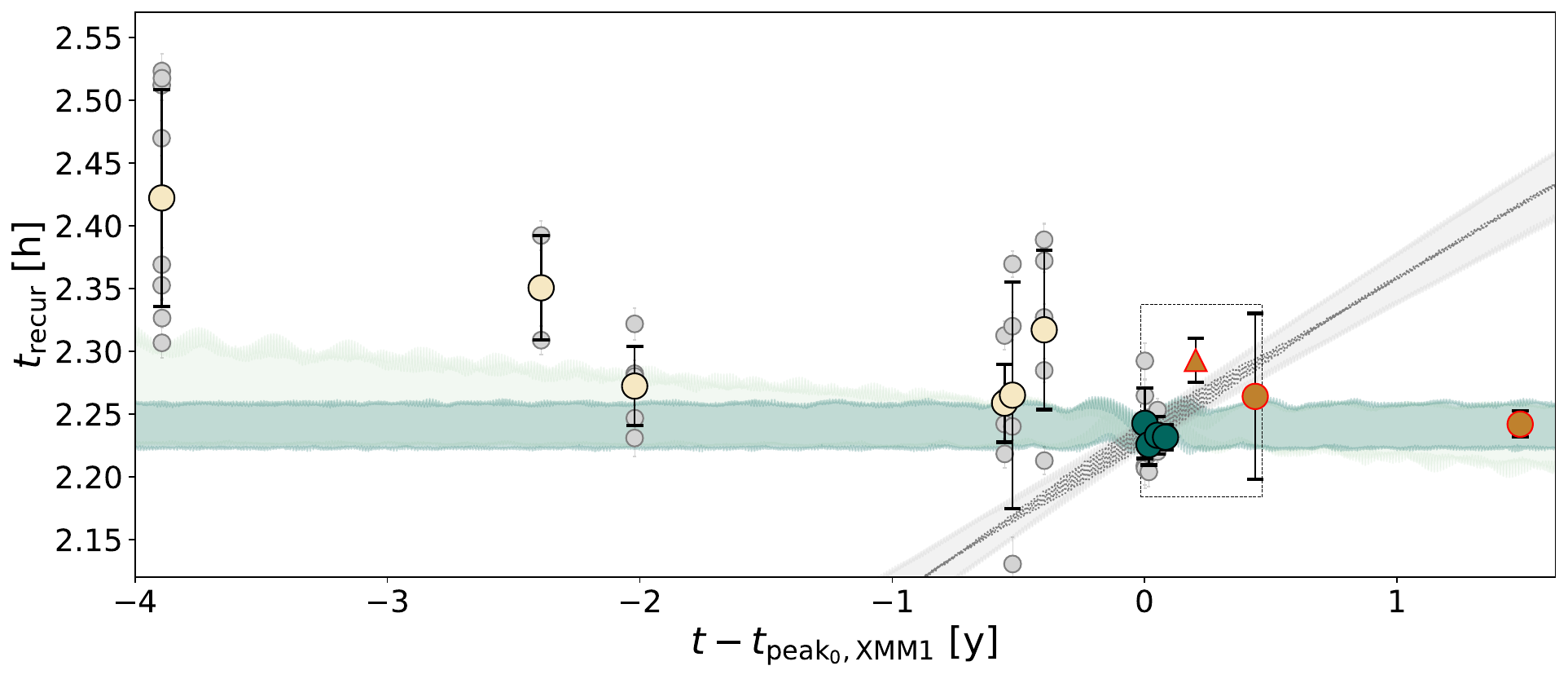}
     \includegraphics[width=0.75\columnwidth]{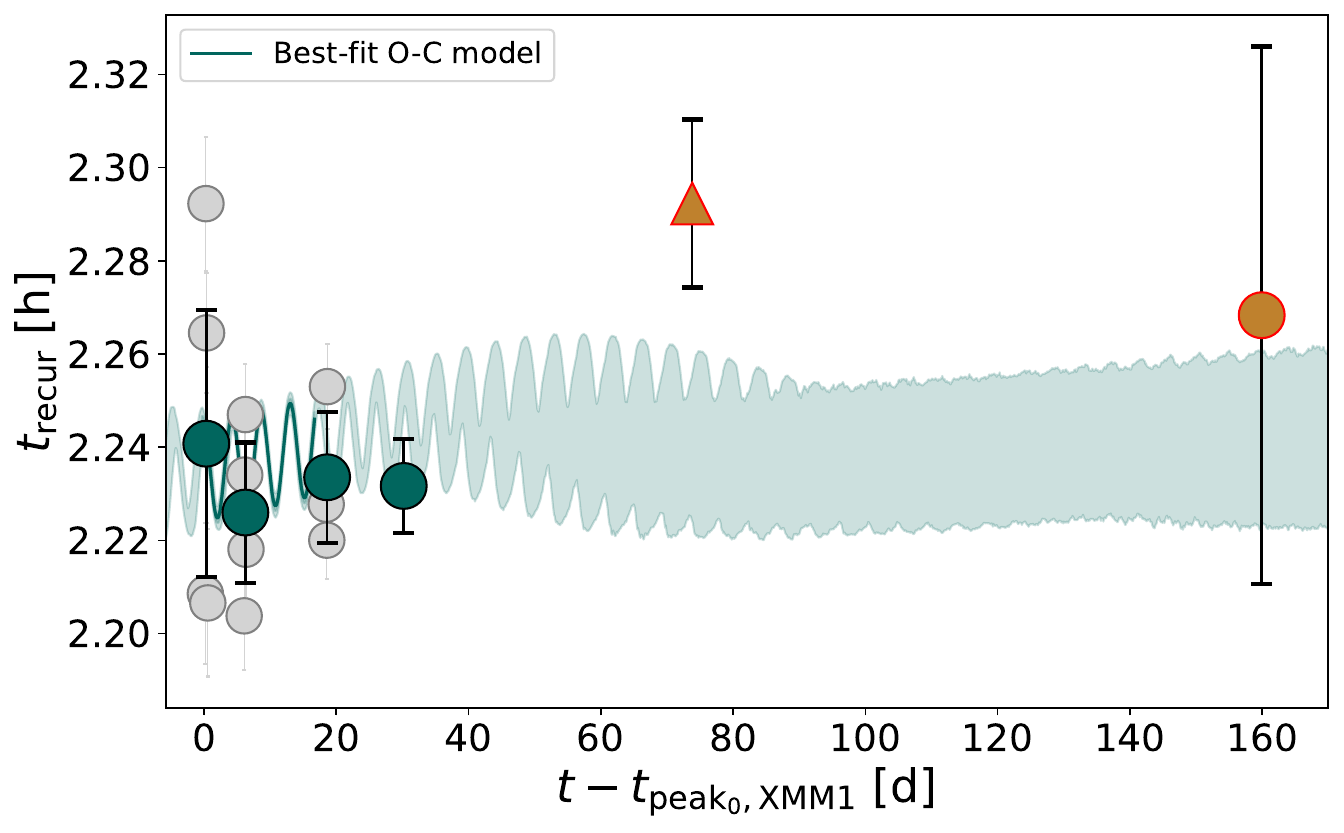}
     \caption{Long-term evolution of the recurrence time ($t_{\rm recur}$), here a proxy of the period of eRO-QPE2, relative in time to the XMM1-4 campaign (green data). Archival 2020-2023 data are shown in yellow \citep{Arcodia+2024:ero2_ticking,Pasham+2024:ero2}, and EP and XMM5 data taken after the XMM1-4 campaign (in addition to recent observations from Guolo et al., in prep) in brown. Filled colors correspond to the per-epoch mean and standard deviation of $t_{\rm recur}$, while individual measurements are shown in gray underneath. The green contour shows the $3\sigma$ uncertainty predictions, extrapolated forward and backward in time, from the best-fit O-C model fitted against the XMM1-4 data alone. The gray contours show the \texttt{lin+pdot+mod} fit, discarded based on the significant underestimate of the archival data $t_{\rm recur}$. The right panel is a zoom-in corresponding to the dashed gray box of the main panel.}
     \label{fig:Trecur}
\end{figure*}

eRO-QPE2 was observed five times after discovery by \emph{XMM-Newton}, and its long-term spectral and timing properties have been reported in \citet{Arcodia+2024:ero2_ticking} and \citet{Pasham+2024:ero2}. The flux of the quiescent accretion disk and the eruptions peak has been found to be consistent within uncertainties across the epochs. Performing the same analysis on the XMM1-5 epochs, we confirm this trend with our new data.

We compared the observed recurrence time $t_{\rm recur}$ from our new data with the literature values in Fig~\ref{fig:Trecur}. In this work, we used the same method to estimate QPE peak times as in \citealp{Arcodia+2024:ero2_ticking}, namely an asymmetric parametric model (see Section~\ref{sec:app_processing}). In addition, we added two archival observations (IDs 0932590101 and 0932590201, taken on 20 Dec. 2023 and 4 Feb. 2024; PI: Wevers), presented in \citealp{Pasham+2024:ero2}, but reprocessed and analysed here following the same procedure as for the rest of the data. The individual consecutive recurrence times are shown in gray in Fig~\ref{fig:Trecur}, with the average and standard deviation shown in yellow for archival data, green for the XMM1-4 data, and brown for ancillary data taken after the XMM1-4 campaign and reported here for the first time. The right panel of Fig~\ref{fig:Trecur} is a zoom-in of the newly reported observations. In dark green, we show the extrapolated O-C best-fit model (see Section~\ref{sec:OC}), and in light green we show the effect of adding a $\dot{P}$ upper limit term ($\dot{P}<2\times10^{-6}$). 
Notably, the span of the best-fit O-C model fitted at the XMM1-4 epoch (green contours) is consistent within uncertainties with the QPE recurrence time observed after the second epoch (i.e. from June 2022 onward; \citealp{Arcodia+2024:ero2_ticking,Pasham+2024:ero2}). 
The recurrence time in 2020 (first epoch in Fig~\ref{fig:Trecur}) is confirmed to be somewhat unique, in that $t_{\rm recur}$ was comparably higher than all other epochs, and it was the only epoch with a significant long-short behavior \citep{Arcodia+2021:eroqpes,Arcodia+2024:ero2_ticking}. Clearly, the O-C solution found at the XMM1-4 epoch (green data and model) does not grasp the full 2020-2025 evolution of eRO-QPE2, as it is expected given that the O-C did not include those data. While the $\dot{P}$ upper limit model relaxes the tension, it is not sufficient to explain the first archival data of eRO-QPE2. A possible explanation for the long-term behavior of eRO-QPE2 on the timescale of years may still be a moderate (consistent with our $\dot{P}<2\times10^{-6}$ upper limit at the July 2024 epoch) dissipative term which is epoch-dependent, for instance the inclination-dependent gas drag period decrease suggested in \citet{Arcodia+2024:ero2_ticking}. A slightly higher $\dot{P}$ term in the past compared to the July 2024 upper limit would qualitatively reproduce the long-term behavior of eRO-QPE2, however this requires further modeling which is beyond the scope of this work. In particular, we note that current QPE disk collision theories do not address the large intra-observation variation in recurrence time (highlighted by the spread of the gray points), which may be due to the currently unmodeled delay between the time of disk crossing and the onset of the QPE emission.


Finally, we compare our analysis in this work with previous attempts inferring orbital parameters from eRO-QPE2 timing data. 
In \citet{Xian+2025:coll}, the authors focus on reproducing the possible $t_{\rm recur}$ decay of eRO-QPE2 (e.g., Fig.~\ref{fig:Trecur}), assuming it is a detection of a period decay. However, comparing average $t_{\rm recur}$ values inferred at different epochs is not a precise trace of the orbital period \citep[e.g., see the discussion in][]{Miniutti+2025:OC,Arcodia+2024:ero2_ticking} and, in fact, our O-C analysis does not detect a negative $\dot{P}$ excluding it at an absolute value of $\lesssim 2 \times 10^{-6}$\,s/s. Thus, the solutions found by \citet{Xian+2025:coll} for eRO-QPE2 are not applicable here. Recently, \citet{Zhou+2024:longterm} fitted the archival 2020-2023 observations with their EMRI trajectory code by modeling both ingresses into and egresses out of the disk as observed eruptions, and reported solutions with semi-major axis $\approx 500 R_g$ and, consequently, black hole masses around $\approx 10^{4.7-4.8}$\Msun. There are several reasons why, similarly to our attempts here (Sect.~\ref{sec:disc_EMRImodels}), we do not consider this numerical solution to be fully reliable. First, it was obtained fitting two observable events per orbit, which is a setup strongly disfavored by the O-C odd/even correlation and the data-model comparison shown in Fig.~\ref{fig:emrimodel_oddeven} obtained with two independent EMRI models. Furthermore, the posterior of the semi-major axis spans $\sim 100-900\,R_g$ within $1\sigma$ in \citet{Zhou+2024:longterm}, thus suggesting the difficulty of finding a single solution for the entire 2020-2023. This is perhaps unsurprising given that in this work two independent EMRI models failed at finding a reliable solution on a much better sampled dataset (Sect.~\ref{sec:app_EMRI}). Finally, in \citet{Zhou+2024:longterm} a random noise component is added as a free parameter to account for the unknown delay between disk crossing and emission. While this component is likely present and surely required, as discussed above in this work as well, in their fit it is allowed to reach 2\,ks in eRO-QPE2, and fitted around $\sim200\,$s (reaching 300\,s at $1\sigma$) and even $\sim 700\,$s in some model runs. However, the O-C delays at these amplitude and short timescales were found to be inconsistent with being dominated by a random distribution, and instead more compatible with a deterministic component (Appendix~\ref{sec:OC_modulations_real} and Fig.~\ref{fig:residuals}). 


\section{Summary and Conclusions}

In this work we reported our timing analysis of eRO-QPE2 with a multi-mission X-ray campaign. 
We performed O-C analysis (Sect.~\ref{sec:OC}), where the only model assumption comes from building the noise model. We found that for models like accretion disk instabilities and red-noise processes, the O-C delays of eRO-QPE2 are reproduced entirely by a QPE recurrence time following a damped random walk, although the fitted parameters are highly uncertain and lack predictive power (Appendix~\ref{sec:app_OC}). Instead, assuming an underlying periodic clock, and testing the O-C delays against white noise 
we found that eRO-QPE2 data are well described by a periodic source 
and a sum of two hierarchical super-period modulations. 

The latter scenario provides numerous constraints for current QPE models. For QPEs as colliding EMRIs, O-C data of eRO-QPE2 are inconsistent with, and disfavor, two observable events per orbit. Our analysis shows that odd and even eruptions are in phase at all the epochs spanned by our baseline (Fig.~\ref{fig:OConly_oddeven},~\ref{fig:OCbestfit_oddeven}, and~\ref{fig:best-fit corner_oddeven}). This result confirms what was found for GSN\,069 by \citet{Miniutti+2025:OC}. 
Furthermore, no reliable solution could be found for models with two eruptions per orbit, even with robust EMRI trajectory models (Sect.~\ref{sec:disc_EMRImodels} and Fig.~\ref{fig:emrimodel_oddeven}). Hence, EMRI disk collision models are only viable if only one event per orbit is observed, be it whether only one crossing per orbit occurs, or whether only one of the two crossings is observed. Interpreting the O-C results in this light obtains the following constraints (Sect.~\ref{sec:discussion}):

\begin{itemize} 
    \item The fitted period is $P\sim8064\,$s\,$\sim 2.24\,$h. The shorter super-period modulation has a period of $\sim4.4$\,d (or $\sim 47$\,P) and an amplitude of $\sim 289\,$s (see Fig.~\ref{fig:best-fit corner}). For the longer modulation, the most conservative estimate would be to consider it a lower limit, namely $\gtrsim 918\,$P at $1\sigma$ ($\gtrsim 600\,$P at $3\sigma$). However, by adding more data to the O-C it appears constrained at $\approx95$\,d (or $\approx 1014\,$P), with an amplitude $\sim3.5\,$ks. 
    
    \item There is no evidence of a strong dissipative term such a period increase or decrease, and we obtain a $3\sigma$ limit on the absolute value of a period decrease at $|\dot{P}| \lesssim 2 \times 10^{-6}\,$s/s. Comparing this limit with the predicted orbital decay due to GW emission rules out IMBH orbiters down to moderate and low eccentricity (we rule out $\gtrsim250\,$\Msun\, at $e\sim0.5$, and $\gtrsim1130\,$\Msun\, at $e\sim0.1$), and high-eccentricity orbiters at all masses including sub solar (Fig.~\ref{fig:pdot_all}). Thus, our campaign excludes most of the required mass and eccentricity parameter space for models with WDs on a very high eccentricity orbit \citep[e.g.,][]{King2020:WD,Wang+2022:wd,Chen+2023:wd}. 



\item In the orbiter-disk collisions framework, the shorter super-period modulation over $\sim4.4$\,d can be interpreted as apsidal precession, which through Eq.~\ref{eq:apsidal} and~\ref{eq:apsidal_mass} implies semimajor axis $a\sim140 R_g$ and $M_{\rm BH}\sim1.5\times 10^5\,$\Msun, and the amplitude of the modulation can be reproduced for $e \sim 0.1$. 

\item The longer modulation over $\approx95$\,d cannot be attributed to EMRI nodal precession, as the observed amplitude of $(3-4)$\,ks is far larger than the maximum time delays from EMRI nodal for the most favorable inclination (edge on disk with low EMRI orbit inclination), which are of order $\sim a/c\approx 100\,$s. 

\item While a possible solution for the longer modulation exists for disk precession (for a given spin, disk surface density profile, radial extent, aspect ratio and viscosity parameter), the same solutions may induce alignment of the disk \citep[e.g., see][]{Franchini+2016:prec} on a timescale shorter than the current disk lifetime lower limit of $\sim5.5\,$y (see top panel of Fig.~\ref{fig:mbhb_all}). Thus, the parameter space for disk precession is narrow, and it will only shrink further over time if more observations will increase the QPE lifetime lower limit. 

\item A possible solution for the longer modulation exists with a hierarchical triple system, as both amplitude and period are reproduced by an outer binary at a separation of $\sim 0.4\,$mpc (thus a factor $\sim400$ larger than the inner EMRI separation, implying a stable system) at any inclination and tertiary masses $M_2$ within a factor $\sim(0.1-1)$ of the EMRI primary ($M_{\rm BH}\sim1.5\times 10^5\,$). No solutions exist with $M_2>M_1$ (see bottom panel of Fig.~\ref{fig:mbhb_all}). 

\item We investigated the parameter space allowed for the EMRI secondary mass $m_\star$ and disk density $\Sigma$, in case of a stellar orbiter and orbital decay due to hydrodynamical gas drag \citep[e.g., ][]{Linial+2023:qpemodel,Yao+2025:simul}. As the O-C $\dot{P}$ implies a maximum disk density, and the QPE integrated energy a minium disk density, the parameter space for disk collisions depends on the cross section of the object. Our constraints (see Fig.~\ref{fig:pdot_drag}, and Fig.~\ref{fig:pdot_drag_stripped}) indicate that main-sequence stars colliding as ``bullets'' are disfavored. Instead, main-sequence stars in which the colliding area is determined by stellar streams, and/or stars with smaller radii compared to main-sequence (e.g. akin to stripped-envelope stars) are allowed. A stellar-mass BH orbiter is also still viable for a system like eRO-QPE2.

 \item We also made use of EMRI trajectory models as a consistency check to vet our O-C interpretation with one event per orbit (Sect.~\ref{sec:disc_EMRImodels}), although no fully reliable solution was found. While all model setups converged to a solution, none was able to recover both short-term and long-term timing data. Perhaps more worryingly, we found that best-fit solutions changed based on both the number of data points used (e.g. censoring some epochs) and the prior volume, which led us to consider any of them unreliable (see Appendix~\ref{sec:app_EMRI} for more details). We suspect that this is due to the application of a model with a large, multi-dimensional, and highly-correlated parameter volume and relatively narrow likelihood peaks. 
 Hence, we advise to use EMRI codes together with more empirical approaches like the O-C analysis performed here, when possible, and encourage caution in the interpretation of any solution. It is also likely that additional physics is required in the current EMRI-disk collision models.

\end{itemize} 

The results found in this work for eRO-QPE2 can be scaled to the full QPE population with some educated assumptions. For instance, short-period systems are considered less pathological than the longer-period QPEs with higher integrated energy per burst \citep[e.g.,][]{Yao+2025:simul,Mummery+2025:collisions}. In the latter case, it will be crucial to understand the nature of the emission and the delay between disk crossing and the start or peak of the X-ray emission. For the QPE sources with more complex timing, we note that if the super-period modulations are not hierarchical, but have comparable periods and/or amplitudes, the expected timing behavior would be more chaotic \citep[see the simulated plots in][]{Chakraborty+2025:qpefit}. Moreover, while a mechanism like apsidal precession is always expected, its amplitude is a function of $a$ and $e$, thus it might not be always detectable. In eRO-QPE2, the longer modulation is the most uncertain component, although from current data it is currently impossible to reconcile with EMRI nodal precession, hard to reconcile with disk precession, while a hierarchical triple system is possible. Clearly, not all QPE sources have to share the same configuration, thus a longer super period does not have to be ubiquitous. 
The $\dot{P}$ upper limit we obtained with this O-C analysis is currently the only reliable data-driven orbital decay constraint for QPE sources. A result that can be generalized is the exclusion of high eccentricity orbiters at all masses (Fig.~\ref{fig:pdot_all}), which formally disfavors partial disruptions of highly-eccentric WDs. Similarly, the constraints shown in Fig.~\ref{fig:pdot_drag} for stellar orbiters, obtained from the tidal radius and the QPE lifetime bounds, can be estimated for all QPE sources. Hence, this work thus offers some legacy constraints from all the current QPE models, which are required to be modified given the current inconsistencies. Naturally, this work also opens the way for entirely new paradigms that can be tested with this multi-mission dataset. 

\begin{acknowledgments}

R.A. is deeply grateful to Z. Arzoumanian, K. Gendreau, and B. Cenko for the positive response to the NICER and \emph{Swift}/XRT ToOs that significantly enriched this work, and the flexibility in maneuvering the complex multi-mission scheduling.

R.A. was supported by NASA through the NASA Hubble Fellowship grant \#HST-HF2-51499.001-A awarded by the Space Telescope Science Institute, which is operated by the Association of Universities for Research in Astronomy, Incorporated, under NASA contract NAS5-26555. 
G.M. acknowledges support from grant n. PID2023-147338NB-C21 funded by Spanish MICIU/AEI/10.13039/501100011033 and ERDF/EU. 
This research benefited from interactions at workshops funded by the Gordon and Betty Moore Foundation through grant GBMF5076.
AF acknowledges financial support from the Unione europea - Next Generation EU, Missione 4 Componente 1CUP G43C24002290001.
A.Mu. acknowledges support from the Ambrose Monell Foundation, the W.M. Keck Foundation and the John N. Bahcall Fellowship Fund at the Institute for Advanced Study.
I.L. is supported by NASA through the NASA Hubble Fellowship grant \#HST-HF2-51581.001-A awarded by the Space Telescope Science Institute, which is operated by the Association of Universities for Research in Astronomy, Incorporated, under NASA contract NAS5-26555, and acknowledges support from a Rothschild Fellowship and The Gruber Foundation, as well as Simons Investigator grant 827103.
M.G. is funded by Spanish MICIU/AEI/10.13039/501100011033 and ERDF/EU grant PID2023-147338NB-C21.
GP acknowledges financial support from the European Research Council (ERC) under the European Union’s Horizon 2020 research and innovation program HotMilk (grant agreement No. 865637) and from the Framework per l’Attrazione e il Rafforzamento delle Eccellenze (FARE) per la ricerca in Italia (R20L5S39T9).
EQ acknowledges support through the European Space Agency (ESA) Research Fellowship Programme in Space Science. A.S. acknowledges the financial support provided under the European Union’s H2020 ERC Consolidator Grant ``Binary Massive Black Hole Astrophysics'' (B Massive, Grant Agreement: 818691) and Advanced Grant ``PINGU'' (Grant Agreement: 101142079).
M.B. acknowledges support from the Italian Ministry for Universities and Research (MUR) program “Dipartimenti di Eccellenza 2023-2027”, within the framework of the activities of the Centro Bicocca di Cosmologia Quantitativa (BiCoQ).

\end{acknowledgments}

%



\software{astropy \citep{2018AJ....156..123A}, 
          swifttools.ukssdc for light curves \citep{Evans+2007:xrtlc,Evans+2009:xrtlc},
          }



\appendix

\section{Details on X-ray data processing and analysis}
\label{sec:app_processing}
\renewcommand{\thefigure}{A.\arabic{figure}}
\renewcommand{\thetable}{A.\arabic{table}}
\setcounter{figure}{0}
\setcounter{table}{0}

Here, we report mode details on the X-ray processing and analysis of the multi-mission dataset used in this work.

\subsection{X-ray data processing}

XMM data were processed using SAS v. 20.0.0 and HEAsoft v. 6.31. Products were extracted from source (and background) regions selecting a circle of 40\arcsec centered on eRO-QPE2 (in a nearby source-free 40\arcsec circle). For all epochs, XMM light curves were extracted in the $0.2-2.0$\,keV range. All good time intervals were kept and since eruptions are soft and background flares are hard, eruptions can still be confidently detected in $0.2-2.0$\,keV range (see also \citealp{Arcodia+2024:ero2_ticking}). We verified a posteriori that the arrival time of the contaminated eruptions is not offset in the presence of background flares, but only more uncertain, by comparing the $0.2-2.0$\,keV with an even softer $0.2-1.0$\,keV extraction. In any case, flares affect only part of XMM1 and XMM3, and out of 21 eruptions detected by XMM in total, only 5 have occurred during a background flare, and only one is significantly affected. For XMM5, the pn camera only recorder half of the first eruption in the exposure, and we used MOS cameras to estimate its peak time.

For XRT, we processed data with the online data analysis tool \citep{Evans+2009:xrt} extracting a light curve between $0.3-2.0\,$keV and initially adopting a per-snapshot binning (more details on the analysis below). NICER data were processed using \texttt{HEAsoft} v6.33 and \texttt{NICERDAS} v12. Data products were reduced using \texttt{nicerl2} with no overshoot or undershoot screening, including both orbit day- and nighttime data, and autoscreening disabled. We then grouped the event files into 200\,s GTIs (using \texttt{nimaketime} and \texttt{niextract-events}), barycenter-corrected the events using \texttt{barycorr}, and extracted spectra for each GTI using \texttt{nicerl3-spect} to follow a time-resolved spectroscopy approach for reliable source light curve extraction. Each spectrum was first fit in a broadband range (0.25-10.0 keV for orbit night/0.38-10.0 keV for orbit day) with the \texttt{SCORPEON} semi-empirical spectral background model leaving solar wind charge exchange line normalizations left free to vary. We then added an absorbed blackbody model (\texttt{tbabs}$\times$\texttt{ztbabs}$\times$\texttt{bbody}) to assess whether the addition of a source component significantly improved the fit. We consider a source ``detection'' any GTI in which the addition of the source improved the background-only fit by $\Delta$C-stat$\geq 25$, a cutoff determined empirically as a suitable boundary between robust detections and background noise. We refer the reader to Section 2.1 of  \citet{Chakraborty+2024:ero1} for further details. For EP, data were processed using standard recipes with the FXT Data Analysis Software Package (FXTDAS) v1.10. Source light curves were extracted from barycenter corrected event files between 0.2 and 5 keV with the Heasoft \textsc{XSELECT} task, using circular regions with 60" radii. The lightcurves from the two telescope modules (FXT-A and FXT-B) were then binned with 200s bins and summed. 

\subsection{QPE arrival times and uncertainties}

Here, we report more details on the estimate of QPE peak times, and the impact of some of the assumptions and analysis techniques on the final results of this work.

\subsubsection{The QPE flare parametric model}

We note that adopting a given parametric model shape over another only impacts arrival times as an offset term, thus it has little impact on our conclusions since both O-C techniques and EMRI trajectory codes express peak times relative to an arbitrary start time. To test this quantitatively, we fitted all eruptions also with a symmetric Gaussian model. We found that the average offsets (with their average uncertainty) between double exponential and Gaussian parametric models are compatible across various epochs ($-151\pm87\,$s, $-168\pm47\,$s, $-134\pm38\,$s, and $-132\pm44\,$s, respectively for XMM1-4), and across eruptions of a single epoch as well, which are all found within a $\sim24\,$s standard deviation (using XMM2 as example). Hence, we confirm the model choice for the QPE eruption profiles has no impact on our timing solutions, and we adopt the double exponential model since it generally performs better than a symmetric profile. 

\subsubsection{Extracting QPE peak times from XRT and NICER data}

For XRT, we note that the online tool, while yielding a series of detections around $\gtrsim 0.02$\,c/s, has also other more ambiguous cases including very low count rates and/or errors larger that the count rate value. To treat this dataset more robustly and systematically, we performed aperture photometry in all the XRT snapshots computing the no-source binomial probability $P_b$ (following the method described in ~\citealp{Arcodia+2024:mbhs}). The source aperture adopted was $20$\arcsec, while background counts were extracted in an annulus with inner and outer radius of $80$\arcsec and $320$\arcsec, respectively. We start the selection of XRT peaks by retaining the 10 snapshots with $P_b<0.003$, 5 of which are within the XMM1-XMM4 campaign. For these, we re-extracted XRT products with different selections of time bins ranging within $\sim300-500\,$s to investigate whether smaller time bins would indicate any intra-snapshot variability, thus a more confident presence of a peak. All 5 snapshots within the XMM1-4, once decomposed in smaller time bins, indicate the presence of the eruptions peak either in a central time bin, or an edge (e.g., Fig.~\ref{fig:xmm1-4_nicerxrt_lcs}). Out of the remaining 5 snapshots with $P_b<0.003$ which are outside the XMM1-4 campaign (which is not used for the O-C analysis), we only retain one that shows clear evidence of a detected peak. 

NICER shares similar orbital constraints, but our processing method adopting spectral analysis in the single good time interval does not suffer from retention of spurious detections as much as the automated XRT online tool. Thus, QPE arrival times can be immediately identified. For both XRT and NICER, we cannot fit eruption profiles due to the more coarse constraint that these data provide (e.g., Fig.~\ref{fig:xmm1-4_nicerxrt_lcs}), and instead assign peak times and errors as follows: if more than one immediately consecutive detection is present (i.e. with no orbital gaps in between) the peak time is assigned to the brightest, otherwise to the single data point. To justify the latter choice, we remind that these detected count rates are compatible with expected values based on the peak fluxes observed by XMM and that eruptions are only detectable for $\sim600-800\,$s by XRT and NICER, thus these detected good time intervals must encapsulate the actual QPE peak. For peak time uncertainties we proceed as follows: if the peak is somewhat resolved (i.e. if 3 or more consecutive bins are detected, and the central one is clearly the brightest) we assign an uncertainty equal to half the time bin size of the light curve, which is 200\,s for NICER and typically 300\,s for XRT (e.g., see the top right and bottom left panels in Fig.~\ref{fig:xmm1-4_nicerxrt_lcs} for examples); if the peak is only partially resolved (i.e. if 2 or more consecutive bins are detected, but the edge one is clearly the brightest), we assign an uncertainty equal to the full time bin size of the light curve (e.g., see the top left panel in Fig.~\ref{fig:xmm1-4_nicerxrt_lcs} for an example); if the eruption is only observed with a single data point the uncertainty on the peak time is 439\,s, which is half the total near-peak duration of the eruptions based on XMM data, and the full error span includes the time for which the source is detectable by XRT and NICER (e.g., see the top middle panel in Fig.~\ref{fig:xmm1-4_nicerxrt_lcs} as an example). In one case ($N_{\rm QPE}=231$ observed by XRT) we increase this to 1000\,s (which fully includes how long eruptions last, while still providing useful constraints) given the slightly lower count rate compared to the expected peak, and the lack of neighboring good time intervals. 

\subsubsection{The energy range}

We tested the impact of extracting light curves with slightly different low energy bounds across missions (i.e. $0.2\,$keV for XMM, $0.3\,$keV for XRT, $0.4\,$keV for NICER). We extracted XMM data of XMM2, as an example, between $0.4-2.0$\,keV and compared with the $0.2-2.0$\,keV light curves. We perform the same analysis described above for XMM data and estimated that the difference between peak times for XMM2 eruption has a mean offset of $\sim13\,$s, with a standard deviation of $\sim 43\,$s, and it is thus consistent with zero. We also note that this negligible offset is much smaller than the typical uncertainties of XRT and NICER arrival times, which are in the range $\sim300-1000$\,s, depending on the eruption (see Table~\ref{tab:OCdata}). Hence, this effect is negligible and does not impact our conclusions.

\subsubsection{QPE peak time versus start time}

We note that taking the peak of the QPEs as representative to study the quasi-periodicity of the source is the most sensible, and agnostic, choice to make in this work. However, for completeness we tested whether adopting the start of the eruptions (as done for some EMRI model testing, e.g. \citealp{Zhou+2025:mass}) would imprint any significant differences in the O-C analysis. In essence, we do not expect a significant impact as QPE durations across epochs were found to be consistent within uncertainties in long-term data of eRO-QPE2 \citep{Arcodia+2024:ero2_ticking,Pasham+2024:ero2}. To confirm this, we took as start time of the XMM eruptions the time at which the flux reaches $1/e^3$ of the peak flux. We computed the difference between start and peak times and found that it is a constant offset, within uncertainties, around $\approx 500\,$s. As mentioned above, offsets do not have any impact in the O-C analysis as all times are relative to an arbitrary start time.

\section{More details on the O-C analysis}
\label{sec:app_OC}
\renewcommand{\thefigure}{B.\arabic{figure}}
\renewcommand{\thetable}{B.\arabic{table}}
\setcounter{figure}{0}
\setcounter{table}{0}

\subsection{Event identification}

\begin{table}[t]
\centering
\caption{QPE arrival times (``O" for observed, ``C" for computed using an estimated period) as seconds elapsed from the start of the XMM1-4 campaign ($t_{\rm 0,XMM1}$), which corresponds to $\rm MJD=60489.6419$. Some eruptions are shown in Fig.~\ref{fig:xmm1-4_lcs} and~\ref{fig:xmm1-4_nicerxrt_lcs}.\\}
\setlength{\tabcolsep}{8pt} 
\renewcommand{\arraystretch}{0.95} 
\centering
\begin{tabular}{rrrrl}
\toprule
$N_{\rm QPE}$ & $\rm O$ [s] & $\rm O_{\rm err}$ [s] & $\rm C$ [s] & Instr.\\
\midrule
0   & 732.69   & 35.87   & 732.69   & xmm \\
1   & 8739.45  & 167.72  & 8788.40  & xmm \\
2   & 16830.53 & 39.51   & 16844.11 & xmm \\
3   & 24781.40 & 37.43   & 24899.82 & xmm \\
4   & 33033.59 & 35.66   & 32955.53 & xmm \\
5   & 41185.92 & 29.17   & 41011.24 & xmm \\
6   & 49251.53 & 52.84   & 49066.95 & xmm \\
7   & 57195.27 & 20.94   & 57122.66 & xmm \\
22  & 177196.76 & 439.30  & 177958.30 & xrt \\
29  & 234335.95 & 439.30  & 234348.27 & xrt \\
34  & 273254.13 & 439.30  & 274626.82 & nicer \\
36  & 289634.68 & 300.00  & 290738.24 & xrt \\
45  & 362394.13 & 439.30  & 363239.62 & nicer \\
65  & 523958.62 & 31.83   & 524353.81 & xmm \\
66  & 531892.37 & 27.13   & 532409.52 & xmm \\
67  & 539934.91 & 27.59   & 540465.23 & xmm \\
68  & 548024.16 & 27.16   & 548520.94 & xmm \\
69  & 556009.37 & 25.62   & 556576.65 & xmm \\
182 & 1466300.33 & 150.00  & 1466871.84 & xrt \\
121 & 975273.71  & 439.30  & 975473.55  & nicer \\
198 & 1595954.82 & 32.10   & 1595763.19 & xmm \\
199 & 1603974.79 & 20.27   & 1603818.90 & xmm \\
200 & 1611967.02 & 21.89   & 1611874.61 & xmm \\
201 & 1620077.86 & 24.75   & 1619930.32 & xmm \\
202 & 1627782.69 & 439.30  & 1627986.03 & nicer \\
204 & 1644134.58 & 200.00  & 1644097.45 & nicer \\
213 & 1716576.54 & 100.00  & 1716598.84 & nicer \\
222 & 1788839.48 & 439.30  & 1789100.22 & nicer \\
229 & 1845185.55 & 200.00  & 1845490.19 & nicer \\
230 & 1852571.47 & 1000.00 & 1853545.90 & xrt \\
323 & 2604787.86 & 29.48   & 2602726.89 & xmm \\
324 & 2612822.99 & 21.27   & 2610782.60 & xmm \\
\bottomrule
\end{tabular}
\label{tab:OCdata}
\end{table}

For our XMM1-4 dataset the association with $N_{\rm QPE}$ was performed with $T_{0,\rm est}$ taken from the fitted peak time of the first eruption of XMM1 (i.e., $732.69\,$s after the start of XMM1, $\rm MJD=60489.6419$, which is the origin of the time axis, e.g. Fig.~\ref{fig:xmm1-4_lcs}), and $P_{\rm est}$ taken from the average peak-to-peak QPE recurrence time of the XMM1, XMM2, XMM3 and XMM4 observations (i.e., $8055.71$\,s). Identification of an observed eruption with $N_{\rm QPE}$ was thus performed by associating a computed (``C") event with the closest observed one (``O"), and we report all identification numbers in Table~\ref{tab:OCdata}, and show the O-C data in Fig.~\ref{fig:OConly}. Data points from different missions are highlighted with symbols that follow those in Fig.~\ref{fig:xmm1-4_lcs} and~\ref{fig:xmm1-4_nicerxrt_lcs}, namely circles for XMM, squares for XRT, stars for NICER. A crucial requirement for this technique to be used, and its results interpreted correctly, is that each event can be associated with an integer $N_{\rm QPE}$ with no ambiguity. All the eruptions of our dedicated XMM1-4 campaign can be associated unambiguously as it was designed with this goal in mind. In practice, this is because gaps in between observations are small enough that the cumulative uncertainty of arrival times does not become larger than a small fraction of the QPE recurrence: the predicted `C' is either clearly overlapping in time with an observed eruption, or it is close enough (i.e. much smaller than half of a QPE cycle). More quantitatively, we note that in both Table~\ref{tab:OCdata} and Fig.~\ref{fig:OConly} most absolute values in the O-C lie around $\approx(0.0-0.2)\,$h, with only a couple of data points around $\sim0.4$\,h and only XMM4 around $\sim0.6$\,h, all much smaller than the $\sim2.3\,$h recurrence. 

\subsection{Modeling the O-C data with red noise} 
\label{sec:app_OCnoise}

In the main text, we reported results obtained for models in which an underlying clock is present (i.e. orbital models) and the underlying noise process is assumed to be white. Here, we present an alternative analysis more appropriate for models in which the QPE event is predicted to be regulated and dominated by red-noise processes (e.g., disk instabilities or accretion related processes). 


We model the stochastic contribution by the so-called Damped Random Walk (DRW) through Gaussian processes (GPs), that allow for the description of the stochastic process through a covariance function and assume the data are drawn from a multivariate Gaussian distribution around a mean function. We refer to \citet{Bertassi+2025:drw} for more details. In short, the GP log-likelihood $L$ is given by:
\begin{equation}
    \log L = -\frac{1}{2} ( \boldsymbol{y}-\boldsymbol{\bar{y}})\boldsymbol{\Sigma^{-1}} (\boldsymbol{y}-\boldsymbol{\bar{y}})^T - \frac{1}{2} |\boldsymbol{\Sigma}| -\frac{n}{2}\log(2\pi)
\end{equation}
with $\boldsymbol{y}$ being the set of observations, $\boldsymbol{\bar{y}}$ being the mean function of the process and $\Sigma$ being the covariance matrix.
For the DRW, the covariance function elements are:
\begin{equation}
    \Sigma_{i,j}=k(x_i,x_j)= \sigma^2e^{-\left(\frac{|x_i-x_j|}{\tau}\right)}
\end{equation}
where $k$ is the covariance function, $x_i,x_j$ are two observation times of the time series, $\sigma$ sets the variance scale of the stochastic process, and $\tau$ is the damping timescale of the process beyond which the data become nearly uncorrelated. The GP mean function is taken to be either zero, or a deterministic component consisting of a linear trend (with coefficients $a_{\rm lin}$ and $b_{\rm lin}$) plus a sinusoidal term (with parameters A, P, $\phi$). We used \texttt{celerite} to compute the GP likelihood and \texttt{raynest} as nested sampling algorithm. The priors assumed for the three models are shown in Table~\ref{tab:drw}.

\begin{table}[t]
\centering
\caption{Prior ranges for the red-noise model testing.} 
\begin{tabular}{lcc}
\toprule
Parameter & DRW & DRW + mod \\ 
\midrule
$a_{\rm lin}$ & -- & $\mathcal{U}(-3,3)$ \\ 

$b_{\rm lin}$ & -- & $\mathcal{U}(-3,3)$ \\ 

$\log A$ & -- & $\mathcal{U}(-2,2)$ \\ 

$\log P$ & -- & $\mathcal{U}(-1,4)$ \\ 

$\phi$ & -- & $\mathcal{U}(0,2\pi)$ \\ 




$\log\tau$ & $\mathcal{U}(-1,5)$ & $\mathcal{U}(-1,5)$ \\ 

$\log\sigma^2$ & $\mathcal{U}(-2,2)$ & $\mathcal{U}(-2,2)$ \\ 

\bottomrule
\end{tabular}
\label{tab:drw}
\end{table}

\begin{figure}[t]
     \centering
     \includegraphics[width=0.99\columnwidth]{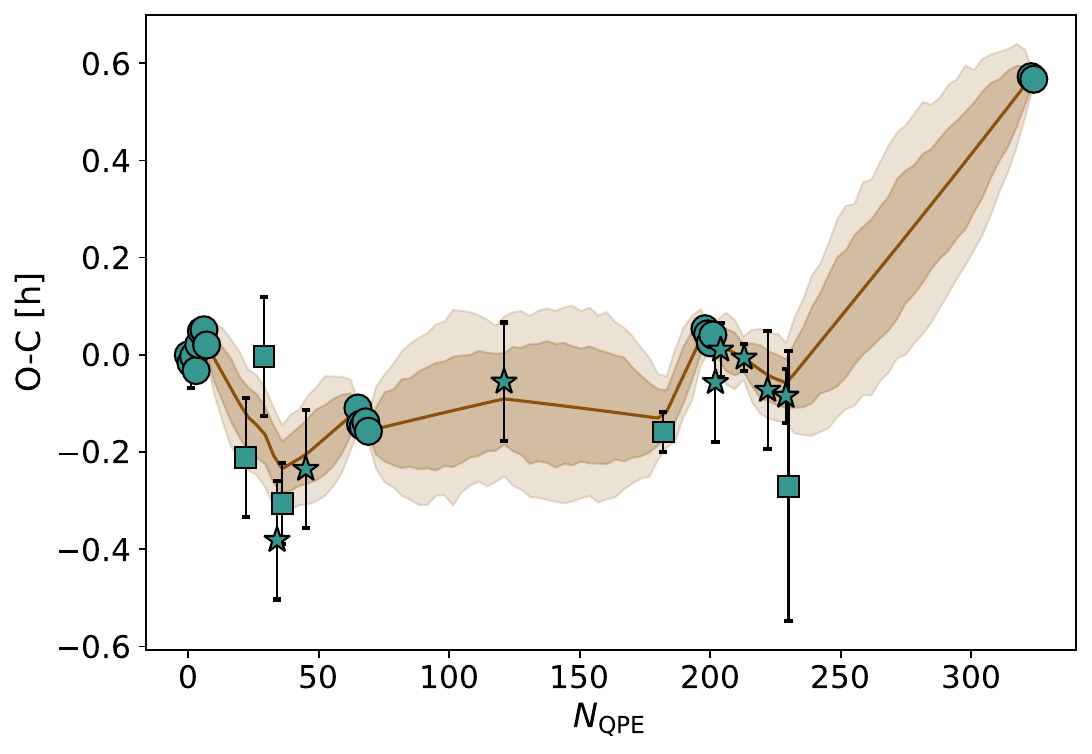}
     \includegraphics[width=0.8\columnwidth]{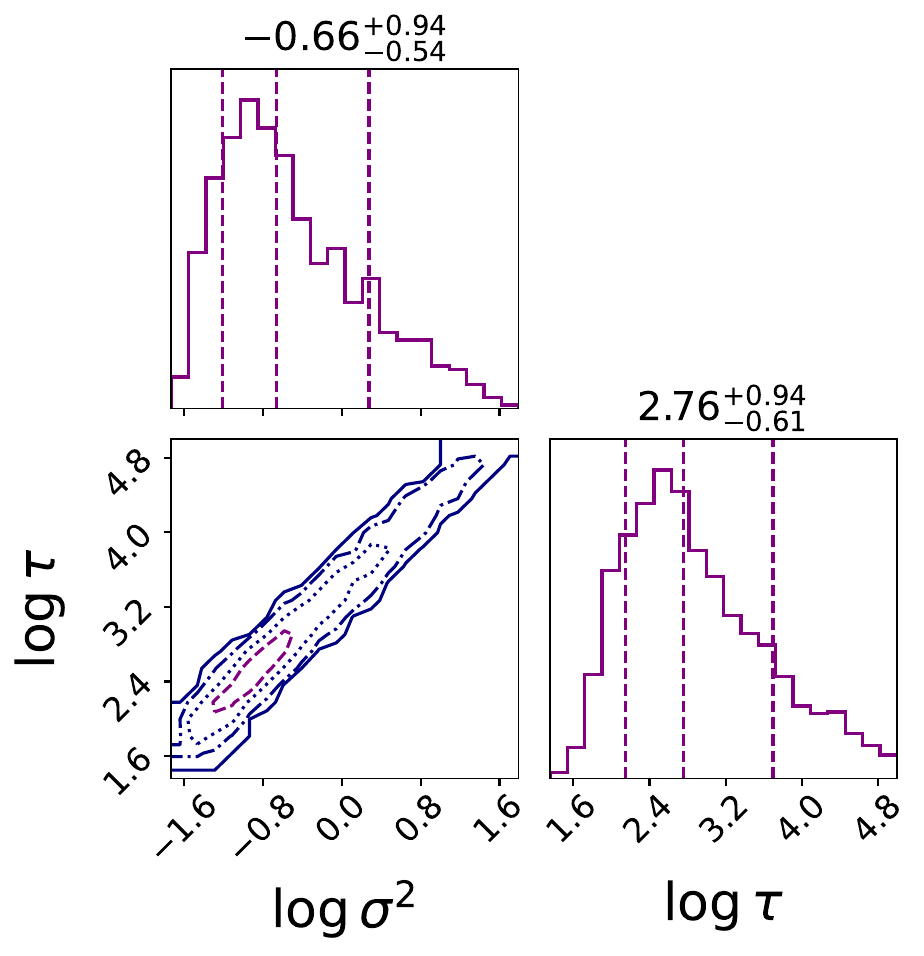}
     \caption{\emph{Top}: same as Fig.~\ref{fig:OCbestfit}, but with the damped random walk model with no deterministic components. \emph{Bottom}: related corner plot.}
     \label{fig:drw}
\end{figure}

We adopted the noise-only model as null and obtained a DRW amplitude $\log\sigma^2 = -0.66^{+0.94}_{-0.54}$ and damping timescale $\log\tau=2.76^{+0.94}_{-0.61}$. We show in Fig.~\ref{fig:drw} the best-fit model predictions for the O-C and the related corner plot. While successful in reproducing the data, the posteriors and model predictions are highly uncertain. Using the non-zero GP mean function to infer the presence of deterministic signals reveals a worse fit statistically, as the null model is preferred with $\Delta \log Z \sim 4$. However, we note that simulations would be required to assess the significance of this model comparison, which is beyond the scope of this work. Furthermore, we note that the fitted $\tau$ spans orders of magnitude and mostly values greater than the total observing baseline, thus predicting that on short timescales QPE recurrence times would be highly correlated. To confirm this, we computed a lag1 autocorrelation coefficient with the 15 QPE recurrence times observed by XMM during the XMM1-4 campaign. However, they appear compatible with zero, thus showing no correlation of consecutive recurrence times. While we note that the number of data points for this autocorrelation is limited, this observable appears in tension with the fitted $\tau$, thus raising further concerns for the DRW-only model. More tests on much better sampled campaigns are required.

\subsection{O-C with deterministic components: model comparison and selection}

\begin{figure*}[t]
     \includegraphics[width=\textwidth]{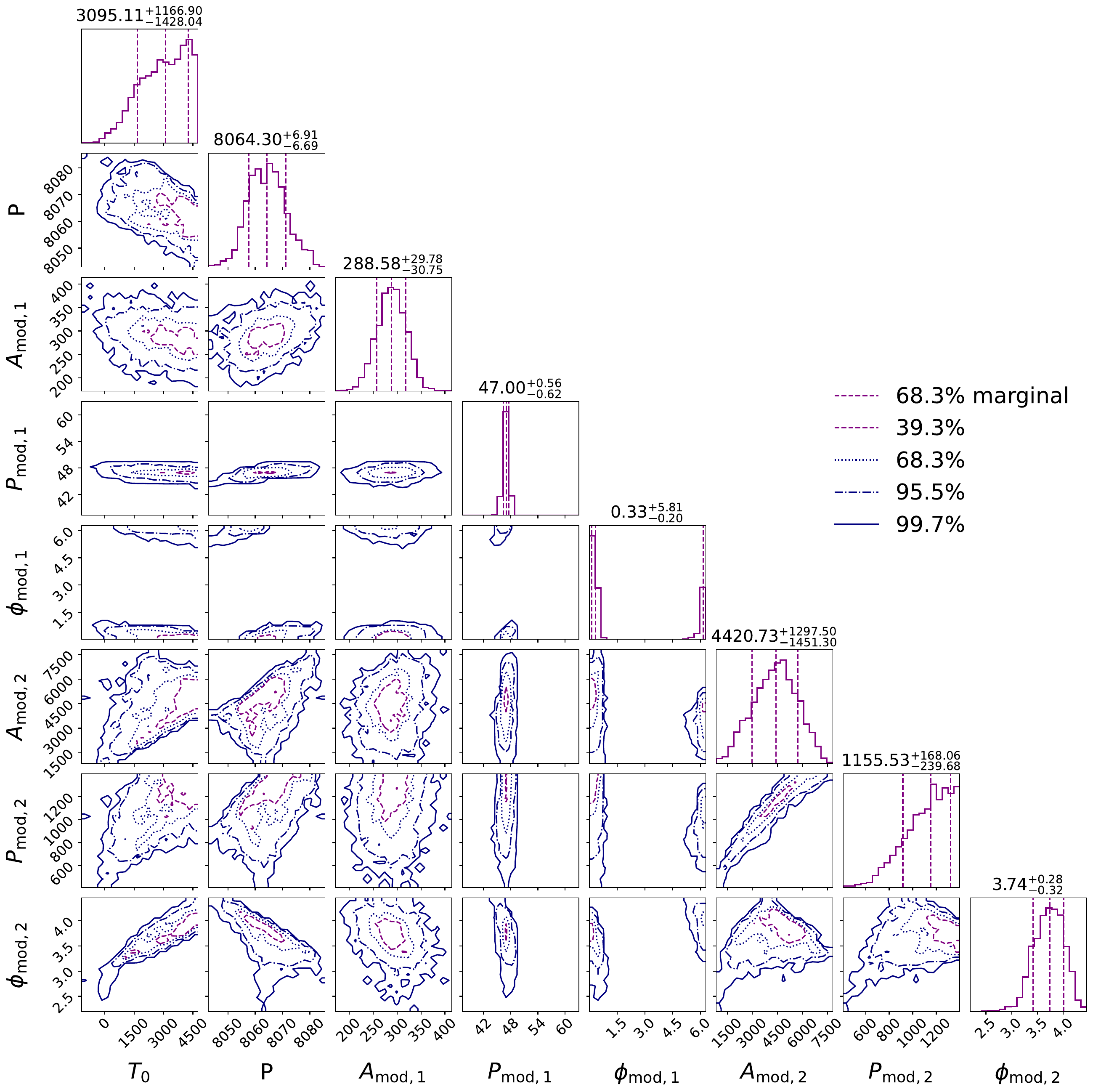}
     \caption{Corner plot of the best-fit model (\texttt{lin+mod$_{\rm1}$+mod$_{\rm2}$}), shown in Fig.~\ref{fig:OCbestfit}.}
     \label{fig:best-fit corner}
\end{figure*}

\begin{figure}[t]
     \centering
     \includegraphics[width=0.99\columnwidth]{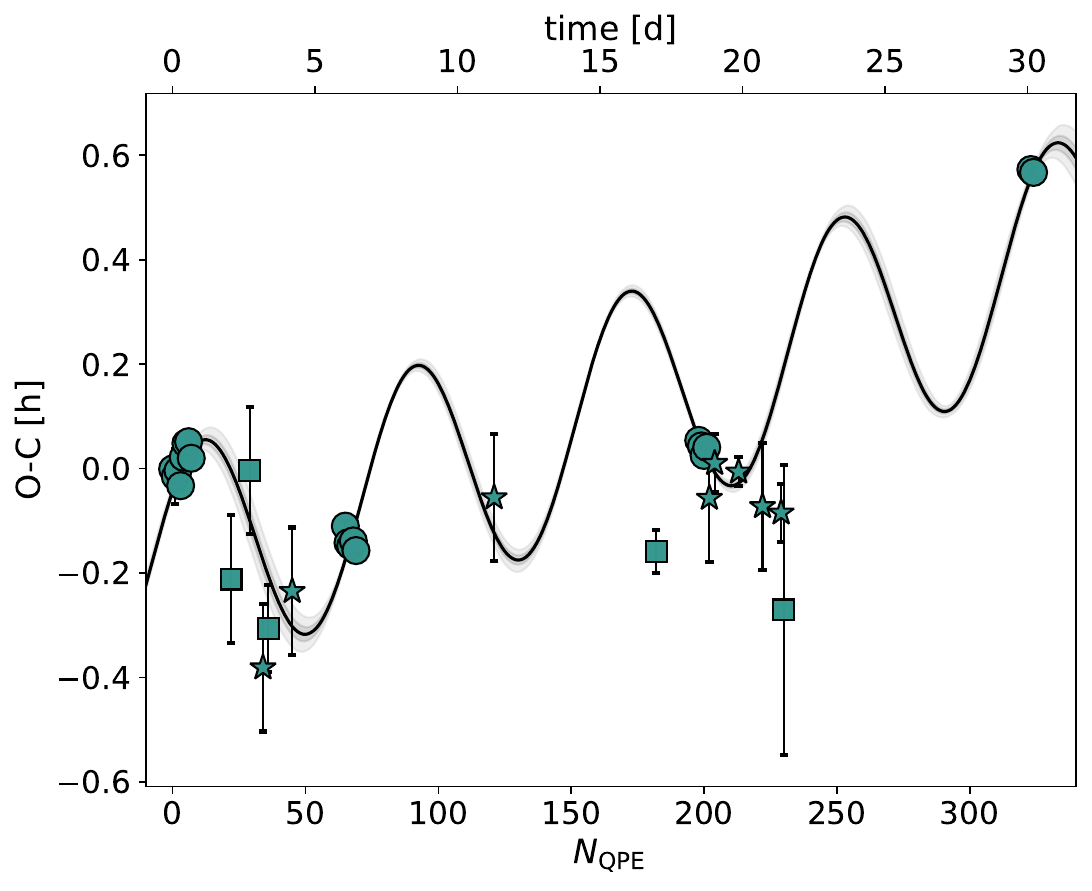}
     \includegraphics[width=0.99\columnwidth]{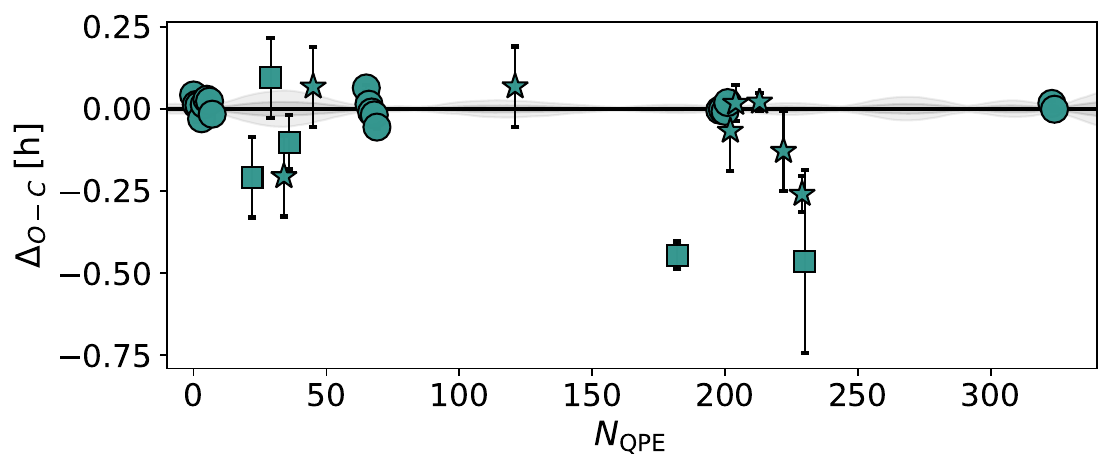}
     \caption{\emph{Top panel}: O-C data fitted by the \texttt{lin+mod} model (the median is shown with a solid line, and $1\sigma$ and $3\sigma$ percentiles as shaded contours), which obtains poor residuals (shown in the bottom panel).}
     \label{fig:OCext}
\end{figure}
\begin{figure}[t]
     \centering
     \includegraphics[width=0.99\columnwidth]{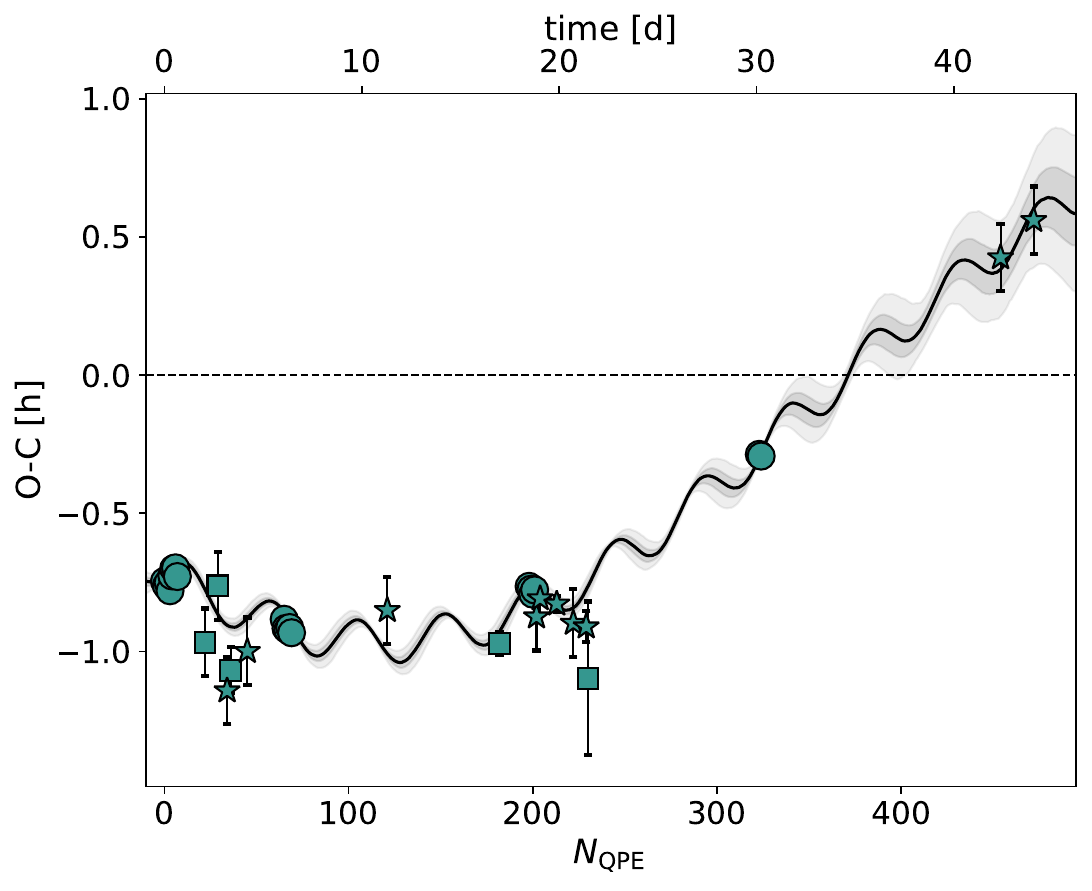}
     \caption{Same as Fig.~\ref{fig:OCext}, but for the `Best +  Aug. data' model. Adding two further eruptions confirms the trend of the longer modulation inferred by the XMM1-4 data alone.}
     \label{fig:OCbest+aug}
\end{figure}

Here, we refer to model comparison and best-fit model selection with deterministic components against white noise. In Sect.~\ref{sec:OC} we mainly presented the adopted best-fit model (\texttt{lin+mod$_{\rm1}$+mod$_{\rm2}$}). In support of the full O-C equation and the discussion on the best-fit parameters reported in Section~\ref{sec:OC} (see, e.g., Fig.~\ref{fig:OCbestfit}), we show here the best-fit corner plot in Fig.~\ref{fig:best-fit corner}. As described in Section~\ref{sec:OC_modulations_real}, to investigate the possible constraint on the longer modulation, we added two further eruptions identified by extrapolating the best-fit model forward in time. Fig.~\ref{fig:OCbest+aug} shows the resulting O-C data and model, which confirm the presence of a modulation on timescales longer than our baseline. Here, we also expand on the preliminary models used to select the above as best-fit model. We first fitted the O-C data with a \texttt{lin+mod} model, namely a single modulation obtained fitting the following equation: $O-C= (T_0-T_{0,\rm est})+ (P-P_{\rm est}) N_{\rm QPE} + A_{\mathrm{mod}} \sin(2\pi N_{\rm QPE}/P_{\mathrm{mod}} + \phi_{\mathrm{mod}})$, which has 5 free parameters ($T_0$, $P$, $A_{\mathrm{mod}}$, $P_{\mathrm{mod}}$, $\phi_{\mathrm{mod,1}}$). The fit yielded poor residuals (see Fig.~\ref{fig:OCext}), as the model is not able to capture both the negative delays of XMM1-3, spanning around $O-C\approx-0.4$\,h, and the positive delay of XMM4. This warranted adding more components. 

\begin{figure}[t]
     \centering
     \includegraphics[width=0.99\columnwidth]{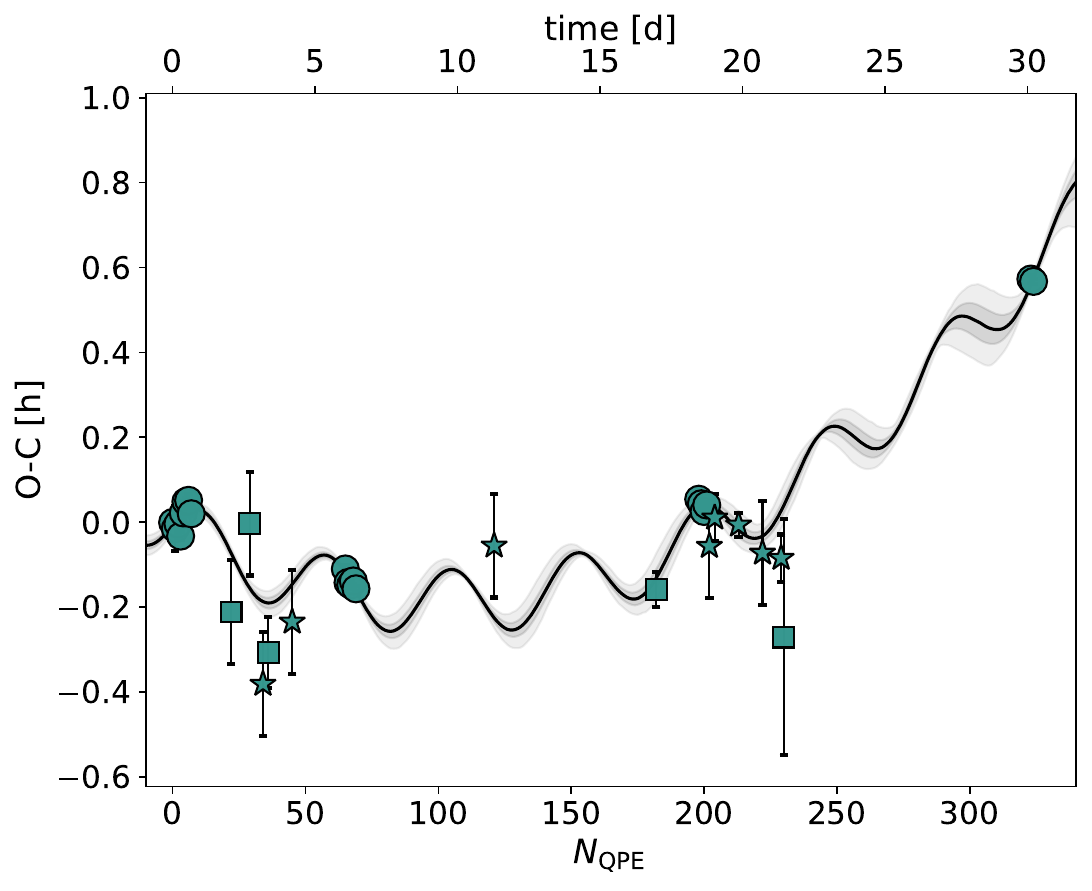}
     \includegraphics[width=0.99\columnwidth]{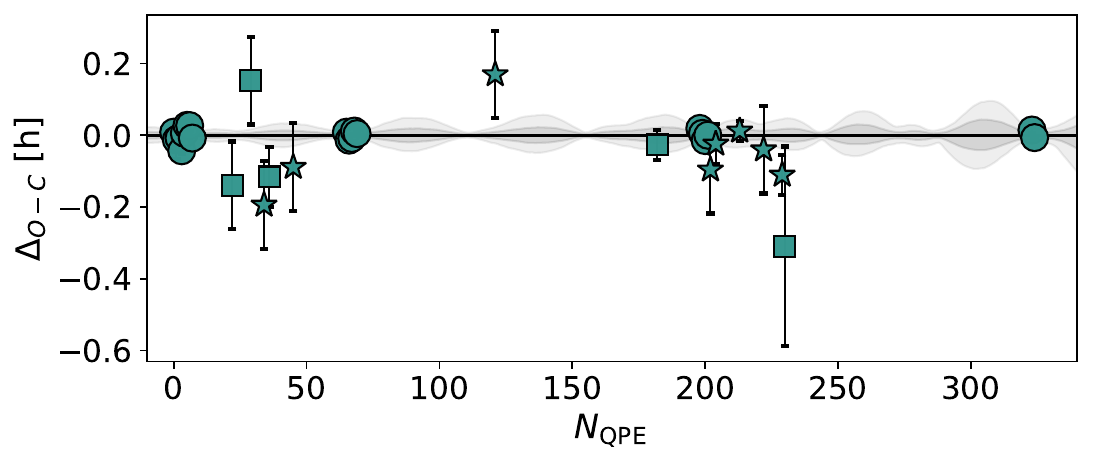}
     \caption{Same as Fig.~\ref{fig:OCext}, but for the \texttt{lin+pdot+mod} model. While a good fit to the XMM1-4 campaign, and while it is not distinguishable from the best-fit model (Fig.~\ref{fig:OCbestfit}), the presence of a strong dissipative term is discarded based on the significant inconsistency with archival data.}
     \label{fig:OCpdotmod}
\end{figure}

We performed two additional fits, one adding a positive dissipative term, $+(1/2\,\dot{\rm P}\,P)N_{\rm QPE}^2$ (\texttt{lin+pdot+mod} model, with positive $\dot{P}$), and one adding a second modulation (\texttt{lin+mod$_{\rm1}$+mod$_{\rm2}$}). Both perform significantly better than the \texttt{lin+mod}, with a difference between the logarithmic Bayesian evidence ($\log Z$) being $\Delta \log Z = 128$ and $\Delta \log Z = 126$, for the \texttt{lin+pdot+mod} and the \texttt{lin+mod$_{\rm1}$+mod$_{\rm2}$} model, respectively, in comparison with the \texttt{lin+mod}. We show the \texttt{lin+pdot+mod} model in Fig.~\ref{fig:OCpdotmod}. While these two models appear equally good at describing the XMM1-4 data, we note that the \texttt{lin+pdot+mod} model requires a strong positive dissipative term ($\dot{\rm P}\approx 1.5\times 10^{5}\,$s/s) which predicts an increasing QPE recurrence time over time. In turn, it predicts that the source had a smaller QPE recurrence time ($t_{\rm recur}$) in the past, namely $\approx 2.1\,$h a year before (in June 2023) and $\approx 2.0\,$h two years before (in June 2022), and so on. Therefore, this model can be immediately discarded based on archival 2020-2023 data of eRO-QPE2, as from June 2022 onward $t_{\rm recur}$ has been observed in the $\sim 2.25-2.30\,$h range \citep{Arcodia+2024:ero2_ticking,Pasham+2024:ero2}, and at even larger values in the 2020 discovery light curve \citep{Arcodia+2021:eroqpes}. We discussed the long-term data of eRO-QPE2 in more detail in Sect.~\ref{sec:literature} and we refer to Fig.~\ref{fig:Trecur}, where the grey contours show the significant discrepancy between the \texttt{lin+pdot+mod} model and the archival data. Thus, the \texttt{lin+pdot+mod} is discarded as a viable description for the O-C data of eRO-QPE2 thanks to the archival observations, and we instead focused on the \texttt{lin+mod$_{\rm1}$+mod$_{\rm2}$} model (shown in the top panel of Fig.~\ref{fig:OCbestfit}). Given that it significantly improved the \texttt{lin+mod} fit and it also obtains good residuals (bottom panel of Fig.~\ref{fig:OCbestfit}) with most data consistent within $1\sigma$ uncertainties, we refrained from adding any more components to avoid overfitting and adopt it as best-fit model. We note that these results, while using simplistic parametric models, are weakly sensitive to the prior volume adopted: varying amplitude prior bounds between tens to tens of thousands of seconds does not impact the fitted values; phases span the maximum bounds $(0,2\pi)$; the bounds of the two modulation periods are separated at $N\sim 100$ (chosen visually from the O-C data), and while this imposes a hierarchy, the fitted values are significantly smaller and larger than this separation bound, for the shorter and longer super-period, respectively. All fitted parameters quoted throughout this work have statistical uncertainties marginalized over all the model parameters. The systematic uncertainties may be large and are unknown at this stage, and a future monitoring campaign with more observed eruptions is needed.

\subsubsection{The robustness of the super-period modulations} 
\label{sec:OC_modulations_real}

\begin{figure}
     \includegraphics[width=0.51\textwidth]{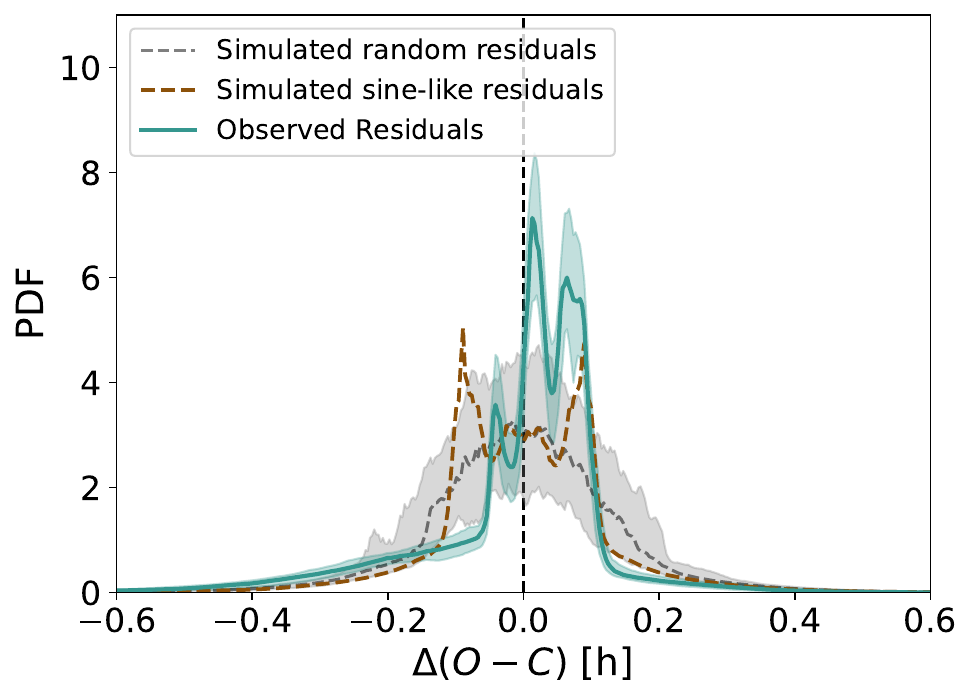}
     \caption{Distribution of O-C residuals after subtracting the linear trend and the longer sinusoidal modulation (e.g., Fig.~\ref{fig:OCbestfit}). The observed residuals are shown in light green and are obtained drawing from the observed uncertainties. Simulated datasets drawn from a random distribution centered at zero, with standard deviation $\sim0.01\,$h (which is the one of the observed residuals) are shown in gray. 
     }
     \label{fig:residuals}
\end{figure}

Here, we evaluated the robustness of the two super-period components fitted by the adopted best-fit model with the deterministic O-C components tested against white noise. 
First, we tested the short-term deterministic sinusoidal component 
analyzing the residual O-C data after subtracting the linear component and the longer modulation from the best-fit model. We note that these residuals only impact the following simulations by providing a realistic standard deviation about the longer modulation, thus it is irrelevant that they were obtained from a fit which includes the shorter modulation too. As a matter of fact, this subtraction approach is the most accurate description of where O-C data would lie about a longer modulation, compared to the poor fit with a single sinusoidal component (Fig.~\ref{fig:OCext}). We randomly extracted 100 fake datasets of O-C residuals using the observed standard deviation ($\sim0.11\,$h), assuming they are randomly distributed as white noise. We added noise to each fake dataset according to the actual measurement errors (i.e. to account for the more uncertain XRT and NICER data points). For each draw we computed an error-weighted kernel density estimate (KDE) and thus obtain a distribution of 100 random KDEs. To compare these simulated random residuals with the true observed ones, we obtained 100 realizations of the observed dataset by perturbing each measurement according to its error bar (again to account for the high uncertainties in XRT and NICER data) and computed an analogous KDE distribution. We show this comparison in the left panel of Fig.~\ref{fig:residuals}. The observed residuals appear inconsistent with the simulated random white-noise distribution, in that they show a much smaller spread and a multi-peaked structure which is, quite intriguingly, peaking around the minimum, maximum and zero, consistent with what sinusoidal residuals would show (we show the simulated sinusoidal distribution in brown in Fig.~\ref{fig:residuals}). 

\begin{figure}[t]
     \includegraphics[width=0.99\columnwidth]{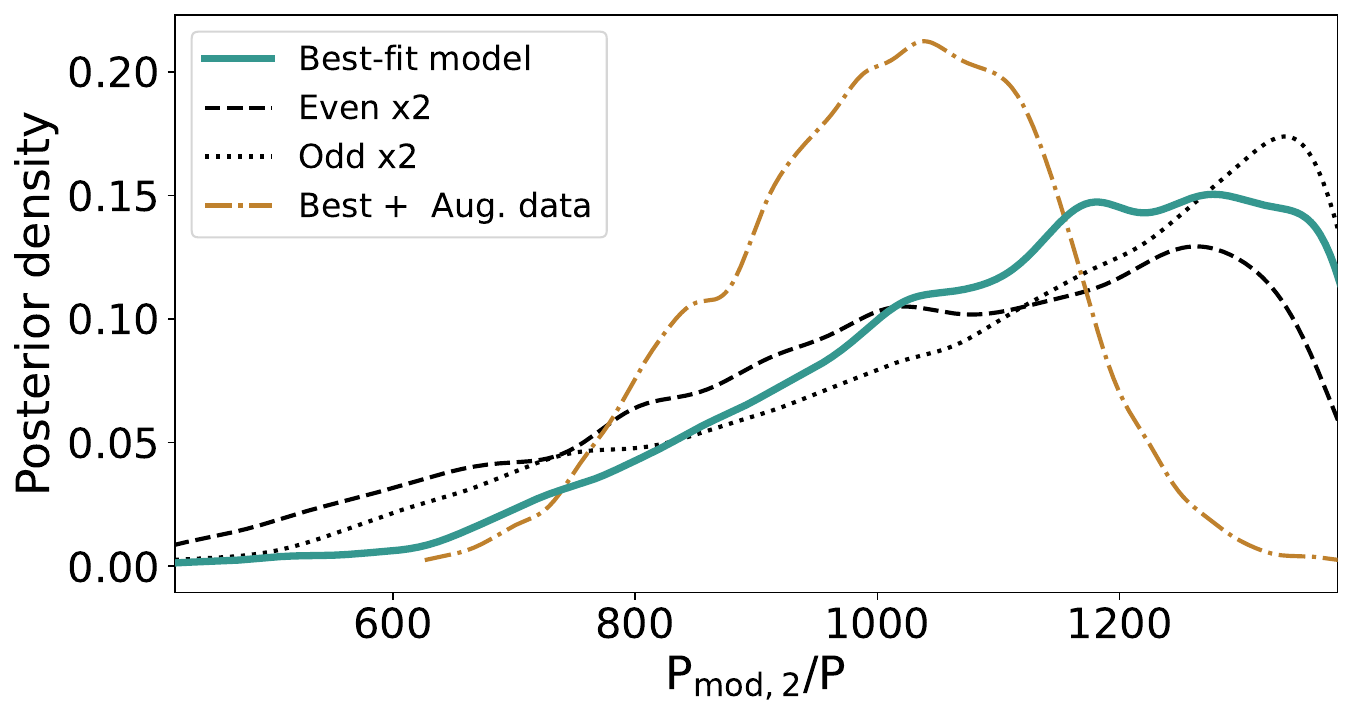}
     \includegraphics[width=0.99\columnwidth]{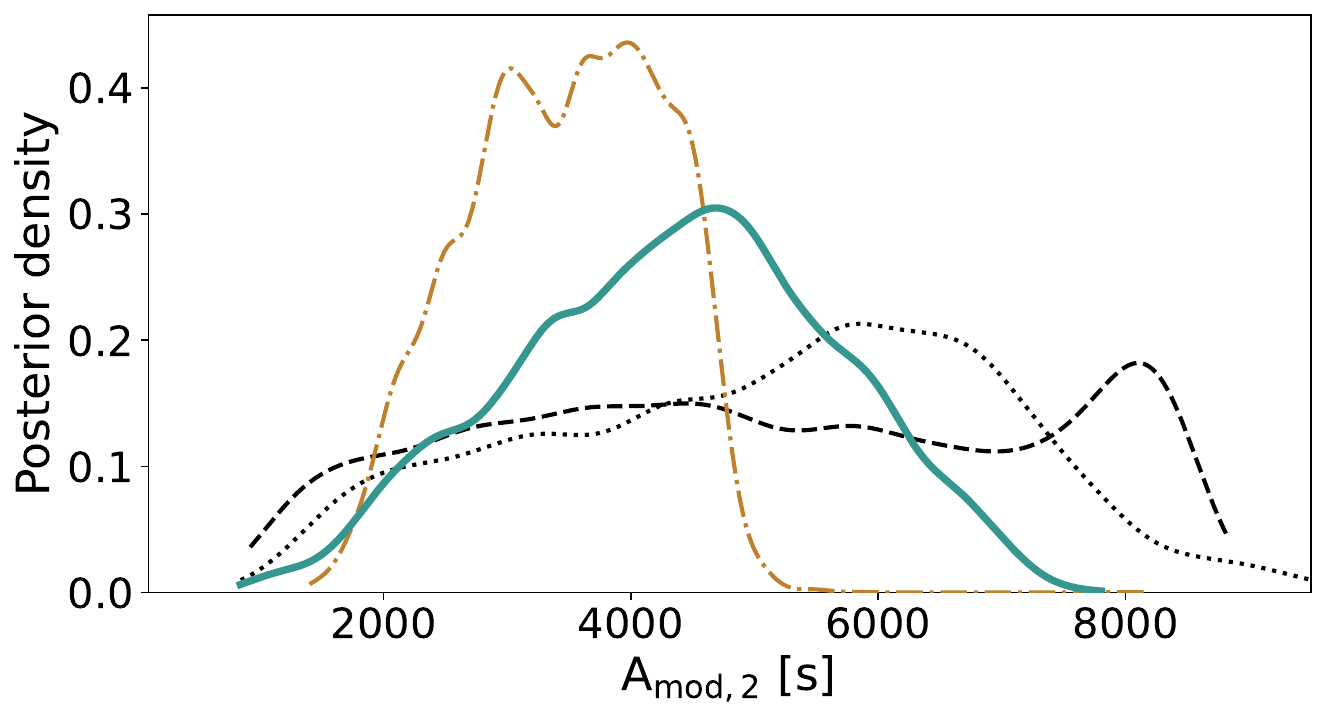}
     \caption{Posteriors of modulation parameters of the O-C fit. The best-fit model (\texttt{lin+mod$_{\rm1}$+mod$_{\rm2}$}) is highlighted in green, with other models in addition to interpret the robustness of the best-fit results. The dot-dashed brown line shows the same best-fit model applied to two further eruptions added after the XMM1-4 campaign.}
     \label{fig:hist_mod}
\end{figure}

Regarding instead the longer modulation (the \texttt{mod$_2$} component), 
we note that while our XMM1-4 campaign is long enough to infer its presence in the data, it is not long enough to probe its full cycle. In our best-fit model (see the corner plot in Fig.~\ref{fig:best-fit corner}), the posterior of $P_{\mathrm{mod,2}}$ is bound at the maximum value allowed (1300), and we deemed unnecessary to extend it any further given that our campaign probes up to $N_{\rm QPE}\sim325$ (and we note a degeneracy with the parameter $T_0$ that should not be bigger than $\sim P/2$). The parameter $A_{\mathrm{mod,2}}$ is also naturally uncertain and spans a few thousand seconds. While it appears a somewhat constrained posterior, one must consider the degeneracy with $P_{\mathrm{mod,2}}$ and its upper bound. We note that NICER, XRT and EP have observed further eruptions after XMM4. They are not part of the main campaign because their $N_{\rm QPE}$ could not be unambiguously identified from the start of the O-C analysis (i.e., using $T_{0,est}$ and $P_{est}$). However, using the XMM1-4 best-fit model to identify further QPEs we were able to add two eruptions observed by NICER around $\sim12.15\,$d and $\sim13.84\,$d after XMM4 (identified with $N_{\rm QPE}=454$ and $472$, respectively). They were identified with updated $T_{0,est}^{ext}=3095.114$\,s and $P_{est}^{ext}=8064.298\,$s from the XMM1-4 best-fit model. Their $O$ ($O_{\rm err}$) values, following the same units and starting point as in Table~\ref{tab:OCdata}, are 3662927.964\,s ($\pm 439.3\,$s) and 3808453.612\,s ($\pm 439.3\,$s). We note that using the original $T_{0,est}$ and $P_{est}$, these two events would have been found halfway between $N_{\rm QPE}=454-455$ and $472-473$, respectively, thus still compatible with the identification using the refined O-C best-fit period. We stress that contrary to the XMM1-4 campaign, this association is model-dependent (as it benefits from the fitted modulations to identify $N_{\rm QPE}$ unambiguously) and we only intend to use it for a consistency check on the longer modulation. Effectively, whether the best-fit model is a good predictor for the following several days, or whether the new solution including the two further bursts changes dramatically. We fit this new dataset and obtain posteriors consistent with the best-fit model for all parameters, within uncertainties. Thanks to the extension of the baseline to 472 events, we also obtain a tighter constraint on the \texttt{mod$_2$} component. Fig.~\ref{fig:hist_mod} shows the posteriors of $A_{\mathrm{mod,2}}$ and $P_{\mathrm{mod,2}}$ with our best-fit model together with the new dataset extending to two further eruptions (`Best + Aug. data'), and the posterior from the analysis of odd and even separately (see Appendix~\ref{sec:app_OC_oddeven}, corrected with a factor two), for a consistency check. Interestingly, the posterior of $P_{\mathrm{mod,2}}$ appear bound at $1016_{-148}^{+119}\,$P, and $A_{\mathrm{mod,2}}$ consequently improves to a more precise constraint ($A_{\mathrm{mod,2}}=3521_{-938}^{+803}\,$s). The fact that with two further eruptions added in the `Best + Aug. data' the longer modulation is consistent suggests that it is a good description of the timing of the eruptions (we show the related O-C data-model comparison in Appendix~\ref{sec:app_OC} and Fig.~\ref{fig:OCbest+aug}). We note that identifying these eruptions with the best-fit model may only bias the identification toward having additional O-C data within $\pm P/2\sim1.1\,$h around the best-fit model (we refer to Fig.~\ref{fig:OCbestfit} for visualization), but does not bias them toward being very close to the prediction as opposed to anywhere else in the $\sim P\sim2.2\,$h interval around it. This is a useful consistency check that confirms that the longer modulation, if it is the most correct description of the data, is not a spurious artifact driven by the XMM4 eruptions, but it is likely a robust feature in the data. 

\begin{figure*}[t]
    \centering
    \includegraphics[width=0.48\textwidth]{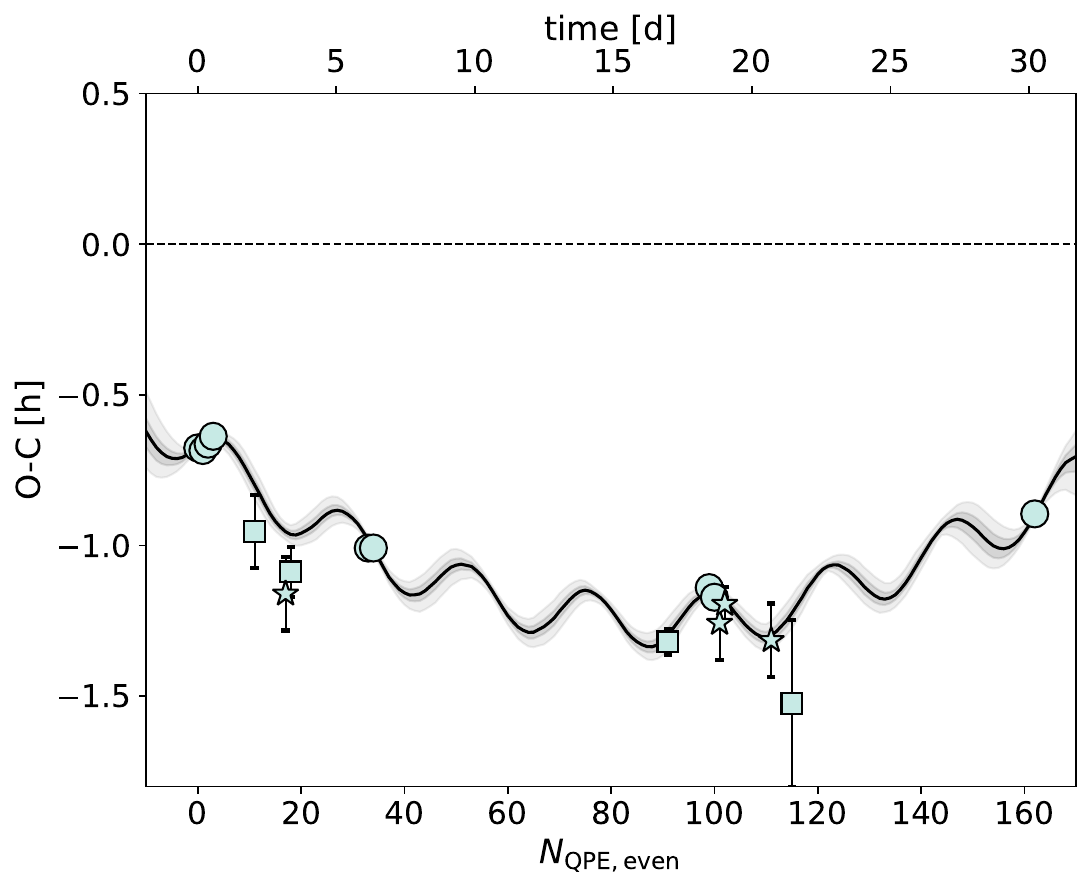}
    \includegraphics[width=0.48\textwidth]{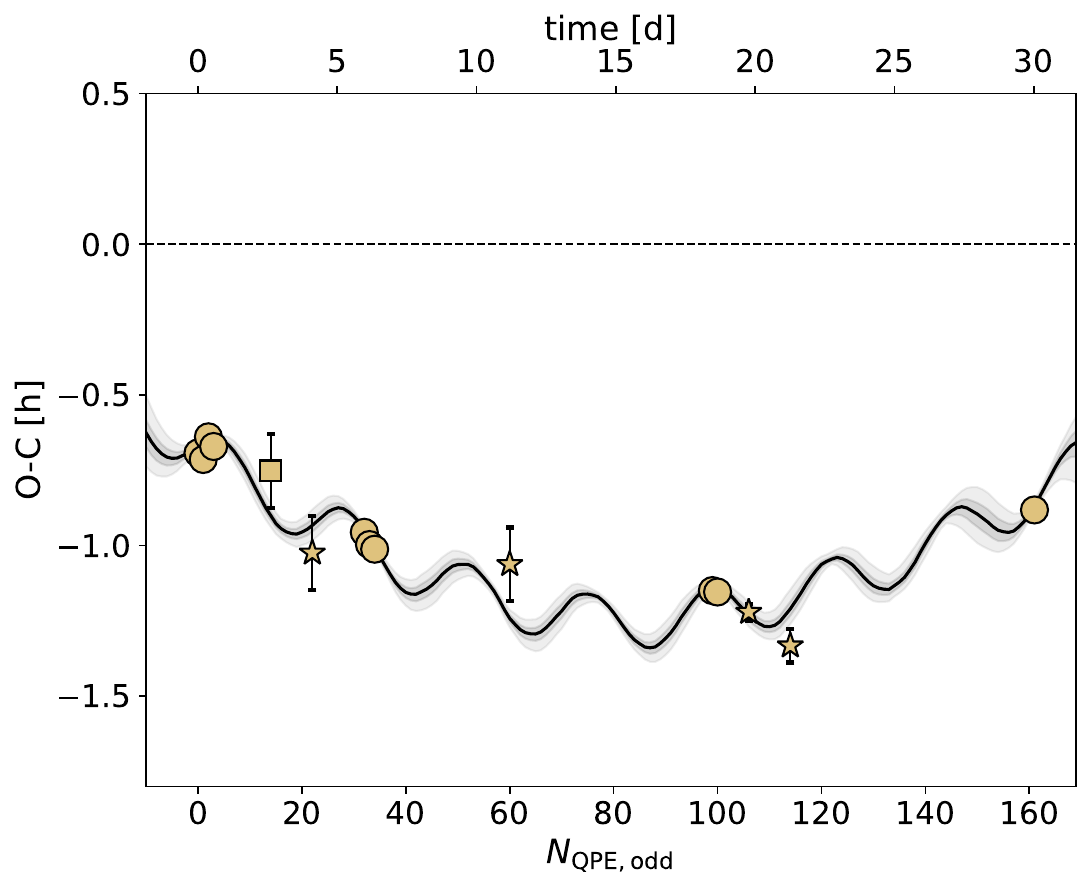}
    \includegraphics[width=0.48\textwidth]{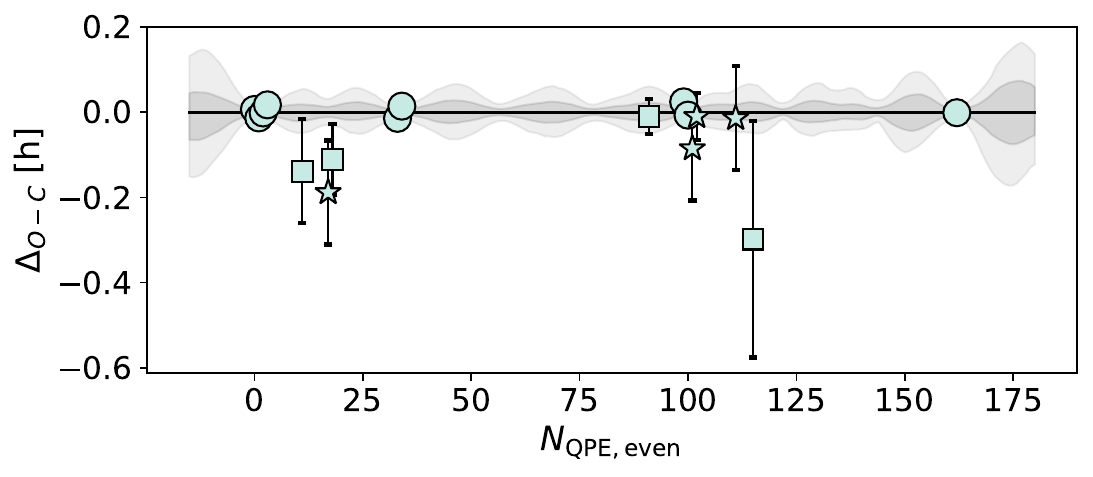}
    \includegraphics[width=0.48\textwidth]{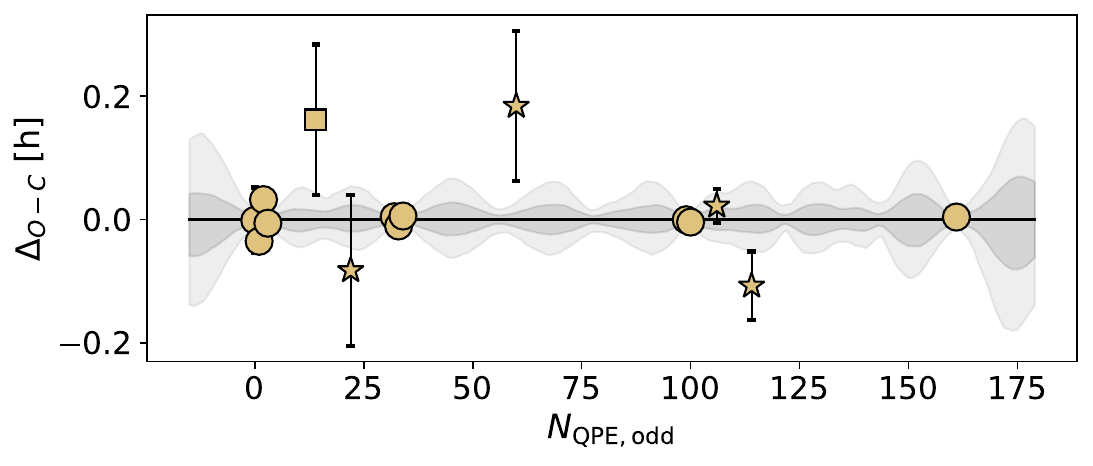}
    \caption{Same as Fig.~\ref{fig:OCbestfit}, but for even (left) and odd (right) eruptions fitted as separate datasets. The parameters $P$, $A_{\mathrm{mod,1}}$, $P_{\mathrm{mod,1}}$, $A_{\mathrm{mod,2}}$, and $P_{\mathrm{mod,2}}$ are fitted in common, while each dataset has its own constant and phase terms. Odd and even eruptions are found in phase on both short and long timescales. The related corner plot is shown in Fig.~\ref{fig:best-fit corner_oddeven}.}
    \label{fig:OCbestfit_oddeven}
\end{figure*}

\subsection{O-C fit with odd and even eruptions separately}
\label{sec:app_OC_oddeven}

The O-C analysis performed on odd and even bursts separately traces the workflow reported in Sect.~\ref{sec:OC}. The burst identification is the same reported in Table~\ref{tab:OCdata}. The O-C event number for odd eruptions is obtained rescaling $N_{\rm QPE}$ from Table~\ref{tab:OCdata} to $N_{\rm QPE,odd} = (N_{\rm QPE} -1)/2$, and those for even eruptions is $N_{\rm QPE,even} = N_{\rm QPE}/2$. $T_{\rm0, est}$ and $P_{\rm est}$ are the same, with the exception of the period now being double that of the fit with all eruptions, namely $16111.42$\,s. The number of free parameters in all the fit changes as follows. Most parameters are fitted jointly (e.g., $P$, $\dot{\rm P}$, $A_{\mathrm{mod}}$, $P_{\mathrm{mod}}$, if present), but constants and phases are not. 

\begin{figure*}[t]
     \includegraphics[width=\textwidth]{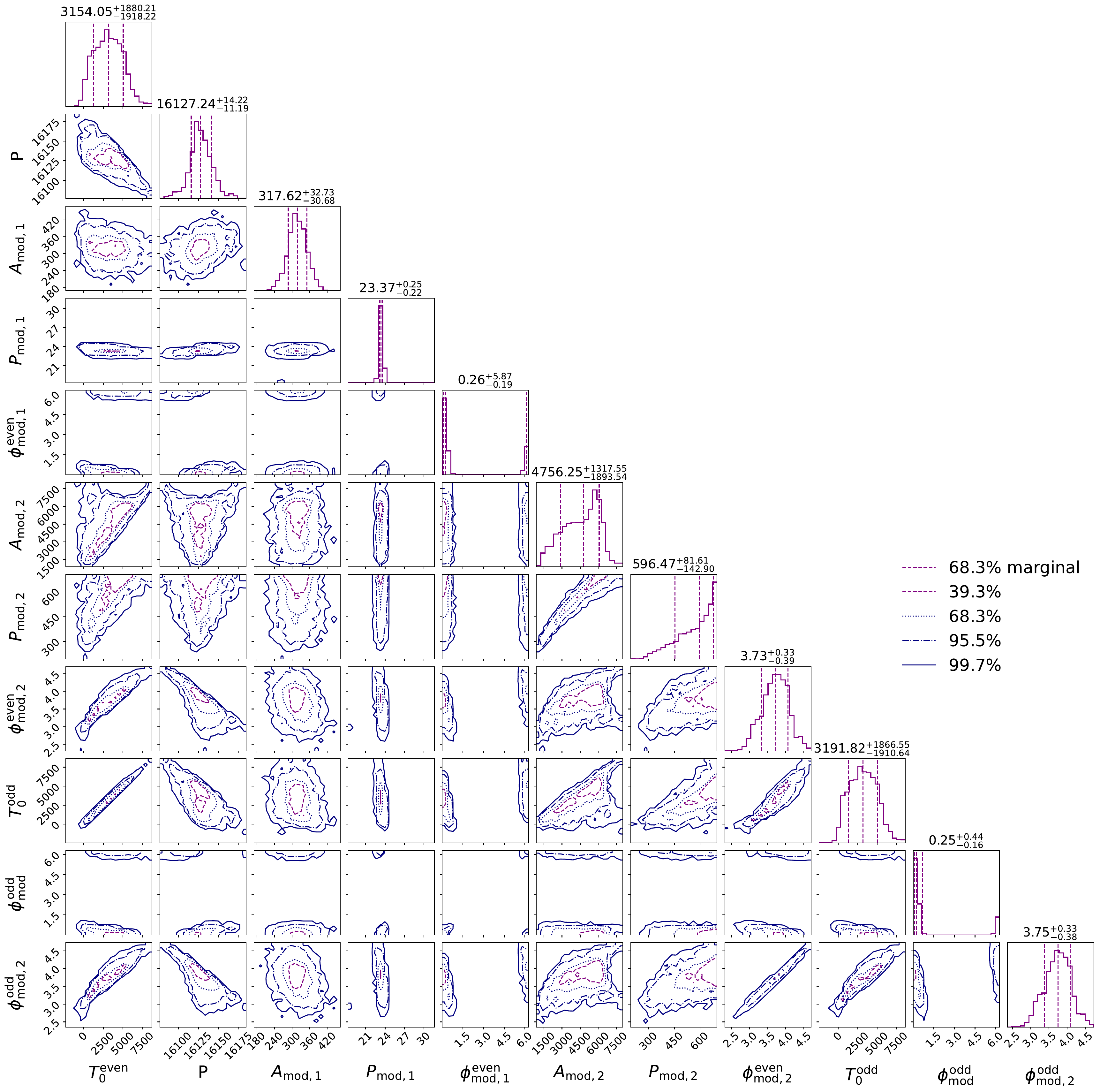}
     \caption{Same as Fig.~\ref{fig:best-fit corner}, but for the fit with odd and even eruptions separately. The best-fit O-C model is shown in Fig.~\ref{fig:OCbestfit_oddeven}.}
     \label{fig:best-fit corner_oddeven}
\end{figure*}

Similarly to the case of all eruptions, for the case of deterministic components the \texttt{lin+mod} is inadequate to reproduce the O-C dataset (Fig.~\ref{fig:OConly_oddeven}), and both \texttt{lin+pdot+mod} and \texttt{lin+mod$_{\rm 1}$mod$_{\rm 2}$} equally improve the fit, with $\Delta \log Z = 113$ and $\Delta \log Z = 114$ with respect to the \texttt{lin+mod} model, respectively. However, the former is discarded as it severely underpredicts the QPE recurrence time of archival 2020-2023 data (see Fig.~\ref{fig:Trecur} for the case of all eruptions). Thus, we adopted the latter as best-fit model, and we show the O-C model in Fig.~\ref{fig:OCbestfit_oddeven} and related corner plot in Fig.~\ref{fig:best-fit corner_oddeven}. The fitted parameters are overall compatible with the fit with all eruptions, modulo a factor two in the periods, for instance the fitted period at the XMM1-4 epoch is $P=16127^{+14}_{-11}\,$s, or $\sim4.48\,$h, which is subtracted from both data and model in Fig.~\ref{fig:OCbestfit_oddeven}. Compatible super-period modulations (modulo the factor two) are fitted for both the \texttt{mod$_{\rm 1}$} and \texttt{mod$_{\rm 2}$} components, namely $P_{\mathrm{mod,1}}=(23.4\pm0.2)\,$P for the faster super period, and $\gtrapprox 454\,$P at $1\sigma$ for the slower. The major result of the fit with odd and even datasets, in relation to their possible interpretation as EMRIs, is that both fitted super-period modulations are consistent with being in phase. It can be see by eye in Fig.~\ref{fig:OCbestfit_oddeven} and from the independent phase values in the corner plot (Fig.~\ref{fig:best-fit corner_oddeven}). More quantitatively, the phase difference between odd and even O-C datasets are $\Delta \phi_{\rm mod, 1} = 0.05 \pm 0.06$ and $\Delta \phi_{\rm mod, 2} = 0.02 \pm 0.02$, respectively. Finally, we added a negative quadratic component, $-(1/2\,\dot{\rm P}\,P)N_{\rm QPE}^2$, to the best-fit to infer an upper limit on $\dot{\rm P}$ (sampling positive values for it). We obtained an upper limit on $\dot{\rm P}<3 \times 10^{-6}\,$s/s, allowing parameters to vary only within the 10th-90th inter-quantile range of the best-fit model.

\begin{figure}
     \includegraphics[width=0.99\columnwidth]{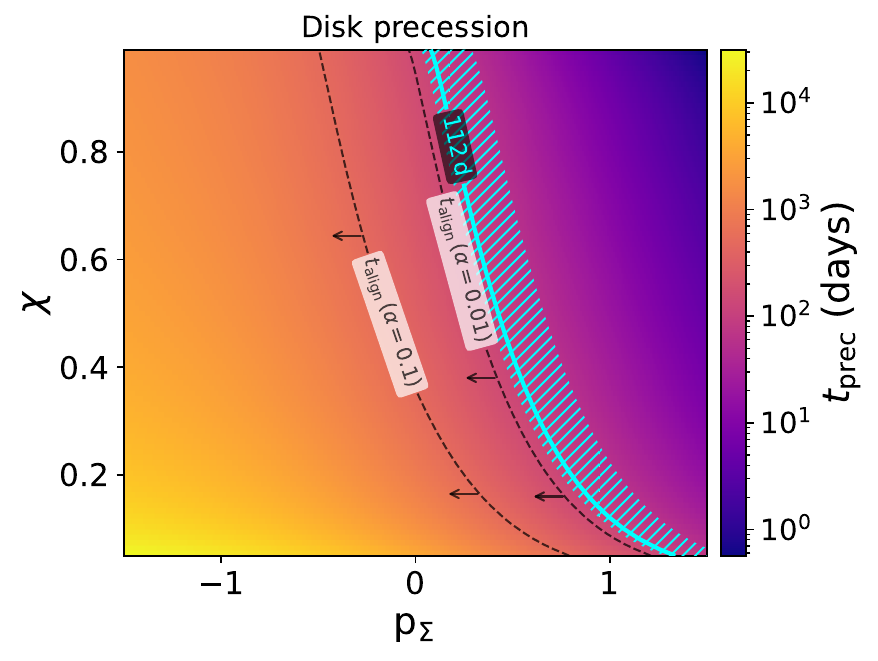}
     \includegraphics[width=0.99\columnwidth]{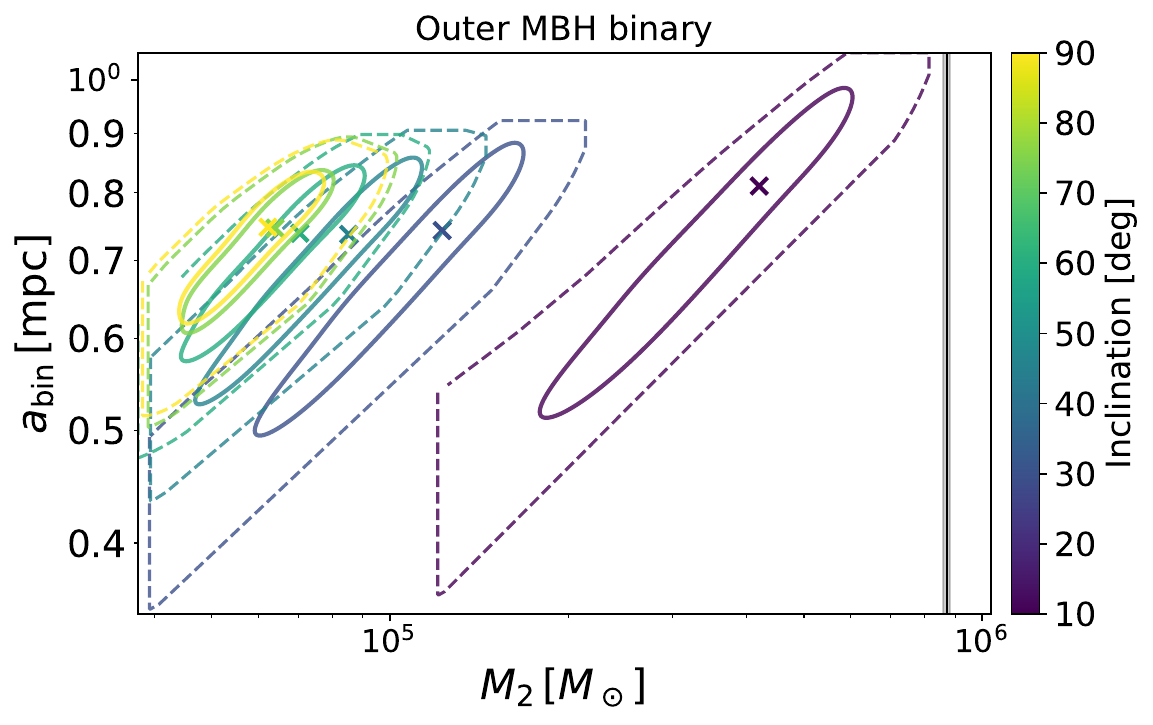}
     \caption{Same as Fig.~\ref{fig:mbhb_all}, but for odd and even eruptions separately, showing that consistent conclusions for the longer modulation term can be obtained.}
     \label{fig:mbhb_oddeven}
\end{figure}

Given the fundamental inconsistency unveiled by the correlated odd and even eruptions, the main body of this articles shows primarily analysis and plots from the case of all eruptions fitted together, appropriate for comparing with single observed events per orbit. However, here we report constraints for EMRI models from the parameters inferred in the fit with separated odd and even eruptions, in case future versions of the QPE=EMRI models will require this setup (appropriate, for instance, if only one parity of QPEs is tracing the crossings). As the ratio between the super-periods parameters and the orbital period is now halved ($P_{\rm mod,1}\sim 23.3$\,P), Eq.~\ref{eq:apsidal_mass} now yields $M_{\rm BH} = (8.7\pm0.1)\times 10^5\,$\Msun. The amplitude of the modulation is $A_{\rm mod,1}\sim 318\,$s, thus compatible with that of the fit with all eruptions, however the estimated $\sim a/c$ time delay for orbit-size delays is larger (it is fewer gravitational radii, but an $R_g$ is now a larger physical size for a larger $M_{\rm BH}$), namely $\sim 297\,s$. This value can be thus achieved also for very low eccentricities. For the longer modulation, EMRI nodal is still an $\sim a/c$ effect thus incompatible with the observed $A_{\rm mod,2}$, and we repeated the tests that led to Fig.~\ref{fig:mbhb_all} with the values inferred for the odd/even case, and generated Fig.~\ref{fig:mbhb_oddeven}. Disk precession is disfavored for with the longer modulation since the parameter space for its solution would imply a disk precession alignment timescale shorter than the QPE lifetime in eRO-QPE2 (top panel of Fig.~\ref{fig:mbhb_oddeven}). For the MBH binary case, no solutions exist for a $M_2>M_1$ for the observed $A_{\rm mod,2}$ and $P_{\rm mod,2}$, and binaries can exist at $M_2<M_1$ for separations $a_{\rm bin}\sim 0.75$\,mpc, which is a factor $\sim260$ larger than the inner EMRI separation $\sim70\,R_g$, thus a stable system. However, the total mass, between a factor one and two of $M_1$, is incompatible with the $M_{\rm BH}-\sigma$ estimate of $\approx 10^5\,$\Msun \citep{Wevers+2024:host}. 

\begin{figure}
     \includegraphics[width=0.99\columnwidth]{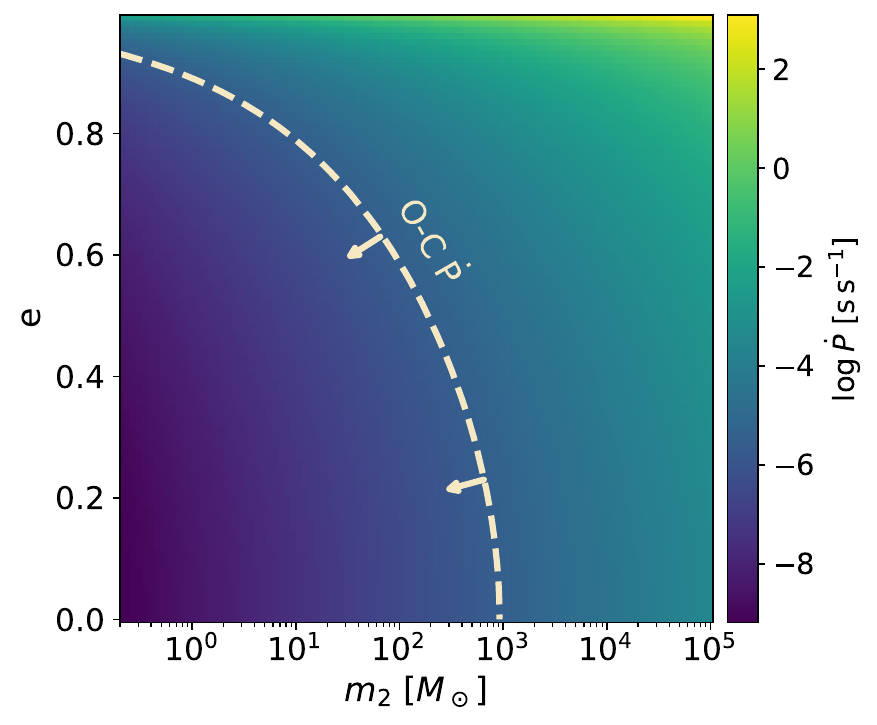}
     \includegraphics[width=0.99\columnwidth]{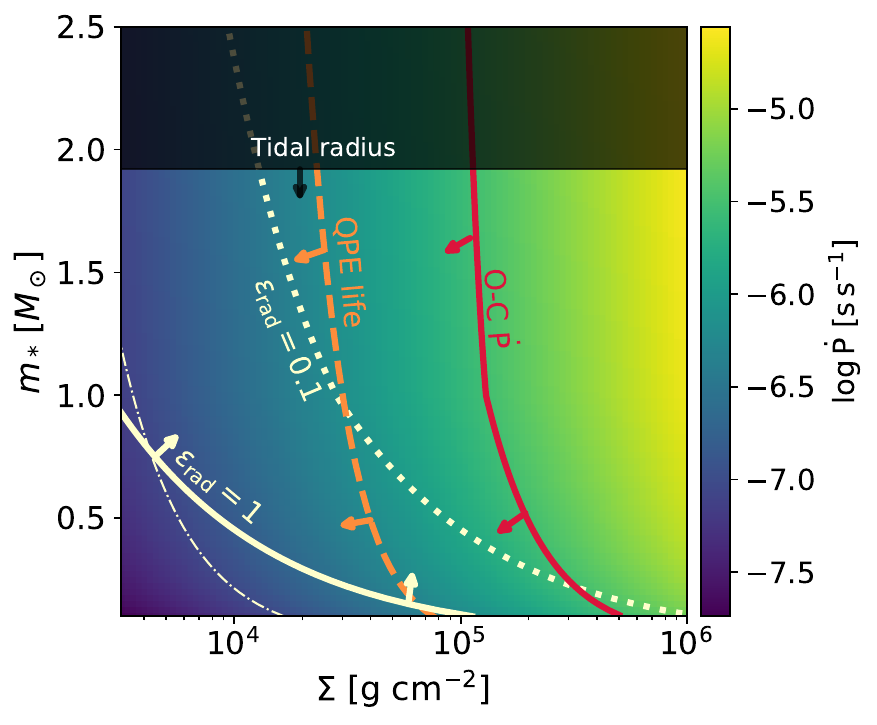}
     \caption{Same as Fig.~\ref{fig:pdot_all} (top) and Fig.~\ref{fig:pdot_drag} (bottom), but for odd and even eruptions separately.}
     \label{fig:pdot_oddeven}
\end{figure}

Finally, the $\dot{P}$ upper limit for the case of odd/even eruptions is $<3\times10^{-6}$, which would still rule out maximum eccentricity at all secondary masses, and most IMBH solutions (top panel of Fig.~\ref{fig:pdot_oddeven}). In the bottom panel of Fig.~\ref{fig:pdot_oddeven} we show the constraints that are obtained from the O-C $\dot{P}$ upper limit and the QPE lifetime in terms of the mass of a star and the density of the accretion disk if the orbital evolution is dominated by gas drag. Compared to Fig.~\ref{fig:pdot_drag}, a main sequence star ``bullet'' has some allowed parameter space even for $\varepsilon\lesssim 0.1$, due to the lower $\Sigma_{\rm min}$ $R=a=70\,R_g$. For $R_{\rm out} \approx 300\,R_g$, appropriate given the estimate from \citet{Wevers+2025:ero2hst}, the range
$\Sigma=(0.5-1)\times 10^5\,$g\,cm$^{-2}$ can be obtained with a range $0.07\lesssim M_{\rm disk,0}/M_{\odot}\lesssim0.17$. Similarly to the case for all eruptions, stellar streams would allow larger cross sections, thus lower $\Sigma_{\rm min}$, thus lower $M_{\rm disk,0}$. Finally, we note that there is a minor difference in some of the slopes as the radius-mass relation breaks to $\sim 0.6$ for supersolar masses (compared to $\sim0.8$ for subsolar). 

\section{More details on the EMRI trajectory models}
\label{sec:app_EMRI}
\renewcommand{\thefigure}{C.\arabic{figure}}
\renewcommand{\thetable}{C.\arabic{table}}
\setcounter{figure}{0}
\setcounter{table}{0}


The first code we used is the publicly available QPE timing analysis code \texttt{QPE-FIT}\footnote{Available at \href{https://github.com/joheenc/QPE-FIT}{https://github.com/joheenc/QPE-FIT}} v. 0.1.11, which performs Bayesian parameter inference for EMRI and SMBH properties using QPE timings under the assumption that they are due to orbiter-disk collisions around a SMBH \citep{Chakraborty+2025:qpefit}. To do so, it assumes a forward model combining an EMRI orbital trajectory $r(t)=(x(t),y(t),z(t))$ with a planar accretion disk having a rigidly precessing normal vector:
\begin{equation}
\hat{d}(t) = \begin{bmatrix}
    \sin\theta_{\rm disk} \cos\Big(2\pi t/T_{\rm disk} + \phi_{{\rm disk},0}\Big) \\
    \sin\theta_{\rm disk} \sin\Big(2\pi t/T_{\rm disk} + \phi_{{\rm disk},0}\Big) \\
    \cos\theta_{\rm disk}
\end{bmatrix},
\end{equation}
where $\theta_{\rm disk}$ is the angle between the disk normal vector and the SMBH spin axis, $T_{\rm disk}$ is the disk nodal precession period, and $\phi_{\rm disk,0}$ is an arbitrary initial phase. In general, $r(t)$ depends on three orbital parameters (semimajor axis $a$, eccentricity $e$, inclination $i$); three initial phase parameters ($\phi_{r,0}$, $\phi_{\theta,0}$, $\phi_{\phi,0}$); and two SMBH parameters which influence the spacetime metric and relativistic timing effects (spin $\chi_\bullet$, mass $\log M_\bullet$). Disk crossings occur when the condition $\vec r(t)\cdot\hat{d}(t)=0$ is satisfied. After computing the disk crossings in the orbiter's frame, their timings are further modified by the Shapiro time delay, a function of $M_\bullet$, and the geometric light travel-time differences, which requires introducing an additional parameter for the observer viewing angle, $\theta_{\rm obs}$. The trajectory $\vec r(t)$ is relatively expensive to compute, as solutions to the geodesic equations in Kerr spacetime are known analytically only in terms of elliptic integrals \citep{vanDeMeent2020:emri_geodesics} which must be solved numerically. Hence, \texttt{QPE-FIT} uses the post-Newtonian orbital solutions for bound Kerr geodesics \citep{Fujita+2009:post_newtonian}, along with GPU-acceleration to simultaneously process large parameter batches in parallel, making the inference problem tractable.

For a given parameter vector $\vec\xi=\{a,e,i,\phi_{r,0},\phi_{\theta,0},\phi_{\phi,0},\chi_\bullet,\log M_\bullet,\theta_{\rm obs},\theta_{\rm disk},T_{\mathrm{disk}},\phi_{\mathrm{disk},0}\}$, \texttt{QPE-FIT} first computes $\vec r(t)$ and $\hat{d}(t)$, then numerically finds the zeros of $r(t)\cdot\hat{d}(t)$, accounts for relativistic/geometric time delays to find the model timings $t_{\mathrm{model},i}$, then combines with the observed QPE timings $\{t_{\mathrm{data},i}\}$ and errors $\{\sigma_{\mathrm{data},i}\}$ to estimate the model likelihood via the chi-squared error:
\begin{align}
\log\mathcal{L}(\vec\xi,\{t_{\mathrm{model},i},\sigma_{\mathrm{model},i}\}) \propto {}\nonumber\\
\qquad -\frac{1}{2}\sum_j^{N_{\rm obs}}\sum_i^{N_{\mathrm{QPE}}}
\bigg(\frac{t_{\mathrm{data},i}-t_{\mathrm{model},i}}{\sigma_{\mathrm{data},i}}\bigg)^2
\end{align}
where $j$ indexes over the $N_{\rm obs}$ total observing windows and $i$ indexes over the $N_{\rm QPE}$ eruptions within each window. Using this log-likelihood function, \texttt{QPE-FIT} performs Bayesian inference to derive posterior probability distributions and the evidence using the nested sampling Monte Carlo algorithm MLFriends \citep{Buchner2019:mlf,Buchner2021:ultranest} using the \texttt{UltraNest} Python package \citep{Buchner2021:ultranest}.

In parallel, we performed an independent fit using an EMRI trajectory model based on the work of \citet{Franchini+2023:qpemodel}, thus named hereafter \texttt{FB23}, but we note that it will be published in Motta et al., (in prep.). Similarly to the \texttt{QPE-FIT} code, we compute the EMRI trajectory $r(t)$ and the time-dependent position of the accretion disk $d(t)$, identify the impact times numerically by solving $r(t)\cdot d(t)=0$, add the time delays, and use the predicted arrival times to evaluate the likelihood function. The system is described by eight free parameters: mass of the MBH $\log M_\bullet$, semi-major axis $a$, eccentricity $e$, orbital inclination $\iota_\mathrm{orb}$, spin $\chi_\bullet$, disk inclination $\iota_\mathrm{disk}$, polar angle of the observer $\theta_\mathrm{obs}$, and a time offset $\Delta t$. The latter encodes the global information on the initial conditions of the system, including the observer’s azimuthal angle, the true anomaly, the longitude of the ascending node, and the argument of the periastron. The accretion disk is assumed to be planar, circular and rigidly precessing, with a power-law density profile $\Sigma(R)$ \citep{Franchini+2016:prec}. It extends from the innermost stable circular orbit out to $R_\mathrm{out}=300R_\mathrm{g}$, and its inclination with respect to the plane perpendicular to the spin remains constant in time, as alignment effects are neglected. Under these assumptions and for a small misalignment angle, the disk precession frequency is computed as the angular-momentum-weighted average of the Lense-Thirring precession frequency $\Omega_\mathrm{LT}$ across the disc. $\Omega_\mathrm{LT}$ at a given disk radius $R$ around a MBH of spin $\chi_\bullet$ reads:
\begin{equation}
\nonumber
    \Omega_\mathrm{LT}(R,\chi_\bullet) = \frac{c^3}{2GM_\bullet} \frac{4\chi_\bullet\left( \frac{R}{R_\mathrm{g}} \right)^{-3/2} - 3\chi_\bullet^2\left( \frac{R}{R_\mathrm{g}} \right)^{-2}}{\chi_\bullet + \left( \frac{R}{R_\mathrm{g}} \right)^{3/2}}
\end{equation}
where $R_\mathrm{g}=GM_\bullet /c^2$ is the MBH gravitational radius. The angular momentum surface density is given by $L(R)=\Sigma(R)\Omega(R)R^2$, with $\Omega(R)$ being the Keplerian orbital frequency of the gas. 

The main difference with \texttt{QPE-FIT} is that the precession period is linked to the MBH spin and the disk properties, providing a stronger constraint to the spin value during the parameter inference. Following \citet{Franchini+2023:qpemodel}, the EMRI trajectory is evolved by integrating the equations of motion within the post-Newtonian framework up to 3.5PN order \citep{Bohe2013:PN, Blanchet2014:PN}, thereby accounting for dissipative effects due to gravitational-wave emission. The equations are numerically integrated over the full temporal baseline of the dataset, and once the crossing times are determined, the code selects the events closest to the observed ones to evaluate the model likelihood. Bayesian inference is performed using the nested sampling algorithm Nessai \citep{Williams2021:nessai}, employing 1000 live points and parallelization over 48 CPU cores. Uniform priors are adopted for all parameters, and the maximum-likelihood estimate is taken as the best-fit solution.

\subsection{Model runs}

\begin{table}[t]
\centering
\caption{Prior ranges for various \texttt{QPE-FIT} runs with two/one flares per orbit (\texttt{QF-2coll}/\texttt{QF-1coll} respectively).} 
\begin{tabular}{lccc}
\toprule
Parameter & Two-flare & One-flare & One-flare \\
 & & (wide) & (narrow) \\
\midrule
$a/R_g$ & $\mathcal{U}(50,1000)$ & $\mathcal{U}(50,1000)$ & $\mathcal{U}(136,187)$ \\
$e$  & $\mathcal{U}(0,0.1)$ & $\mathcal{U}(0,0.5)$ & $\mathcal{U}(0,0.5)$ \\
$i$ & $\mathcal{U}(0^\circ,180^\circ)$ & $\mathcal{U}(0^\circ,180^\circ)$ & $\mathcal{U}(0^\circ,180^\circ)$ \\
$\phi_{r,0}$ & $\mathcal{U}(0,2\pi)$ & $\mathcal{U}(0,2\pi)$ & $\mathcal{U}(0,2\pi)$ \\
$\phi_{\theta,0}$ & $\mathcal{U}(0,2\pi)$ & $\mathcal{U}(0,2\pi)$ & $\mathcal{U}(0,2\pi)$ \\
$\phi_{\phi,0}$ & $\mathcal{U}(0,2\pi)$ & $\mathcal{U}(0,2\pi)$ & $\mathcal{U}(0,2\pi)$ \\
$\chi_\bullet$ & $\mathcal{U}(0,0.998)$ & $\mathcal{U}(0,0.998)$ & $\mathcal{U}(0,0.998)$ \\
$\log(M_\bullet/M_\odot)$ & $\mathcal{U}(4.5,6.5)$ & $\mathcal{U}(4.5,6.5)$ & $\mathcal{U}(5,5.2)$ \\
$\theta_{\rm{obs}}$ & $\mathcal{U}(0,\pi)$ & $\mathcal{U}(0,\pi)$ & $\mathcal{U}(0,\pi)$ \\
$\theta_{\rm disk}$ & $\mathcal{U}(0^\circ,20^\circ)$ & $\mathcal{U}(0^\circ,20^\circ)$ & $\mathcal{U}(0^\circ,20^\circ)$ \\
$T_{\rm disk}/P_{\rm orb}$ & $\mathcal{U}(200,1000)$ & $\mathcal{U}(100,2000)$ & $\mathcal{U}(1000,2000)$ \\
$\phi_{\rm disk,0}$ & $\mathcal{U}(0,2\pi)$ & $\mathcal{U}(0,2\pi)$ & $\mathcal{U}(0,2\pi)$ \\
\bottomrule
\end{tabular}
\label{tab:qpefit_priors}
\end{table}

\begin{table}[t]
\centering
\caption{Prior ranges for the \texttt{FB23} runs with two flares per orbit (\texttt{FB23-2coll}).} 
\begin{tabular}{lc}
\toprule
Parameter & Two-flare \\
\midrule
$a/R_g$ & $\mathcal{U}(50,300)$ \\
$e$  & $\mathcal{U}(0,0.9)$ \\
$i_{\rm{orb}}$ & $\mathcal{U}(0^\circ,90^\circ)$ \\
$\chi_\bullet$ & $\mathcal{U}(0,1)$ \\
$\log(M_\bullet/M_\odot)$ & $\mathcal{U}(3,8)$ \\
$\theta_{\rm{obs}}$ & $\mathcal{U}(0,\pi)$ \\
$i_{\rm disk}$ & $\mathcal{U}(0^\circ,90^\circ)$ \\
$\Delta t/h$ & $\mathcal{U}(0,500)$ \\
\bottomrule
\end{tabular}
\label{tab:fb23_priors}
\end{table}

\begin{figure}
     \includegraphics[width=0.99\columnwidth]{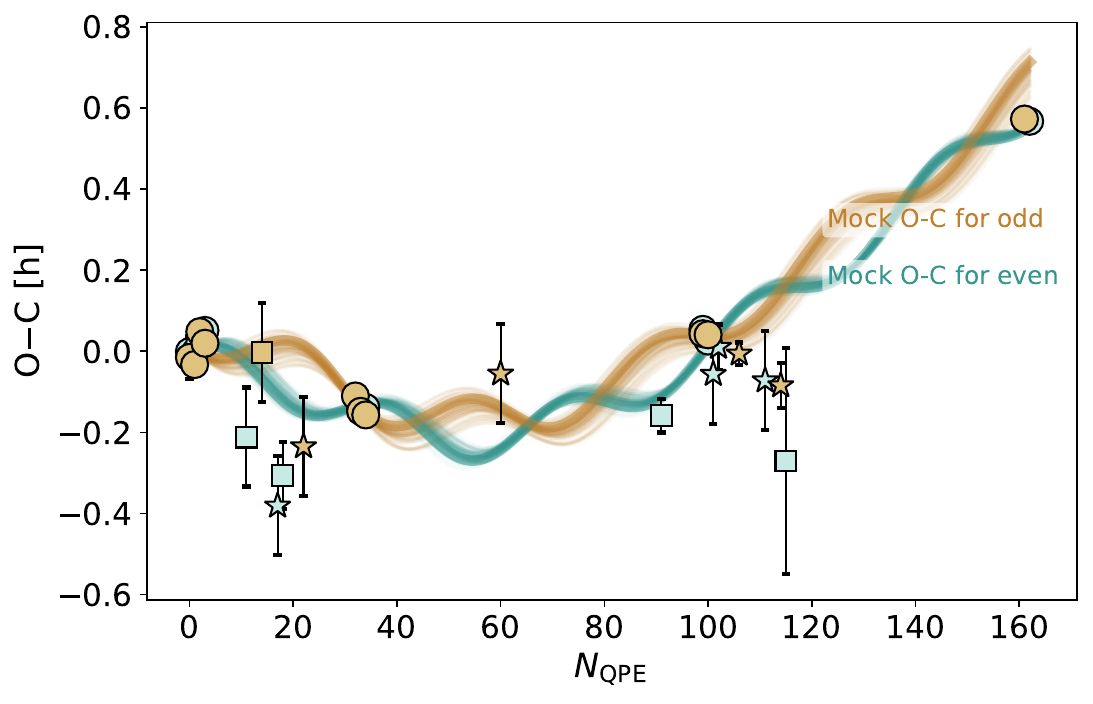}
     \includegraphics[width=0.99\columnwidth]{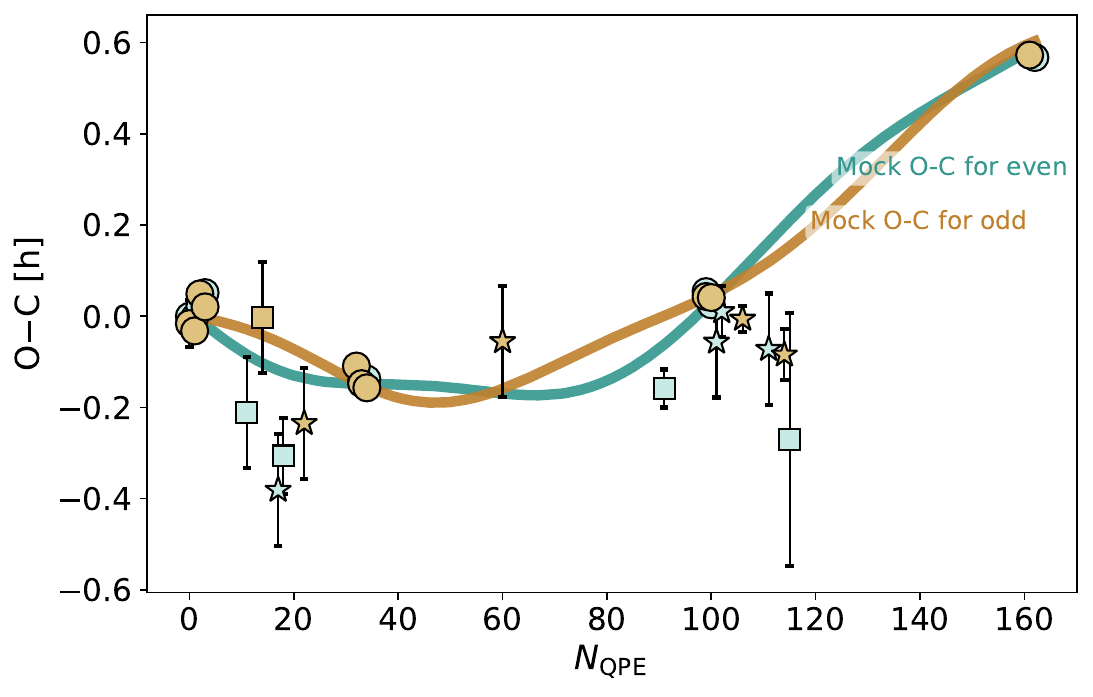}
     \caption{O-C data for odd and even eruptions (from Fig.~\ref{fig:OConly_oddeven}) overlaid with the mock O-C delays from EMRI trajectory best-fit models (\texttt{QPE-FIT} run \texttt{QF-2coll} at the top, the analogous fit with the \texttt{FB23} model at the bottom, \texttt{FB23-2coll}) assuming both odd and even eruptions trace disk crossing. In both cases, the EMRI model generates anti-correlated odd and even delays on the apsidal precession timescale as expected (even if data are correlated). The fits circumvent this inconsistency by fitting the XMM data (circles), which are the most and those with the most accurate uncertainties, at the crossing between odd and even modes of the modeled O-C. Thus, these solutions are not reliable.}
     \label{fig:emrimodel_oddeven}
\end{figure}

When fitting the QPE timings we first assumed a model in which two observable eruptions are generated by the ascending/descending passages of the orbiter through the accretion disk, as is usually invoked in EMRI models (e.g. \citealt{Xian+2021:collisions,Linial+2023:qpemodel,Franchini+2023:qpemodel,Zhou+2024:qpemodel}). We performed this sampling run with both \texttt{QPE-FIT} and \texttt{FB23} models, and name the runs \texttt{QF-2coll} and \texttt{FB23-2coll}, respectively. The priors of this sampling run are given in the second column of Table~\ref{tab:qpefit_priors} for \texttt{QPE-FIT}, and the second column of Table~\ref{tab:fb23_priors} for \texttt{FB23-2coll}. We applied the faster \texttt{QPE-FIT} to the full XMM1-5 dataset, and the fit converged to a solution with parameters $a\sim 91 R_g$, $e\sim 0.04$, $\log M_\bullet\sim 5.8$. For the \texttt{FB23-2coll} run, we applied this setup to the XMM1-4 dataset given the much longer run times in integrating over the full temporal baseline in the \texttt{FB23} model. The fit converged to a solution with parameters $a\sim 205 R_g$, $e\sim 0.03$, $\log M_\bullet\sim 5.2$. Notably, they are two incompatible solutions, but we note that the semimajor axis agrees within a factor 2 in physical sizes. Each of them is at face value reasonable for a system like eRO-QPE2 when compared to the O-C analysis performed on odd/even eruptions separately (Appendix~\ref{sec:app_OC_oddeven}) and $M_{\rm BH}$ values in the literature \citep{Wevers+2024:host,Wevers+2025:ero2hst}. To investigate more in detail their comparison with the empirical O-C data, we overplotted the predicted O-C curves from the best-fit model posteriors to the observed O-C delays in Fig.~\ref{fig:emrimodel_oddeven}. Both solutions have in common that XMM data are fitted at the crossing points between odd and even predicted curves, which is what ultimately led us to consider them unreliable. We refer to the main text (Sect.~\ref{sec:disc_EMRImodels}) for more details on this.

Given that our analysis disfavored EMRI models with two observable collisions per orbit, we have provided a possible interpretation of our O-C results assuming one event per orbit in Sect.~\ref{sec:discussion}. Here, we report out attempts to fit timing data of eRO-QPE2 with \texttt{QPE-FIT}, the fastest of the two algorithms used here, to test whether more robust EMRI trajectory models would find a compatible solution (\texttt{QF-1coll}). In summary, no solution stands the test of either data censoring (varying the application to XMM1-4 data or XMM1-5 data) nor significant prior volume changes. Hence, no fully reliable solution is found for the case of one collision per orbit either. We report four main setups, although many more tests have been done to test the algorithm robustness to data censoring and prior volume.

\begin{figure}
     \includegraphics[width=0.99\columnwidth]{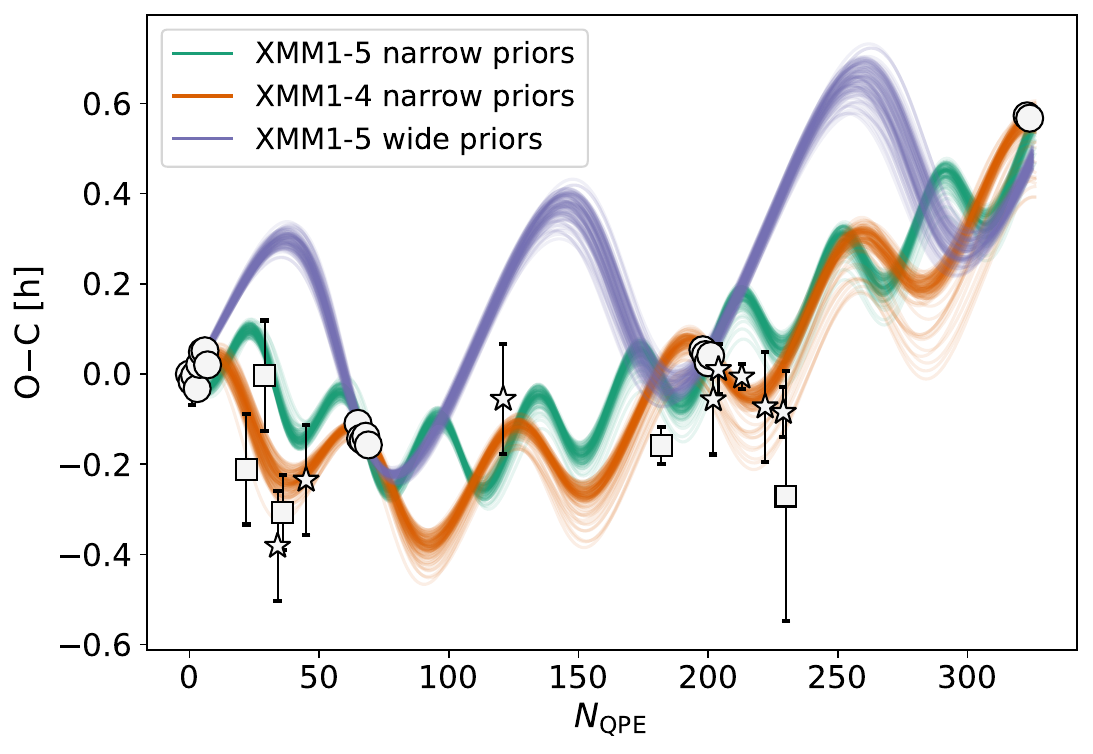}
     \caption{O-C data overlaid with the mock O-C delays from (\texttt{QPE-FIT}. Three separate runs for the one-impact \texttt{QF-1coll} flavor are shown, as described by the legend. The best-fit models from the 6-month XMM1-5 timing data (purple and green lines) do not reproduce the short-term variability shown by XRT and NICER data, with either wide or narrow priors (see Table~\ref{tab:qpefit_priors}). The best-fit model of the XMM1-4 dataset (orange), while reproducing relatively well the O-C delays with narrow priors, does not reproduce the censored data point XMM4 (not in the plot).}
     \label{fig:mockOC_oneimpact}
\end{figure}

We first attempted to find solutions with a wide prior range, choosing bounds on the semimajor axis, eccentricity, SMBH mass, and and disk precession period that were agnostic to the empirical fits (second column of Table~\ref{tab:qpefit_priors}). These fits terminated at a $\log Z=-399\pm0.5$, converging to parameters $a\approx (352\pm 3) R_g$, $e\approx 0.34\pm0.01$, $\log M_\bullet\approx 4.6\pm0.01$, and $T_{\rm disk}\approx (1386\pm106)P_{\rm orb}$. It is immediately apparent that this fit does not recover the same apsidal period as the empirical fits, instead showing a shortest oscillation timescale of $a(1-e^2)/(3R_g)\approx 104P_{\rm orb}$. More quantitatively, this solution ignores the short-term timing variation shown by intra-XMM observations in the XMM1-4 epoch, mainly shown by NICER and XRT data. Perhaps more worryingly, once only the XMM1-4 data are fitted (thus censoring any data post XMM4) the solution changes. To investigate whether more stability to data censoring would be obtained with a narrow parameters volume, we restricted the prior bounds to values compatible with the O-C solution (third column of Table~\ref{tab:qpefit_priors}). A solution is found with $\log Z=-321\pm 0.4$, with $\Delta \log Z = 78$ compared to the runs with the wider bounds, indicating that the multi-modality and extremely small posterior/prior volume ratio of this problem presents a challenge for even robust statistical inference algorithms such as nested sampling. The main fitted parameters are $a = (141\pm0.5) R_g$, $e=0.12\pm0.01$, $\log M_{\rm BH}/M_\odot = 5.194\pm0.002$, and $P_{\rm disk}/P=1364^{+53}_{-211}$. While they appear in remarkable agreement with the O-C parameters (corner plot in Fig.~\ref{fig:best-fit corner}), once the predicted O-C delays are superimposed to the observed one (similarly to the previous test done with the two-impact scenario, Fig.~\ref{fig:emrimodel_oddeven}), again the medium-term variability highlighted by XRT and NICER data is largely ignored. Again quite worryingly, censoring data post XMM4, changes the fitted parameters to $a \sim  184 R_g$, $e\sim0.16$, $\log M_{\rm BH}/M_\odot \sim 5.0$. We show the super-position between mock and observed O-C delays from the narrow prior runs in Fig.~\ref{fig:mockOC_oneimpact}, as an example. Similar inconsistencies are obtained for the wider prior volume. Finally, we note that we explored different volume sampling algorithms within the \texttt{Ultranest} code with no change.

\bibliography{sample631}{}
\bibliographystyle{aasjournal}



\end{document}